\documentstyle[12pt]{article}
\normalbaselines

\begin{document}
\setcounter{secnumdepth}{5}
\setcounter{tocdepth}{5}
\renewcommand{\theequation}{\mbox{\arabic{section}.\arabic{equation}}}

\begin{flushright}

{\raggedleft NTZ 16/1993\\}

{\raggedleft hep-th/9312137\\[4.cm]}

\end{flushright}

\renewcommand{\thefootnote}{\fnsymbol{footnote}}
\begin{center}
{\LARGE\baselineskip1.1cm A Functional Integral Equation for the
Complete Effective Action in\break Quantum Field Theory \\[2.cm]}
{\large K.\ Scharnhorst\footnote[1]{E-mail:
kjsch @ qft.physik.uni-leipzig.d400.de}}\\[0.3cm]
{\small Universit\"at Leipzig\footnote[2]{temporary address}

Naturwissenschaftlich-Theoretisches Zentrum

Augustusplatz 10/11

D-04109 Leipzig

Federal Republic of Germany}\\[3.cm]
\thispagestyle{empty}
\end{center}
\renewcommand{\thefootnote}{\arabic{footnote}}

\newpage

\  \\

\vspace{1.cm}
\renewcommand{\baselinestretch}{1.5} {\it  }

\begin {abstract}
\noindent
Based on a methodological analysis of the effective
action approach certain conceptual foundations of
quantum field theory are reconsidered to establish
a quest for an equation for the effective action.
Relying on the functional integral formulation of
Lagrangian quantum field theory a functional
integral equation for the complete effective action
is proposed which can be understood as a certain
fixed point condition. This is motivated by a critical
attitude towards the distinction artificial from
an experimental point of view between classical
and effective action. While for free field theories
nothing new is accomplished, for interacting theories
the concept differs from the established
paradigm. The analysis of this new concept is concentrated
on gauge field theories treating QED as the prototype
model. An approximative approach to the functional
integral equation for the complete effective action
of QED is exploited to obtain certain nonperturbative
information about the quadratic kernels of the action.
As particular application the approximative calculation
of the QED coupling constant $\alpha$ is explicitly studied.
It is understood as one of the characteristics
of a fixed point given as a solution of
the functional integral equation proposed.
Finally, within the present approach the vacuum energy
problem is considered and possible implications on the
induced gravity concept are contemplated.\\[0.6cm]
\end{abstract}
\renewcommand{\baselinestretch}{1.} {\it  }

\newpage
\section*{Contents}

\contentsline {section}{\numberline {1}Introduction}{\pageref{AS}}
\contentsline {section}{\numberline {2}An Equation for the Complete
Effective Action}{\pageref{BS}}
\contentsline {subsection}{\numberline {2.1}Do We Need an Equation?
--- Methodological Considerations}{\pageref{BS1}}
\contentsline {subsection}{\numberline {2.2}Proposing an Equation
for the Complete Effective Action}{\pageref{BS2}}
\contentsline {subsection}{\numberline {2.3}Exploring the
Equation}{\pageref{BS3}}
\contentsline {section}{\numberline {3}Gauge Field Theories}{\pageref{CS}}
\contentsline {subsection}{\numberline {3.1}Ward-Takahashi
Identities}{\pageref{CS1}}
\contentsline {subsection}{\numberline {3.2}Schwinger-Dyson
Equations}{\pageref{CS2}}
\contentsline {section}{\numberline {4}QED --- An Approximative
Approach to the Equation for the Complete Effective Action}{\pageref{DS}}
\contentsline {subsection}{\numberline {4.1}The Approximative
Approach in General}{\pageref{DS1}}
\contentsline {subsection}{\numberline {4.2}Designing an
Approximation Strategy}{\pageref{DS2}}
\contentsline {subsubsection}{\numberline {4.2.1}Consequences of
Gauge Invariance for the Kernel of the Fer\discretionary {-}{}{}mion
Action $\Gamma _I^F$}{\pageref{DS21}}
\contentsline {subsubsection}{\numberline {4.2.2}Requirements on
the Kernel of the Gauge Field Action $\Gamma _I^G$}{\pageref{DS22}}
\contentsline {subsubsection}{\numberline {4.2.3}The Approximation
Strategy in Ideal, and in Practice}{\pageref{DS23}}
\contentsline {subsection}{\numberline {4.3}Bringing the
Approximation Strategy to Work: Explicit Calculation}{\pageref{DS3}}
\contentsline {subsubsection}{\numberline {4.3.1}Performing the
Functional Integration}{\pageref{DS31}}
\contentsline {subsubsection}{\numberline {4.3.2}The Integral
Equation for the Kernel of the Fermion Action}{\pageref{DS32}}
\contentsline {paragraph}{\numberline {4.3.2.1}Solving the
Integral Equation in the Asymptotic UV Region}{\pageref{DS321}}
\contentsline {paragraph}{\numberline {4.3.2.2}Solving the
Integral Equation in the Asymptotic IR Region}{\pageref{DS322}}
\contentsline {subsubsection}{\numberline {4.3.3}The Fixed Point
Condition for the Kernel of the Gauge Field Action and the
Approximative Calculation of the QED Coup\-ling Constant
$\alpha $}{\pageref{DS33}}
\contentsline {section}{\numberline {5}The Vacuum Energy,
and Related Problems}{\pageref{ES}}
\contentsline {section}{\numberline {6}Discussion and
Conclusions}{\pageref{FS}}
\contentsline {section}{Appendix A}{\pageref{GS}}
\vspace{-0.3cm}
\contentsline {section}{Appendix B}{\pageref{HS}}
\vspace{-0.3cm}
\contentsline {section}{References}{\pageref{IS}}
\vspace{-0.3cm}
\contentsline {section}{Figures}{\pageref{JS}}

\newpage

\section{\label{AS}Introduction}
\setcounter{equation}{0}

Physical reality can be approached by means of quantum
field theory from different perspectives. This in
particular depends on the kind of information one
is interested to extract in order to solve a problem under
consideration but it is also influenced by the
individual view toward the fundamental difficulties
met in present day standard quantum field theory
(and its generalized concepts like string theory).
To a large extent, these different approaches reflect technical
difficulties to fully (in particular, nonperturbatively)
understand quantum field theoretical models rather than really
differences in concept on a fundamental level.
However, few pioneers of quantum field theory
like {\sc Dirac} \cite{dira1},\cite{dira2} and
{\sc Feynman} \cite{feyn1},\cite{feyn2}
in particular pointing to the UV divergency
problem always maintained the view that the right
theory has not yet been found. This attitude has
apparently not received majority support in time
but in this respect it does not seem to exist any
majority opinion at all \footnote{For a description of
the attitude in one large part of the community see ref.\
\cite{shir}, e.g..}. From this state of affairs we
feel free to draw justification for a reconsideration of certain
conceptual foundations of quantum field theory
constituting the purpose of the present paper.\\

Notwithstanding above mentioned problems,
it seems to exist wide agreement that the scattering
matrix can be considered as the fundamental object for
describing a particular quantum field theoretic model.
This amounts to saying that full knowledge of the complete
scattering matrix is considered equivalent to the solution
of a quantum field theory and all interesting information,
at least in principle, can be extracted from it.
Construction of the scattering matrix can be attempted by
different methods. For instance, the so-called S-matrix
theory as studied in the 1950s in reaction to the
emergence of the divergency problem
in Lagrangian quantum field theory was designed
to find the (finite) scattering matrix from rather
general fundamental principles like causality, unitarity,
Lorentz invariance using dispersion techniques without
making reference to any Lagrangian underlying the theory
(see \cite{brow},\cite{eden}, e.g.). However,
although quite general and interesting
results have been obtained principles applied turned out
not restrictive enough to completely fix the scattering
matrix for realistic theories.
Nowadays, after the successful re-emergence of (renormalizable) Lagrangian
quantum field theory at the end of the 1960s description
of the scattering matrix is supplied in a standard way in
terms of the effective action of the theory
considered \cite{b}. In this sense, we may view the
effective action as the genuine fundamental object
of interest and will concentrate on its study in this article.\\

Historically, beyond the S-matrix theory already mentioned
attempts to cure UV divergencies by nonlocal
field theories played a particular role since the emergence
of the divergency problem in the 1930s (for a review including
references see \cite{efim1},\cite{efim2}; also \cite{efim3}).
Although it has been recognized
early that nonlocal field theories may be accompanied by
new, perhaps even more unpleasant difficulties, so with
unitarity and (macro-)causality, theoretical thinking in
this direction never ceased to exist. Most prominent,
present day string theory although much more
ambitious can be viewed as a particular way of
giving preference to a special kind of nonlocality \cite{elie}.
In recent years, few papers were again dealing with
nonlocal quantum gauge field theories \cite{part}--\cite{corn}
(to mention only this subject)
where in part the nonlocalities introduced are understood
as regulators. Although having a different aim than fighting
UV divergencies, also the average action concept proposed
recently should be mentioned here \cite{wett1},\cite{wett2}.
In principle, the drawback of all these nonlocal
approaches however consists in the arbitrariness in the choice
of the nonlocality introduced. So far, no unique recipe
starting from first principles has been proposed.\\

However, the dominant paradigm in the field remains
local renormalizable Lagrangian quantum field theory
(Throughout the paper we will denote it by the term
standard quantum field theory.). But, also there
nonlocality is a well-known phenomenon because it is a feature
of the effective action that can be derived for
any quantum field theory (either local or nonlocal)
and which also serves (in most cases) as generating
functional of the one-particle-irreducible (1PI) Green
functions. In general, the effective action is
attributed different meanings by different authors.
Few regard the effective action as some low energy
representation of a quantum field theory obtained
by integrating out certain (massive) degrees
of freedom, while others consider the effective
action as a full fledged description of the model
under investigation from which arbitrary S-matrix
elements (related to any observation one might be
able to perform) can be derived. We will stick here
to the latter view. To us, very pragmatically the
effective action is that object which contains all
the information ever to be measured under certain
defined circumstances and there is no other
(independent) object linking theory to physical
reality. The shape of the effective action of
course may depend on some of these circumstances (external
conditions, e.g.). A similar point of view has recently
been described with respect to the gravitational
effective action by {\sc Vilkovisky} \cite{vilk1}.
The effective action concept we have in mind aims
at quantum field theoretic models, especially
those which are realistic like QED, and assumes that
certain sectors of physical reality can be described
in a consistent way independently of each other.
It is therefore quite different from the TOE ('theory of everything')
concept often related to superstring theory.\\

In short, the program of the present article can be
described by saying that we intend to find a
concept which allows to determine the structure of
the (highly complex) observable 'effective action'
without making reference to any other quantity not
accessible to observation. In particular, the approach to quantum
field theory will be based on a critical
attitude towards the distinction
artificial from an experimental point of view
between the so-called classical action and the
effective action. This way we will be lead
to propose an equation for determining the (finite)
effective action, which can be understood
as a certain fixed point condition. It will be an equation for
functionals of fields (actions) and is therefore designed to remove
(to a certain extent --- the field content has to be
prescribed as usual) the arbitrariness in the
choice of the Lagrangian standing at the beginning
of any field theory. Such however can only be expected
to happen for interacting theories, where our approach differs
from the established paradigm. For free field theories,
where this is not the case,
nothing new is accomplished in this respect. As technical
tool we rely on the functional integral formulation of
Lagrangian quantum field theory which seems to be
the appropriate and most convenient language for
the description of our concept. While nonlocality will be
an inherent feature of our approach in most cases,
it is by no means the conceptual starting point
of the present investigation. Of course, the program as
just sketched is an abstract one. However, once we have
proposed the general concept it will simply serve us
as a guiding line for finding an appropriate
approximative approach to perform explicit calculations
(in this article: in QED as the prototype gauge field theory).\\

In the past decade the effective action concept has
received interest from the point of view of its
invariant geometrical formulation. This is
an important step in ensuring the physical relevance
of the effective action because its physical
consequences should not depend on the particular
choice of coordinates for the field variables. Initial work in this
direction traces back to {\sc Vilkovisky} \cite{vilk2},\cite{vilk3}
and {\sc DeWitt} \cite{dewi}, for a recent discussion
of the geometrical effective action see \cite{camb},
for a review including further references \cite{buch}.
For the purpose of the present article (to reduce complexity
of the considerations) we simply bypass the subject and
maintain that always those field coordinates are
applied in terms of which the formalism takes it
naive (non-geometrical) shape. Furthermore, for
gauge field theories, a main concern of the unique
(geometrical) effective action concept, we find that
generalized Landau gauge is the only sensible gauge.
Inasmuch as for gauge field theories the geometrical
effective action has been found to agree with the
naive one (calculated by means of the standard background field method)
exactly for generalized Landau gauge we
feel free to ignore the subject also there
\cite{frad}--\cite{nach}.\\

The outline of the article is as follows. In chapter 2
we explain the general concept in some length.
This is done in three
steps. In section 2.1 based on a methodological analysis
we establish a quest for an equation for the complete
effective action. While section 2.2 serves to suggest
a particular answer to this question by imposing
a certain fixed point condition in terms of a
functional integral equation section 2.3
discusses some features of this equation, among
others the relation between standard quantum field
theory and the present approach. Chapter 3
then applies the concept to gauge field theories.
Specifically, there we formulate the functional
integral equation for the complete effective action
of QED and then in sections 3.1 and 3.2 Ward-Takahashi identities
and Schwinger-Dyson equations are discussed respectively.
Chapter 4 contains the major body of the explicit
calculation performed. The model under investigation
is QED in 4D Minkowski (Euclidean) space. While section 4.1 spells
out what kind of approximative approach to the functional
integral equation for the complete effective action of QED
is applied in general section 4.2 and its subsections
serve to establish a more concrete approximation strategy
suited for explicit calculation and are concentrating on
the quadratic kernels of the action. Subsections 4.2.1
and 4.2.2 discuss certain general requirements on the
quadratic kernels of the fermion and gauge field actions
respectively while subsection 4.2.3 explains the
approximation strategy finally chosen.
Section 4.3 also split into several subsections
then presents the explicit calculation in some detail.
Subsection 4.3.1 contains technical details of the
functional integration performed. While for the quadratic kernel
of the gauge field action we rely on a certain Ansatz
subsection 4.3.2 establishes an integral equation for the quadratic
kernel of the fermion action. This integral equation is then
approximatively solved in subsections 4.3.2.1 and 4.3.2.2 in
the asymptotic UV and IR regions respectively. This analysis
yields certain nonperturbative information about the quadratic
kernel of the fermion action.
In the final subsection 4.3.3 of chapter 4 as particular
application of the present method the approximative calculation
of the QED coupling constant $\alpha$ is explicitly studied.
It is understood as one of the characteristics
of a fixed point given as a solution of
the functional integral equation proposed.
Certain technical details of the
calculation described in chapter 4 are deferred
to two Appendices at the end of the article.
Chapter 5 shortly discusses the vacuum
energy problem for QED on the 1-loop level. Final
consideration then is devoted to the relevance of the proposed
approach to the induced gravity concept. The article
closes in chapter 6 with a discussion of some aspects of the results
obtained.\\

\newpage

\section{\label{BS}An Equation for the Complete Effective
\hfill\break Action}
\subsection{\label{BS1}Do We Need an Equation? \hfill\break
\hspace*{0.4cm} --- Methodological Considerations}
\setcounter{equation}{0}

As introductory step let us
begin with displaying key elements of the standard
formulation of the effective action.
We consider Lagrangian quantum field theory
in flat (Minkowski) space-time and in this chapter we
use scalar field theory to pursue the discussion. Hereby,
it is understood that generalization to more complicated
theories (in particular, gauge field theories) can be performed
merely by standard means.\\

Construction starts with the generating functional of Green
functions
\parindent0.cm

\begin{equation}
\label{BA1}
Z[J] =\ C \int D\phi\ \ {\rm e}^{\displaystyle\ i\Gamma_0 [\phi]\
+\ i \int dx J(x) \phi (x)}
\hspace{1.5cm},\hspace{0.5cm}\\
\end{equation}

where $\Gamma_0 [\phi]$ is the so-called classical action of
the theory and $C$ some fixed normalization constant. Then,
the generating functional of the connected Green functions is

\begin{equation}
\label{BA2}
W[J]\ =\ -i \ln Z[J] \hspace{1.5cm}.\hspace{0.5cm} \\
\end{equation}

The effective action $\Gamma [\bar\phi]$ which also is
the generating functional of the one-particle-irreducible
(1PI) Green functions is obtained as the first Legendre
transform of $W[J]$.

\begin{equation}
\label{BA3}
\Gamma [\bar\phi]\ =\ W[J] - \int dx J(x) \bar\phi (x) \\
\end{equation}

Here,

\begin{equation}
\bar\phi (x)\ =\ {\delta W[J]\over \delta J(x)} \\
\end{equation}

is understood which in turn leads to

\begin{equation}
\label{BA4}
{\delta \Gamma [\bar\phi]\over \delta \bar\phi (x)}\ =\ -\ J(x)\\
\end{equation}

in analogy to the classical field equation for $\Gamma_0 [\phi]$.
Equivalently, using above relations following formula for the
effective action can be considered as the defining one

\begin{equation}
\label{BA5}
{\rm e}^{\displaystyle\ i \Gamma [\bar\phi]}\ \ =\ C
\int D\phi\ \ {\rm e}^{\displaystyle\ i\Gamma_0 [\phi + \bar\phi]\
+\ i \int dx J(x) \phi (x)}
\hspace{1.5cm} ,\hspace{0.5cm}\\
\end{equation}

where the r.h.s.\ of above equation has to be
calculated at a current $J(x)$ which is a functional
of $\bar\phi $ and given by eq.\ (\ref{BA4}).
Therefore, as the r.h.s.\ is a functional of
both $J$ and $\bar\phi $ eqs.\ (\ref{BA4}), (\ref{BA5}) have to
be understood as functional integro-differential
equations for determining the (off-shell) effective
action and give an implicit definition only. But, as we
will argue below eq.\ (\ref{BA5}) is not an
equation in the narrow sense of the meaning of the word,
instead it rather should be called a formula.\\

\parindent1.5em
The latter point is barely discussed in the
literature and shall now be considered from a
methodological point of view. Observe that eq.\ (\ref{BA1})
defines a map
$g_1: \Gamma_0 [\phi ]\longrightarrow Z[J]$ from the
class of functionals called classical actions to the
class of functionals $Z$. Furthermore, we have
mappings $g_2: Z[J]\longrightarrow W[J]$ (eq.\ (\ref{BA2}),
single-valued up to the uninteresting for the present purpose
fixing of the sheet of the Riemann surface) and
$g_3: W[J]\longrightarrow \Gamma [\bar\phi]$
(eq.\ (\ref{BA3})). These three maps together define a
map $g_3\circ g_2\circ g_1 = f: \Gamma_0 [\phi ]
\longrightarrow \Gamma [\bar\phi] $ (eq.\ (\ref{BA5}))
from the set of so-called classical actions to the set of
effective actions. In total, this map is unique up to
the renormalization problem which can always be treated
in the present context by applying an appropriate
regularization procedure for properly handling
the divergencies. Inasmuch as this map $f$ is
constructed explicitly eq.\ (\ref{BA5}) is not a
genuine equation with possibly a variety of solutions
but rather expresses the image $\Gamma $ of $\Gamma_0$
with respect to the map $f$ --- it is a formula.\\

Above consideration justifies following view. Once the
functional integral measure is constructed (and typically
this is done for a whole class of classical actions and then
fixed forever) the classical action $\Gamma_0$ uniquely
determines the corresponding effective action
$\Gamma $. In other words, the effective action
does not contain more information than (implicitly)
contained in the classical action (supplemented by the
functional integral measure). This point is usually not
stressed in studying concrete models due to the
calculational complexity involved. Although it is of next
to no practical (i.e., calculational) relevance
therefore, it involves important
methodological implications. The most important one
consist in the fact that the effective
action does not appear as object in its own right
but as a derived quantity only. Mere reformulations
of the calculational tools used to determine the effective
action, like Schwinger-Dyson equations, e.g., do not
change this character.\\

Before proceeding further let us
mention that formulas (\ref{BA1})--(\ref{BA5}) reflect
two features of modern quantum field theory. On one
hand side, they stand for the convincing success of
quantum field theory as witnessed in the last few
decades, of a theory providing us with
operational instruments
producing numbers which agree with measurement to a
degree not seen elsewhere in physics (or in any other
science). On the other hand, they also stand for the
fundamental conceptual difficulties inherent to local quantum
field theory. The most important of them is manifesting
itself in terms of the well-known ultraviolet
divergencies. Although there exist two
or three (different) mainstream opinions with respect to this issue
(and some other, related ones)
numerous dissenting ones can also be
found. From this observation one may conclude that
research apparently has not yet lead to any generally accepted concept
explaining and removing the problems in a finally
convincing manner as judged from physics as an inherently
consistent building combining theory and experiment.
This amounts to saying that search
in different directions seems justified and even certain
doubt in the foundations of quantum field theory should
not be rejected at once. With this in mind, in what
follows we will apply the point of view that perhaps
even certain foundations of quantum field theory are
not understood up to their end and we will see whether
we can throw different light on them. In this context,
as outlined in the Introduction, we will focus on the
effective action which we consider as the appropriate
object to be studied.\\

Let us ask for principles effective actions should be
governed by in general. While we have no problem in giving
principles they should obey, like Lorentz invariance, CPT
invariance, e.g., the answer to the question what they are
in detail determined by in view of considerations given
further above reduces to saying that they are uniquely given
as image of the corresponding classical action by means of the
map $f$ containing information about the functional integral measure.
This way, the question is traced back to the uncertainty
in classical field theory what Lagrangian to choose.
Although, one does not necessarily need to worry about
this point here we will. Basically, we prescribe an
effective action in terms of some low
energy information rather than to find it from independent
(quantum) principles not exhausted by fixing the classical
action. And, if we are honest, at best we may say that our
prescription is approximately right.\\

One may now confront the methodological insight
obtained so far with the deductive idea often
applied in theoretical physics that the special
case (here the classical action) should be derived
from the more general one (here the effective
action) and not the other way around. In this sense,
the complete effective action is the genuine
fundamental object to be studied.
If so, up to further investigation,
one is willing to allow
that the complete effective action might be an object in
its own right \footnote{Of course, any
effective action has a certain classical
limit but coincidence of its classical limit
with that of another effective action does not
necessarily entail identity of both effective actions then.}
then one has to find a method of determining the complete
effective action differing from the established method \footnote{
Certainly, also such a different method which does not
start with the classical action may, at the end,
lead to the conclusion that classical actions and
effective actions are related to each other one-to-one,
but then this is a result of the method and not the
starting point.}. There are not so many methods available
and to use an equation for determining the complete effective
action seems to be an approach natural within theoretical
physics. Therefore, above view leads to the task to
find such an equation for the complete effective action.\\

To the best of the authors knowledge such a question has
not been raised so far in the existing literature.
Independent of the kind of further answer to it, it
should be emphasized that in view of the fundamental role
of the effective action in quantum field theory it
deserves one. Even rejection of the question (e.g., by
closely sticking to the established formalism) has
important methodological consequences as we have
demonstrated above.\\

Search for an equation for the complete effective action needs
to be ruled by a couple of principles. First, solutions of such
an equation should be able to reproduce standard quantum
field theoretic result with the required accuracy in order
to stay in line with experiment. Obviously,
this leaves not much room for an answer differing
from the known one. Second, the formalism
connected with such an equation should sufficiently differ
from standard quantum field theory in order to be able to
remove known problems, at least in part. And third,
any sensible search for an equation for the complete
effective action should take into account that
the eventual result needs to be sufficiently
general in order to be applicable to various
situations and has to be restrictive enough
at the same time in order to allow to
derive from it concrete information.\\

While the call for an equation for the complete effective action
still might be shared by a number of researchers and
probably represents the least disputable part of the
present investigation, to reach agreement with respect
to an eventual answer to it very likely will be much
more difficult. In the following section we are going
to propose an answer which then shall be investigated
in some further detail.\\

\subsection{\label{BS2}Proposing an Equation for
the Complete Effective \hfill\break Action}

Basically, there are two different routes to find the
particular answer on the question put forward in the
preceding section we prefer by proposing
a specific  equation for the complete
effective action. One way is to discuss certain
principles to be built in and then to write down
an equation which embodies these. The other way
which we will choose now is heuristically to
motivate an equation which then will be analyzed with
respect to its conceptual content.\\

Let us consider the map $f: \Gamma_0 [\phi ]
\longrightarrow \Gamma [\bar\phi] $ defined in
section \ref{BS1} mapping so-called classical actions
to effective actions. Although it is not necessarily
well defined for the domain of classical actions
(which are local functionals in general)
we will not change the map $f$ itself but instead
we will now extend the domain of this map.
For this purpose it suffices to mention that the set
of so-called classical actions can be considered as
a sub-set of the class of effective actions. From now
on we understand the map $f$ as a mapping of the set
of effective actions into itself.\\

On the basis of formulas given in the preceding section
we will now explicitly define the map $f$ for the
extended domain. Again, we define the generating
functional of Green functions by
\parindent0.cm

\begin{equation}
\label{BB1}
Z_n[J_n] =\ C\  {\rm e}^{\displaystyle\ -i\Gamma_{n-1} [0]}\
\int D\phi\ \ {\rm e}^{\displaystyle\ i\Gamma_{n-1} [\phi]\
+\ i \int dx J_n(x) \phi (x)}
\hspace{1.5cm},\hspace{0.5cm}\\
\end{equation}

where as in eq.\ (\ref{BA1}) $C = C(\mu)$ is some
fixed dimensional normalization constant depending
on an arbitrary mass parameter $\mu$ and compensating
the dimension of the functional integral measure $D\phi$.
Changes in $\mu$ correspond to changes in the
normalization of the vacuum energy connected with
$\Gamma [0]$. In extending the domain of the map $f$
we have introduced an additional normalization factor
$\exp(\ -i\Gamma_{n-1} [0])$ (This is not a major point
but worth to be appreciated from a conceptual point of
view.). Classical actions typically are normalized to
obey $\Gamma_0 [0] = 0$. Then, eq.\ (\ref{BA5}) tells us
that $\Gamma [0]$ is completely originated by vacuum
fluctuations governed by the classical action
$\Gamma_0$ (up to some normalization of the vacuum
energy fixed for a whole class of actions).
By including the additional normalization factor this
principle is generalized to the map $f$ acting in the
extended domain and admits calculation of the vacuum
energy as usual \footnote{Having in mind standard quantum field
theory, of course, here we refer to vacuum energy
modifications under external conditions.}.\\

\parindent1.5em
The generating functional of the connected Green functions is

\begin{equation}
\label{BB2}
W_n[J_n]\ =\ -i \ln Z_n[J_n] \hspace{1.5cm}.\hspace{0.5cm} \\
\end{equation}

\parindent0.em
The generating functional of the 1PI Green functions
(the image of $\Gamma_{n-1}$) is given by

\begin{equation}
\label{BB3}
\Gamma_n [\bar\phi_n]\ =\ W_n[J_n] - \int dx J_n(x) \bar\phi_n (x)
\hspace{1.5cm},\hspace{0.5cm}\\
\end{equation}

where

\begin{equation}
\bar\phi_n (x)\ =\ {\delta W_n[J_n]\over \delta J_n(x)} \\
\end{equation}

and consequently

\begin{equation}
\label{BB4}
{\delta \Gamma_n [\bar\phi_n]\over \delta \bar\phi_n (x)}\ =\ -\ J_n(x)
\hspace{1.5cm} .\hspace{0.5cm}\\
\end{equation}

The generalization of eq.\ (\ref{BA5}) reads
\begin{equation}
\label{BB5}
{\rm e}^{\displaystyle\ i \Gamma_n [\bar\phi_n]}\ \ =\ C\
{\rm e}^{\displaystyle\ -i\Gamma_{n-1} [0]}\
\int D\phi\ \ {\rm e}^{\displaystyle\ i\Gamma_{n-1} [\phi + \bar\phi_n]\
+\ i \int dx J_n(x) \phi (x)}
\hspace{0.3cm} ,\hspace{0.3cm}
\end{equation}

where the r.h.s.\ of above equation is again to be
calculated at a current $J_n(x)$ which is a functional
of $\bar\phi_n $ given by eq.\ (\ref{BB4}), and
eqs.\ (\ref{BB5}), (\ref{BB4}) are acting as functional
integro-differential equations for determining
$\Gamma_n$ (and the same accompanying comment as in
section \ref{BS1}). The map
$g_3\circ g_2\circ g_1 = f: \Gamma_{n-1}
\longrightarrow \Gamma_n$ is explicitly given by eqs.\
(\ref{BB1}) ($g_1: \Gamma_{n-1}\longrightarrow Z_n$),
(\ref{BB2}) ($g_2: Z_n\longrightarrow W_n$), and (\ref{BB3})
($g_3: W_n\longrightarrow \Gamma_n$).\\

\parindent1.5em

Consider now iterations of the map $f$ leading to
some discrete series of effective actions
$\ldots \stackrel{f}{\longrightarrow} \Gamma_{n-1}
\stackrel{f}{\longrightarrow}\Gamma_{n}
\stackrel{f}{\longrightarrow}\Gamma_{n+1}
\stackrel{f}{\longrightarrow}\ldots\ $. Eventually
this still can be combined with a certain truncation
procedure, e.g., acting on the obtained effective
action after each application of the map $f$. It
is worth noting that the successive calculation of higher
loop contributions to the effective action in
standard quantum field theory is such an iteration
and truncation procedure. However, for the present
purpose we do not consider any truncation procedure.
Obviously, the most interesting question one may ask
with respect to the iterations of the map $f$ is
whether it has any fixed point. It should be expected
that the fixed point condition for the map $f$ is not
trivially fulfilled for any arbitrary action and should
distinguish certain (complete) effective actions. Now, we
propose that the fixed point condition for the map
$f$ defined above yields the equation for the
complete effective action we are looking for.\\

The equation for the complete effective action which
is equivalent to the fixed point condition for the
map $f$ reads

\begin{equation}
\label{BB6}
{\rm e}^{\displaystyle\ i \Gamma [\bar\phi]}\ \ =\ C\
{\rm e}^{\displaystyle\ -i\Gamma [0]}\
\int D\phi\ \ {\rm e}^{\displaystyle\ i\Gamma [\phi + \bar\phi]\
+\ i \int dx J(x) \phi (x)}
\hspace{1.5cm} ,\hspace{0.5cm}\\
\end{equation}

\parindent0.cm
where

\begin{equation}
\label{BB7}
J(x) \ =\ -\ {\delta \Gamma [\bar\phi]\over \delta \bar\phi (x)}
\hspace{1.5cm} .\hspace{0.5cm}\\
\end{equation}

Eqs.\ (\ref{BB6}) and (\ref{BB7}) together define a
genuine functional integro-differential equation for
determining the complete (off-shell) effective action $\Gamma$ of a
quantum field theory. Of course, this equation needs to
be supplemented by additional information to
specify the particular conditions
under which it should be solved. Accumulated experience
in quantum field theory tells us that in general
solutions of eq.\ (\ref{BB6}) -- if there exists any
at all -- should be expected to be nonlocal and
nonpolynomial functionals $\Gamma$ of the field
$\phi$. Optimistically, one might think that above
equation for the complete effective action
is sufficiently restrictive in the case
of interacting theories to enable us not
only to find the structure of the effective
action but hopefully also to determine
dimensionless parameters it contains
(e.g., coupling constants and mass ratios).
What concerns its applicability, so the eventual range of theories
remains to be explored. But it seems, that at least
any theory which cannot be understood as being induced by
some more fundamental one should be subject to the concept.\\

\parindent1.5em

\subsection{\label{BS3}Exploring the Equation}

Before we will analyze eq.\ (\ref{BB6}) from the
conceptual side let us ask whether it
has any solution at all. The answer is that any
free field theory solves eq.\ (\ref{BB6}) (In saying
so, of course, we neglect the vacuum energy problem.).
For free field theories the formulation proposed in
section \ref{BS2} completely agrees with the
standard formulation of quantum field theory
displayed in section \ref{BS1}. However, the former
obviously differs from the latter for interacting
theories. In the future it remains to be seen
whether there exists any interacting field theory
which solves eq.\ (\ref{BB6}).\\

Now, we will study eq.\ (\ref{BB6}) with respect
to its methodological content. Observe, that the
proposed equation for the complete effective action is
exclusively expressed in terms of an observable
(at least, in principle) quantity
namely the complete effective action which should be
finite, of course. This specifies the concept
of renormalizable quantum field theory by relying
on observable objects only (Bare and dressed
quantities agree here.). In this context, one
may wonder whether the conceptual distinction
between classical action and effective action
is really a productive one. Although any
theoretician may extract the classical limit
from any solution of eq.\ (\ref{BB6}) one may
justified ask what does this tell an experimental
physicist. In reality, vacuum fluctuations cannot
be switched off (at best, they can be modified) and
the experimentally relevant quantity is the effective
action. Rather, the experimental physicist is interested
in the leading (low energy, long distance, low intensity)
terms of the derivative expansion of the effective
action but these do not necessarily coincide with
what is called the classical action although they
will contain it in most cases. In view of our equation
for the complete effective action also of limited sense is to
ask which effective action term is induced and
which is not because eq.\ (\ref{BB6}) is a
self-consistency condition.\\

Continuing above consideration, it should be
mentioned that already in standard quantum field
theory there is no difference in principle between
a certain mode of vacuum fluctuations
and macroscopic (external)
fields. This is reflected by the insight that the
effective action has a dual nature, namely on one
hand side it is considered as action governing the
behaviour of macroscopic (external) fields and at
the same time it is the generating functional of
1PI Green functions playing herewith a central
role in describing vacuum fluctuations. In
addition, any particular mode of vacuum
fluctuations is acting in the background of
all of them and merely experiences their total effective
impact as described by the complete effective action.
Therefore, the path integral construction should
not rely on the classical action governing the
weight of each path (mode) as is done in standard
quantum field theory but the weight of each path
(mode) should be determined by the complete effective action
expressing the vacuum properties in total. Of course,
this involves a certain self-referentiality which
finds its adequate formulation in terms of a genuine
equation. Concluding this we may say that eq.\
(\ref{BB6}) is the theoretical expression of the dual
nature of the complete effective action being
effective action and generating functional at the
same time. In other words, vacuum fluctuations are
governed by one and the same action like
macroscopic phenomena.\\

Having obtained certain insight into principles
embodied in the proposed equation for the complete effective
action in the following let us turn to eventual
methods of its solution. To expect any final
answer on this right now clearly would not
be realistic, instead few aspects which
come to mind immediately should be discussed
only. Although there is no quick answer at hand to
the question, one may ask whether the map $f$ has
something like a contraction property in a certain
neighborhood of a solution of eq.\ (\ref{BB6}). If
this is the case one could attempt its solution by
iteration. With this concept in mind we will see how
the relation of standard quantum field theory to the
present formulation can be described. The standard
formulation of quantum field theory can be viewed as
first iteration of the map $f$ starting from a certain
low energy (local) approximation (the so-called classical
action) to the complete effective action. This can be considered as
natural starting point which is expected to be close to
a fixed point of the map $f$ for 'experimental' reasons.
However, it is clear that
in view of eq.\ (\ref{BB6}) even the 'complete' (assuming
we had summed up usual perturbation theory) effective
action of standard quantum field theory given by eq.\
(\ref{BA5}) is not the complete one in the sense of
eq.\ (\ref{BB6}) but remains just an approximation.
The approximation method represented by standard
local quantum field theory works reasonably good in
lower spacetime dimensions, with
considerable effort in 4 dimensions, but it becomes badly
defined for most theories in higher dimensions. So,
one may consider the properties of a theory with
respect to renormalization as information about
the possible quality of an approximate solution
of eq.\ (\ref{BB6}) obtained from some local Ansatz
by iteration of the map $f$.
Quantization of a classical theory can be understood
as method for approximately solving eq.\ (\ref{BB6}).
However, simple extrapolation of the classical
Lagrangian to arbitrary high energies leads to
the well-known UV divergencies. \\

For practical (i.e., calculational) purposes the
map $f$ is not a very convenient one. Instead,
one may use a somewhat simpler map $\tilde f$
which differs from $f$ but, as one may see easily
from eq.\ (\ref{BB6}), it has one and the same set of
fixed points like $f$. This simpler map
$\tilde f : \Gamma_{n-1}\longrightarrow \Gamma_n $
can be given by the following formula.
\parindent0.em

\begin{equation}
\label{BC3}
{\rm e}^{\displaystyle\ i \Gamma_n [\bar\phi]}\ \ =\ C\
{\rm e}^{\displaystyle\ -i\Gamma_{n-1} [0]}\
\int D\phi\ \ {\rm e}^{\displaystyle\ i\Gamma_{n-1} [\phi + \bar\phi]\
+\ i \int dx J_{n-1}(x) \phi (x)}
\end{equation}

The advantage of this formula is that it provides
us with a compact and explicit representation
of the $\tilde f$-image of $\Gamma_{n-1}$.
However, in general an image of this
map $\tilde f$ will not have the property to be
generating functional of 1PI Green functions.\\

\parindent1.5em

Concluding this section, let us express our view that
the proposed equation for the complete effective action embodies
a couple of features which seem reasonable and interesting
from a physical point of view and also offers a guiding
line for a re-evaluation of the established technical approach
to quantum field theory and eventually its appropriate
modification. From now on we simply will take eq.\ (\ref{BB6}) as
granted and consider it as starting point for further analysis.\\

\newpage

\section{\label{CS}Gauge Field Theories}
\setcounter{equation}{0}

In the present (and in the following) chapter
we are going to study the
equation for the complete effective action derived
in chapter \ref{BS} in the case of gauge field
theories. Although we will have in mind gauge field
theories in general here we restrict ourselves to QED
and comment only the case of non-Abelian gauge
theories. In doing so it is understood that the
Faddeev-Popov procedure used in standard quantum
field theory for defining the functional integral measure
can be applied in a slightly generalized way also in
the present context, in particular, taking into account
that in general solutions of eq.\ (\ref{BB6}) are nonlocal
and the gauge condition to be chosen will be, for convenience,
nonlocal likewise.\\

We start by defining the generalized map $f$ for QED.
The generating functional Z of the Green functions is
\parindent0.cm

\begin{eqnarray}
&&\hspace{-1.cm}Z_n[J_n,\bar\eta_n,\eta_n]\ =\ C\
{\rm e}^{\displaystyle\ -i\Gamma_{n-1} [0,0,0]}
\ \int D\left[a_{\mu}\right]\ D\psi D\bar\psi\
{\rm e}^{\displaystyle\ i\Gamma_{n-1} [a,\psi,\bar\psi]}
\ \cdot\hfill\nonumber\\
\vspace{0.2cm}\nonumber\\
\label{C1}
&&\cdot\ {\rm e}^{\displaystyle\  i\Gamma_{gf}[a] +
\ i \int d^4x \left[ J_{n\mu}(x) a^{\mu} (x)
+ \bar\eta_n (x) \psi (x) +  \bar\psi (x) \eta_n (x)\right]}
\hspace{0.3cm},\hspace{0.3cm}
\end{eqnarray}

where

\begin{eqnarray}
\label{C2a}
\Gamma_{gf}[a] &=& -\ {1\over 2 \lambda}\ \int d^4y\ \left(F[a;y]\right)^2
\ \ \ ,\\
\vspace{0.3cm}\nonumber\\
\label{C2b}
&&\hspace{1.cm}F[a;y]\ =\ \int d^4x\ n_\mu(y-x)\ a^\mu(x)\ \ \ .
\end{eqnarray}

As usual, $\Gamma_{gf}$ is a gauge breaking term containing a linear,
homogeneous functional $F$ of $a_{\mu}$ (for the moment $n_\mu$ is any
arbitrary but appropriately chosen vector-valued distribution)
and the brackets in $D\left[a_{\mu}\right]$
(eq.\ (\ref{C1})) are thought to indicate that the
Faddeev-Popov determinant has to be taken into account
\footnote{It is an almost trivial factor for Minkowski space QED, but
already at finite temperature it becomes important. In
addition, always having in mind possible generalization
to non-Abelian gauge theories it serves as reminder for
this complication then to be considered.}. $\Gamma_{n-1}$ is
out of the class of gauge invariant effective actions.\\

\parindent1.5cm

Then, the  $W$-functional is given by

\begin{equation}
\label{C3}
W_n[J_n,\bar\eta_n,\eta_n]\ =\
-i \ln Z_n[J_n,\bar\eta_n,\eta_n] \hspace{1.5cm},\hspace{0.3cm}
\end{equation}

\parindent0.cm
and the image of $\Gamma_{n-1}$ is

\begin{eqnarray}
&&\hspace{-0.5cm}\Gamma_n [A_n,\Psi_n,\bar\Psi_n]\ =\ \hfill\nonumber\\
\vspace{0.2cm}\nonumber\\
\label{C4}
&&=\ W_n[J_n,\bar\eta_n,\eta_n] - \int d^4x \left[ J_{n\mu}(x) A_n^{\mu} (x)
+ \bar\eta_n (x) \Psi_n (x) + \bar\Psi_n (x) \eta_n (x)\right]
\hspace{0.3cm}.\hspace{0.3cm}
\end{eqnarray}

Again, we have the relations

\begin{eqnarray}
\label{C5}
A_{n\mu} (x)\
&=\ {\displaystyle{\delta W_n[J_n,\bar\eta_n,\eta_n]
\over \delta\ J_n^{\mu}(x)}}\ \ \ ,\hspace{0.5cm}
{\displaystyle{\delta \Gamma_n [A_n,\Psi_n,\bar\Psi_n]
\over \delta\ A_n^{\mu} (x)}}\ &=\ -\ J_{n\mu}(x)
\hspace{0.5cm},\hspace{0.5cm}\\
\vspace{0.2cm}\nonumber\\
\label{C7}
\Psi_n (x)\
&=\ {\displaystyle{\delta W_n[J_n,\bar\eta_n,\eta_n]
\over\delta\ \bar\eta_n (x)}}\ \ \ ,\hspace{0.5cm}
{\displaystyle{\delta \Gamma_n [A_n,\Psi_n,\bar\Psi_n]
\over \delta\ \Psi_n (x)}}\ &=\ \bar\eta_n (x)
\hspace{0.5cm},\hspace{0.5cm}\\
\vspace{0.2cm}\nonumber\\
\label{C9}
\bar\Psi_n (x)\
&=\ -\ {\displaystyle{\delta W_n[J_n,\bar\eta_n,\eta_n]
\over\delta\ \eta_n (x)}}\ \ \ ,\hspace{0.5cm}
{\displaystyle{\delta \Gamma_n [A_n,\Psi_n,\bar\Psi_n]
\over \delta\ \bar\Psi_n (x)}}\
&=\ -\ \eta_n (x) \hspace{0.5cm}.\hspace{0.5cm}
\end{eqnarray}

\parindent1.5em
Performing now shifts in the integration variables we find

\begin{eqnarray}
&&\hspace{-0.7cm}{\rm e}^{\displaystyle\ i
\Gamma_n [A_n,\Psi_n,\bar\Psi_n]}\ \
=\hfill\nonumber\\ \vspace{0.3cm}\nonumber\\
&&=\ C\  {\rm e}^{\displaystyle\ -i\Gamma_{n-1} [0,0,0]}\
\int D\left[a_{\mu}\right]\ D\psi D\bar\psi\
\ {\rm e}^{\displaystyle\
i\Gamma_{n-1} [a + A_n,\psi + \Psi_n,\bar\psi + \bar\Psi_n]}\ \
\cdot \nonumber\\
\vspace{0.3cm}\nonumber\\
\label{C11}
&&\hspace{0.8cm}\cdot\ {\rm e}^{\displaystyle\ i\Gamma_{gf}[a + A_n]
+\ i \int d^4x \left[ J_{n\mu}(x) a^{\mu} (x)
+ \bar\eta_n (x) \psi (x) +  \bar\psi (x) \eta_n (x)\right]}
\end{eqnarray}

\parindent0.em
describing the map $f$ from the gauge invariant effective
action $\Gamma_{n-1}$ to its image $\Gamma_n$. From
the discussion leading to the background field method
in gauge field theories we know that $\Gamma_n$ is in general
not gauge invariant because as one easily recognizes
from eq.\ (\ref{C11}) the
shift in the gauge field integration interferes
with the gauge fixing term for the quantum fluctuations
\footnote{Further features, met in non-Abelian gauge field
theories, we may disregard here.}.
This is remedied in standard quantum field theory by
starting in eq.\ (\ref{C1}) with a modified gauge fixing term
$\Gamma_{gf}[a - A]$ and the field $A_\mu$ is fixed
to obey $A_\mu = A_{n\mu}$ (cf.\ \cite{c} and references therein).
But, in our approach the application of this procedure
would entail that the map $f$ (in particular, the gauge
condition for the quantum fluctuations) had to be
modified in each iteration step in dependence on
the actual shape (gauge) of
$A_{n\mu}$, i.e., of $F[A_n - A;y]$.
While in standard quantum field theory $A_{n\mu}$ can be
understood as some fixed background field
(essentially, this makes the background field
method acceptable) our situation is worse
because $A_{n\mu}$ also contains pieces of arbitrary
vacuum fluctuations to be integrated over later on.
There is only one safe way to ensure that the gauge
for $A_{n\mu}$ and that for the vacuum fluctuations $a_\mu$ do
not interfere in a gauge dependent way
(i.e., that the shift in the argument
of the gauge field integration does not interfere
with the gauge fixing term), namely one has to
choose for $A_{n\mu}$ the gauge

\begin{eqnarray}
\label{C11a}
F[A_n;y]\ =\ 0
\end{eqnarray}

If $A_{n\mu}$ is a sum of independent pieces condition
(\ref{C11a}) applies to each component because $F$ is
linear and homogenous. Now, as already mentioned in
general $A_{n\mu}$ contains pieces of vacuum fluctuations
to be integrated over in further iterations,
consequently we have to impose condition (\ref{C11a})
also onto these vacuum fluctuations. This argument of
course applies to each iteration step of the map $f$
and therefore the only consistent gauge is the generalized Landau
gauge $\lambda = 0$. So, a 'sharp' gauge has to be
imposed on all gauge fields, on external fields as
well as on vacuum fluctuations, i.e., the whole system
of functional relations is bound to one definite gauge.
Of course, the gauge functional $F$ can be chosen
as convenience may require and the full gauge invariant
effective action $\Gamma_n$ consequently is obtained
by letting $F$ vary. The conclusion that only the generalized
Landau gauge leads to sensible and invariant results
well agrees with investigations dealing with the
concept of the unique (geometrical) effective
action \cite{frad}--\cite{nach}.\\

\parindent1.5em
 From eq.\ (\ref{C11}) we read off now the equation for the
complete (gauge invariant) effective action of QED.

\begin{eqnarray}
&&\hspace{-1.5cm}{\rm e}^{\displaystyle\ i \Gamma [A,\Psi,\bar\Psi]}\ \
=\hfill\nonumber\\ \vspace{0.3cm}\nonumber\\
&&=\ C\  {\rm e}^{\displaystyle\ -i\Gamma [0,0,0]}\
\int D\left[a_{\mu}\right]\ D\psi D\bar\psi\
\ {\rm e}^{\displaystyle\
i\Gamma [a + A,\psi + \Psi,\bar\psi + \bar\Psi]}\ \
\cdot \nonumber\\
\vspace{0.3cm}\nonumber\\
\label{C12}
&&\hspace{1.0cm}\cdot\ {\rm e}^{\displaystyle\ i\Gamma_{gf}[a]
+\ i \int d^4x \left[ J_{\mu}(x) a^{\mu} (x)
+ \bar\eta (x) \psi (x) +  \bar\psi (x) \eta (x)\right]}\\
\vspace{0.5cm}\nonumber\\
&&\hspace{7.cm}F[A;y]\ =\ 0\ ;\ \ \lambda\ \longrightarrow\ 0\nonumber
\end{eqnarray}

\parindent0.em
In any explicit calculation we will always leave the gauge
parameter $\lambda$ unfixed because this allows to better
keep track of terms involved, and in the final results one
may simply set $\lambda = 0$ then to find the correct answer.\\

\parindent1.5em
Having defined above the notation we are prepared now to study
in the following Ward-Takahashi identities and Schwinger-Dyson
equations within the present formulation of QED.\\

\subsection{\label{CS1}Ward-Takahashi Identities}

In standard QED derivation of Ward-Takahashi identities
merely relies on the fact that the classical action is
gauge invariant. Therefore, generalization of this
consideration to the present formulation is
straightforward and reasoning proceeds without any major formal
difference to the standard approach. Here, for convenience
we will closely follow ref.\ \cite{d}, sect.\ 7.4., as an
appropriate textbook treatment.\\

Consider in eq.\ (\ref{C1}) an infinitesimal gauge transformation

\parindent0.cm
\begin{equation}
\label{CA1}
a_{\mu}\ \longrightarrow\ a_{\mu} + \partial_{\mu}\Lambda\ \ ,
\hspace{0.5cm}\psi\ \longrightarrow\ \psi - ie\Lambda\ \psi \ ,
\hspace{0.5cm}\bar\psi\ \longrightarrow\ \bar\psi + ie\Lambda\ \bar\psi\ \ .
\end{equation}

Then, we obtain in first order of $\Lambda (x)$ (remember
that $F$ was chosen as a linear functional)

\begin{eqnarray}
&&\hspace{-1.5cm}\left\{\ {1\over\lambda} \ _x\partial_{\mu}
\int d^4y\ {\delta F[A_n;y]\over\delta\ A_{n\mu}(x)}\
F\left[-i{\delta\over\delta J_n};y\right]
-\ \partial_{\mu} J_n^{\mu}(x)\ - \right. \hfill\nonumber\\
\vspace{0.5cm}\nonumber\\
\label{CA2}
&&\left. -\ e \left( \bar\eta_n(x)\ {\delta\over\delta\bar\eta_n(x)}
- \eta_n(x)\ {\delta\over\delta\eta_n(x)}\right)\right\}\
Z_n[J_n,\bar\eta_n,\eta_n]\ =\ 0 \hspace{0.5cm}.\hspace{0.5cm}
\end{eqnarray}

By means of eqs.\ (\ref{C3})-(\ref{C9}) above equation yields

\begin{eqnarray}
&&\hspace{-1.5cm}{1\over\lambda} \ _x\partial_{\mu}
\int d^4y\ {\delta F[A_n;y]\over\delta\ A_{n\mu}(x)}\ F[A_n;y]
\ +\ \partial_{\mu} {\displaystyle{\delta
\Gamma_n [A_n,\Psi_n,\bar\Psi_n]\over \delta\ A_{n\mu} (x)}}\
+ \hfill\nonumber\\
\vspace{0.5cm}\nonumber\\
\label{CA3}
&&+\ ie\ \Psi_n\ {\displaystyle{\delta
\Gamma_n [A_n,\Psi_n,\bar\Psi_n]\over \delta\ \Psi_n (x)}}\
-\ ie\ \bar\Psi_n\ {\displaystyle{\delta
\Gamma_n [A_n,\Psi_n,\bar\Psi_n]\over \delta\ \bar\Psi_n (x)}}\
=\ 0 \hspace{0.5cm}.\hspace{0.5cm}
\end{eqnarray}

 From this equation different Ward-Takahashi identities
can be derived. As standard example let us consider the following.
Taking functional derivatives of eq.\ (\ref{CA3}) with
respect to $\bar\Psi_n (z^\prime)$, $\Psi_n (z)$ and
then setting $\bar\Psi_n = \Psi_n = A_{n\mu} = 0$
one finds

\begin{eqnarray}
&&\hspace{-1.3cm} \ _x\partial_{\mu}\ {\displaystyle{\delta^3
\Gamma_n [A_n,\Psi_n,\bar\Psi_n]\over \delta\bar\Psi_n (z^\prime)
\ \delta\Psi_n (z)\ \delta A_{n\mu} (x)}}\
\Bigg\vert_{\bar\Psi_n = \Psi_n = A_{n\mu} = 0}\ =
\hfill\nonumber\\
\vspace{0.5cm}\nonumber\\ \label{CA4}
&&\hspace{-1.2cm}=\ ie\ \left\{\delta^{(4)} (x-z^\prime)\
{\displaystyle{\delta^2\Gamma_n [0,\Psi_n,\bar\Psi_n]\over
\delta\bar\Psi_n (z^\prime)\ \delta\Psi_n (z)}}\
-\ \delta^{(4)} (x-z)\ {\displaystyle{\delta^2
\Gamma_n [0,\Psi_n,\bar\Psi_n]\over \delta\bar\Psi_n (z^\prime)\
\delta\Psi_n (z)}}
\right\}_{\bar\Psi_n = \Psi_n = 0}\hspace{-0.8cm}.\
\end{eqnarray}

With

\begin{eqnarray}
&&\hspace{-1.5cm}\int d^4x\ d^4z\ d^4z^\prime\
{\rm e}^{i(p^\prime z^\prime - pz - qx)}\ {\displaystyle{\delta^3
\Gamma_n [A_n,\Psi_n,\bar\Psi_n]\over \delta\bar\Psi_n (z^\prime)
\ \delta\Psi_n (z)\ \delta A_n^{\mu} (x)}}\
\Bigg\vert_{\bar\Psi_n = \Psi_n = A_{n\mu} = 0}\
=\hfill\nonumber\\
\vspace{1.1cm}\nonumber\\ \label{CA5}
&&\hspace{5.cm}=\ e\ (2\pi )^4\
\delta^{(4)} (p^\prime - p - q)\ \tilde\Gamma_{n\mu}(p,q,p^\prime)
\end{eqnarray}

and

\begin{equation}
\label{CA6}
\int d^4z\ d^4z^\prime\ {\rm e}^{i(p^\prime z^\prime - pz)}\
{\displaystyle{\delta^2
\Gamma_n [0,\Psi_n,\bar\Psi_n]\over\delta\Psi_n (z)
\ \delta\bar\Psi_n (z^\prime)}}\ \Bigg\vert_{\bar\Psi_n = \Psi_n = 0}\
=\ (2\pi )^4\ \delta^{(4)} (p^\prime - p)\ \tilde S_n^{-1}(p)
\end{equation}

eq.\ (\ref{CA4}) yields the well-known Ward-Takahashi identity

\begin{equation}
\label{CA7}
q^{\mu}\ \tilde\Gamma_{n\mu}(p,q,p + q)\
=\ \tilde S_n^{-1}(p + q)\ -\ \tilde S_n^{-1}(p)
\hspace{0.5cm}.\hspace{0.5cm}
\end{equation}

We have seen that each image $\Gamma_n$ of the map $f$
respects the Ward-Takahashi identity (\ref{CA7}) which
is a consequence of the gauge invariance of its counter
image $\Gamma_{n-1}$ (Beyond this property the counter
image $\Gamma_{n-1}$ does not show up explicitly.).
This in particular is also true for any solution of
eq.\ (\ref{C12}).\\

\parindent1.5em
Now, one may convince oneself that also in non-Abelian
gauge field theories the derivation of generalized Ward
identities (i.e., Slavnov-Taylor identities;
cf.\ \cite{b}, sect.\ IV.7) remains unchanged and
they also hold at each step of any iteration of the
map $f$. Violation of these (generalized) Ward
identities (so, if anomalies occur) means that
the equation for the complete effective action of
such a theory will not have any solution. To see this note that
the existence of an anomaly would entail that the
image $\Gamma_n = f(\Gamma_{n-1})$ of an action
has a different behaviour than its counter image $\Gamma_{n-1}$,
so blocking any attempt to solve the equation.
In this sense, the well-known model builders requirement of
anomaly cancellation (cf.\ \cite{d}, sect.\ 9.10., e.g.)
can be understood as solvability
condition for the functional integral equation for the
complete effective action of a theory under consideration
\footnote{Of course, as in standard quantum field theory this
concerns only dynamical fields.}.\\

\subsection{\label{CS2}Schwinger-Dyson Equations}

Let us start the study of Schwinger-Dyson equations with
a comment concerning the nature of these equations in
standard quantum field theory. They represent a chain
of hierarchical equations connecting (1PI) Green functions
of a theory and can be seen as a formulation standing
in a certain equivalence
to the functional integral representation given (for a
scalar theory) by eqs.\ (\ref{BA1})-(\ref{BA3}). However,
as we have argued in section \ref{BS1} the effective
action $\Gamma$ is merely the image of the classical
action $\Gamma_0$ with respect to the map $f$ and
therefore Schwinger-Dyson equations are formulas
to be viewed as a device for tackling the calculational
complexity met in explicitly determining the effective
action $\Gamma$, rather than genuine equations (Whether beyond
this they also admit other solutions should not be
further considered here.). From this it is clear that
Schwinger-Dyson equations can be understood as a kind of
representation of the map $f$ and they can
also be formulated for the map $f$ acting in the extended
domain of effective actions in general. Only, if we impose
the fixed point condition for the map $f$ Schwinger-Dyson
equations turn out to be genuine equations corresponding to
the equation for the complete effective action (\ref{BB6}).\\

In the present section we study QED Schwinger-Dyson
equations for the map $f$ acting in the extended domain
of (gauge invariant) effective actions. Only at the end
we will specialize the result to the fixed point condition
for the map $f$. For convenience, in deriving
Schwinger-Dyson equations here we follow the textbook
treatment given in ref.\ \cite{e}, sect.\ 10.1, as far as possible.\\

\parindent0.cm
First we exploit the gauge field integration.
 From eq.\ (\ref{C1}) we find

\begin{eqnarray}
&&\hspace{-2.cm}\left\{ J_{n\mu}(x)\ +\
{\displaystyle{\delta \Gamma_{gf}\over\delta A_n^{\mu}(x)}}
\left[ -i{\delta\over\delta J_n} \right]\ \right.+\
\hfill\nonumber\\
\vspace{0.5cm}\nonumber\\ \label{CB1}
&&+\ \left. {\displaystyle{\delta \Gamma_{n-1}\over\delta A_n^{\mu}(x)}}
\left[ -i{\delta\over\delta J_n},-i{\delta\over\delta \bar\eta_n},
i{\delta\over\delta \eta_n} \right]\ \right\} \
Z_n[J_n,\bar\eta_n,\eta_n]\ =\ 0 \hspace{0.3cm}.\hspace{0.3cm}
\end{eqnarray}

Splitting $\Gamma_{n-1}$ into a free (quadratic) and an interaction
part (denoted by $\Gamma^{(int)}_{n-1}$) we obtain

\begin{eqnarray}
&&\hspace{-1.5cm}
- {\displaystyle{\delta\Gamma_n [A_n,\Psi_n,\bar\Psi_n]
\over\delta\ A_n^{\mu} (x)}}\
-\ {1\over\lambda} \int d^4x^\prime\
{\delta F[A_n;x^\prime]\over\delta\ A^{n\mu}(x)}\
F\left[A_n;x^\prime\right]\ + \hfill\nonumber\\
\vspace{0.5cm}\nonumber\\
&&\hspace{-1.5cm}
+ \int d^4x^\prime\ D_{n-1 \mu\nu}^{-1} (x-x^\prime )\
A_n^{\nu} (x^\prime) + \hfill\nonumber\\
\vspace{0.5cm}\nonumber\\ \label{CB2}
&&\hspace{-1.5cm}
+\ {\rm e}^{\displaystyle -i W_n[J_n,\bar\eta_n,\eta_n]}\ \
{\displaystyle{\delta\Gamma_{n-1}^{(int)}\over\delta A_n^{\mu}(x)}}
\left[ -i{\delta\over\delta J_n},-i{\delta\over\delta\bar\eta_n},
i{\delta\over\delta\eta_n} \right]\
{\rm e}^{\displaystyle\ i W_n[J_n,\bar\eta_n,\eta_n]}\ \ =\ 0
\end{eqnarray}

with

\begin{equation}
\label{CB3}
{\displaystyle{\delta^2 \Gamma_{m} [A,0,0]\over
\delta A^{\mu}(x)\ \delta A^{\nu}(x^\prime)}}\
\Bigg\vert_{A = 0}\ =\  D_{m\ \mu\nu}^{-1} (x-x^\prime )
 \hspace{0.3cm}.\hspace{0.3cm}
\end{equation}

Taking a functional derivative with
respect to $A_{n\nu}$ and setting $\bar\eta_n = \eta_n = J_n = 0$
(and equivalently $\bar\Psi_n = \Psi_n = A_{n\mu} =0$)
eq.\ (\ref{CB2}) yields

\begin{eqnarray}
&&\hspace{-1.2cm}
- D_{n\ \mu\nu}^{-1} (x-x^\prime )\ +\ D_{n-1\ \mu\nu}^{-1} (x-x^\prime )\
 -\ {\rm e}^{\displaystyle -i W_n[0,0,0]}\ \int d^4z\ D_{n-1\ \nu\lambda}^{-1}
(x^\prime -z)\ \cdot\hfill\nonumber\\
\vspace{0.5cm}\nonumber\\ \label{CB4}
&&\hspace{-1.2cm}\cdot\
{\displaystyle{\delta \Gamma_{n-1}^{(int)}\over\delta A_n^{\mu}(x)}}
\left[ -i{\delta\over\delta J_n},-i{\delta\over\delta\bar\eta_n},
i{\delta\over\delta\eta_n} \right]\
{\displaystyle{\delta\over\delta J_{n\lambda}(z)}}\
{\rm e}^{\displaystyle\ i W_n[J_n,\bar\eta_n,\eta_n]}\
\Bigg\vert_{\bar\eta_n = \eta_n = J_n = 0}\ =\ 0\ .
\end{eqnarray}

Finally, the fixed point condition for the map $f$ leads
to the following Schwinger-Dyson equation.

\begin{equation}
\label{CB5}
{\displaystyle{\delta\ \Gamma^{(int)}\over\delta A^{\mu}(x)}}
\left[ -i{\delta\over\delta J},-i{\delta\over\delta\bar\eta},
i{\delta\over\delta\eta} \right]\
{\displaystyle{\delta\over\delta J^{\nu}(z)}}\
{\rm e}^{\displaystyle\ i W[J,\bar\eta,\eta]}\
\Bigg\vert_{\bar\eta = \eta = J = 0}\ =\ 0
\end{equation}

Let us now exploit the fermionic integration. Likewise
we obtain from eq.\ (\ref{C1})

\begin{equation}
\label{CB6}
\left\{ \eta_n (x)\ +\
{\displaystyle{\delta \Gamma_{n-1}\over\delta\bar\Psi_n(x)}}
\left[ -i{\delta\over\delta J_n},-i{\delta\over\delta \bar\eta_n},
i{\delta\over\delta \eta_n} \right]\ \right\} \
Z_n[J_n,\bar\eta_n,\eta_n]\ =\ 0 \hspace{0.3cm}.\hspace{0.3cm}
\end{equation}

Again, splitting $\Gamma_{n-1}$ into a free (quadratic)
and an interaction part we find

\begin{eqnarray}
&&\hspace{-1.5cm}
- {\displaystyle{\delta \Gamma_n [A_n,\Psi_n,\bar\Psi_n]
\over \delta\ \bar\Psi_n (x)}}
\ +\ \int d^4x^\prime\ S_{n-1}^{-1} (x-x^\prime )\ \Psi_n (x^\prime)
\ + \hfill\nonumber\\
\vspace{0.5cm}\nonumber\\ \label{CB7}
&&\hspace{-1.5cm}+\ {\rm e}^{\displaystyle -i W_n[J_n,\bar\eta_n,\eta_n]}\
{\displaystyle{\delta \Gamma_{n-1}^{(int)}\over\delta\bar\Psi_n (x)}}
\left[ -i{\delta\over\delta J_n},-i{\delta\over\delta\bar\eta_n},
i{\delta\over\delta\eta_n} \right]\
{\rm e}^{\displaystyle\ i W_n[J_n,\bar\eta_n,\eta_n]}\ \ =\ 0
\end{eqnarray}

with

\begin{equation}
\label{CB8}
{\displaystyle{\delta^2 \Gamma_{m} [0,\Psi,\bar\Psi]\over
\delta\Psi (x^\prime)\ \delta\bar\Psi (x)}}\
\Bigg\vert_{\bar\Psi = \Psi = 0}\ =\  S_{m}^{-1} (x-x^\prime )
 \hspace{0.3cm}.\hspace{0.3cm}
\end{equation}

Taking a functional derivative with respect to $\Psi_n$ and
setting $\bar\eta_n = \eta_n = J_n = 0$ (and equivalently
$\bar\Psi_n = \Psi_n = A_{n\mu} = 0$) eq.\ (\ref{CB7}) yields

\begin{eqnarray}
&&\hspace{-1.5cm}
-\ S_n^{-1} (x-x^\prime )\ +\ S_{n-1}^{-1} (x-x^\prime )
\ -\ {\rm e}^{\displaystyle -i W_n[0,0,0]}\ \int d^4z\
S_{n-1}^{-1} (z-x^\prime )\ \cdot\hfill\nonumber\\
\vspace{0.5cm}\nonumber\\ \label{CB9}
&&\hspace{-1.5cm}\cdot\
{\displaystyle{\delta\Gamma_{n-1}^{(int)}\over\delta\bar\Psi_n(x)}}
\left[ -i{\delta\over\delta J_n},-i{\delta\over\delta\bar\eta_n},
i{\delta\over\delta\eta_n} \right]\
{\displaystyle{\delta\over\delta\eta_n (z)}}\ \
{\rm e}^{\displaystyle\ i W_n[J_n,\bar\eta_n,\eta_n]}\ \
\Bigg\vert_{\bar\eta_n = \eta_n = J_n = 0}\hspace{-0.4cm} =\ 0
\ .\
\end{eqnarray}

And, the fixed point condition for the map $f$ leads to
this Schwinger-Dyson equation.

\begin{equation}
\label{CB10}
{\displaystyle{\delta\Gamma^{(int)}\over\delta\bar\Psi (x)}}
\left[ -i{\delta\over\delta J},-i{\delta\over\delta\bar\eta},
i{\delta\over\delta\eta} \right]\
{\displaystyle{\delta\over\delta\eta (z)}}\ \
{\rm e}^{\displaystyle\ i W[J,\bar\eta,\eta]}\ \
\Bigg\vert_{\bar\eta = \eta = J = 0}\ \ =\ 0
\end{equation}

\parindent1.5em
The Schwinger-Dyson equations (\ref{CB4}), (\ref{CB5}),
(\ref{CB9}), (\ref{CB10}) cannot be
further studied unless the interaction part of the effective
action has been specified, at least in a certain approximation.
This in particular also concerns the final transition to
relations between 1PI Green functions which hinges on this
information. Therefore, presently it remains open how useful
this kind of representation of the map $f$ will be in future
investigations.\\

\newpage

\section{\label{DS}QED --- An Approximative Approach to the Equation for the
Complete Effective Action}
\setcounter{equation}{0}

Besides structural investigation of the equation for
the complete effective action of most interest appears
to be whether the proposed approach is enabling us to
extract concrete information for specific models not
at all or not easily obtainable by established standard methods.
We will focus here on QED as a realistic physical
theory at the same time also being of major theoretical
interest as simple prototype of a gauge field theory.
The aim of this chapter is to demonstrate that the
present approach indeed admits explicitly to find
certain information about the complete effective
action of QED that, in addition, can be
seen to be of nonperturbative nature.
Of course, the concrete study of the equation for the complete
effective action of QED (eq.\ (\ref{C12})) cannot be
expected to be rigorous for the time being. It will
be necessary to apply an approximation which however
in certain respect should circumvent some of the
problems appearing in standard quantum field theory.
In particular, as far as possible we will take
care that no inappropriate approximation giving
rise to UV divergencies is introduced.
Although most approximations
we will exploit in this chapter can be expected to be
reasonable for small values of the QED coupling constant $\alpha$,
the explicit calculation we will undertake
has to be understood in the first place as a model game to test
in principle the calculational accessibility of the concept
proposed. As a particular application of the new concept we
will explicitly study how to determine the coupling
constant $\alpha$ (i.e., the theoretical value
of the fine structure constant)
being understood as one of the characteristics of a fixed
point of the map $f$. This is done using certain simple
approximations (capable of future improvement)
which at the end however turn out somewhat too simple yet to
succeed numerically. \\

The approximative approach in general relied on in the present chapter
will be as follows.\\

\subsection{\label{DS1}The Approximative Approach in General}

We will study one iteration of the map $f$
starting from a certain Ansatz $\Gamma_I$ which is
mapped by means of $f$ to its image $\Gamma_{II}$.
The gauge invariant Ansatz for $\Gamma_I$ is chosen
as a natural generalization of the so-called classical action
$\Gamma_0$ (to obtain this replace $d_I, a_I, b_I$ by delta
functions) which is the starting point for standard
QED perturbation theory.

\parindent0.em
\begin{eqnarray}
\label{D1}
\Gamma_I [A,\Psi,\bar\Psi]\ &=&\ \Gamma_I^{G} [A]\ +\
\Gamma_I^{F} [A,\Psi,\bar\Psi]\\
\vspace{0.3cm}\nonumber\\ \label{D2}
\Gamma_I^{G} [A]\ &=&\ {1\over 2}\ \int d^4x\ d^4x^\prime\
A^\mu(x)\ \cdot\nonumber\\ &&\hspace{1.5cm}\cdot\
\left[ g_{\mu\nu}\ _x\Box - \ _x\partial_\mu \ _x\partial_\nu\right]
d_I\left(x-x^\prime\right)\ A^\nu(x^\prime)\\
\vspace{0.3cm}\nonumber\\ \label{D3}
\Gamma_I^{F} [A,\Psi,\bar\Psi]\ &=&\ \int d^4x\ d^4x^\prime\ \ \bar\Psi (x)
\ \ {\rm e}^{\displaystyle\ ie \int^{x^\prime}_x dy_\mu\ A^\mu(y)}\ \
\cdot\nonumber\\
\vspace{0.3cm}\nonumber\\
&&\hspace{-1.cm}\cdot\ \left[ a_I\left(x-x^\prime\right)
\ \left( i\not\hspace{-0.07cm}\partial_{x^\prime}
- e \not\hspace{-0.13cm}A(x^\prime) \right)\
-\ m\ b_I\left(x-x^\prime\right) \right]\Psi (x^\prime)
\end{eqnarray}

$m$ is the electron mass, $d_I$, $a_I$, $b_I$ are
functions (distributions) arbitrary for the moment and
the gauge functional $F$ appearing in eq.\ (\ref{C11})
is to be chosen later in a way appropriate and convenient
for explicit calculation
\footnote{For future purposes we are
introducing the notation

\begin{eqnarray*}
\Gamma^F_I [A,\Psi,\bar\Psi]&=&\int d^4x\ d^4x^\prime\ \ \bar\Psi (x)\
S_I^{-1}[A](x,x^\prime)\ \Psi (x)\\
\vspace{0.3cm}\\
S_I^{-1}[0](x,x^\prime)\ =\ S_I^{-1}(x-x^\prime)\ &=&\ i\
a_I\left(x-x^\prime\right)\not\hspace{-0.07cm}\partial_{x^\prime}\
-\ m\ b_I\left(x-x^\prime\right)\hspace{0.5cm}.
\end{eqnarray*}
In general, we will alternatively write $l(x)$ or
$l(r)$, $r = -m^2 x^2 $ for one and the same
function what however will not lead to any confusion in
the context used respectively (All functions $l(x)$ we are studying
depend on $x$ via $x^2$ only. $l$ stands here for
$d$, $a$, $b$.). Fourier transforms
are defined for $l(x)$ by

\begin{eqnarray*}
l(x)\ =\ \int {d^4p\over (2\pi)^4}\ \ {\rm e}^{\ ipx}\ \ \tilde l(p)
\end{eqnarray*}
and equivalently we use the notation $\tilde l(p)$ and
$\tilde l(s)$, $s = - {p^2\over m^2}$ for one
and the same function respectively.}.
Furthermore, the line integration in the phase factor
in eq.\ (\ref{D3}) is understood to be performed along a
straight line connecting starting and end point.
Eq.\ (\ref{D3}) is written in such shape as to keep
contact with standard QED ($\tilde a\ =\ \tilde b\ \equiv\ 1$)
as close as possible.\\

\parindent1.5em
Finally, the equation for the complete effective action
(\ref{C12}) will be taken into account in such a way
that we require at the end
$d_I = d_{II}$, $a_I = a_{II}$, $b_I = b_{II}$, at least
in some approximation. All new structures of $\Gamma_{II}$
not appearing in the Ansatz $\Gamma_I$ will be viewed as
induced ones within this approximation and remain beyond
the scope of present interest.\\

It should find mention that an Ansatz similar to
eq.\ (\ref{D3}) (with $a_I = b_I$)
has unsuccessfully been explored earlier within the
framework of nonlocal QED by
{\sc Chr\'etien} and {\sc Peierls} \cite{f} (see
also \cite{peie}). For a discussion and an
explanation of the failure of the attempt turn to
\cite{scharn2}.
With reference to \cite{f}, the action (\ref{D3})
has also recently been studied in a
different context (effective Lagrangians
in nuclear theory) than ours \cite{ohta}--\cite{tern}.\\

\subsection{\label{DS2}Designing an Approximation Strategy}

After having spelt out above what general kind of
approximative approach we are going to rely
on we need now to translate it into operational
terms which are fundamental to the explicit
calculation we are aiming at. So far,
$d_I$, $a_I$, $b_I$ are understood as
completely arbitrary and clearly it is difficult
to perform an explicit calculation based on such
a general Ansatz. Therefore, below first we will
discuss whether the mostly general Ansatz for
$d_I$, $a_I$, $b_I$ can sensibly be restricted
to a certain subclass the final solution can be
searched in. Of particular interest is whether
these distributions can adequately be modelled
by means of local operators. Let us start with the consideration
of $a_I$, $b_I$ characterizing the fermion action
$\Gamma_I^F$.\\

\subsubsection{\label{DS21}Consequences of Gauge Invariance for the Kernel
of the Fer\-mion Action $\Gamma_I^F$}

One of the crucial solvability conditions of eq.\
(\ref{C12}) is that the map $f$ should not violate
gauge invariance. This in particular entails that
the map $f$ must not induce any mass term for the
gauge field $A_\mu$. Even a finite non-vanishing coefficient of
such a mass term is not allowed not to speak about
infinite ones which are pushed aside in standard QED
by applying a gauge invariant regularization. Inasmuch
as here we are aiming at finite solutions of the equation
for the complete effective action (i.e., some approximation
to it) even in a gauge non-invariant regularization
scheme (like cut-off regularization) mass terms should
not survive after lifting the regularization.\\

In the following we study restrictions arising from gauge
invariance on the possible behaviour of the so far arbitrary kernel
$S^{-1}_I$ of the fermion action $\Gamma_I^F$.
In order to look for a mass term of the gauge field $A_\mu$
we restrict ourselves to the class of constant gauge potentials
$A_\mu (x)\ =\ e^{-1}k_\mu\ \equiv\ const.$ the consideration of which
is sufficient for this purpose. For this simple background
$\Gamma_{II}$ is given by the determinant of
$S^{-1}_I$ in the presence of the constant background
$k_\mu$ which can be viewed in momentum space representation
as a constant external momentum. Because we cannot assume
from the very beginning that the result in eq.\ (\ref{DF2})
below will be finite (This is related to the vacuum energy
problem which we will not consider in this chapter.) we
are barred from simply using a shift $p\longrightarrow p-k$ (which
would make vanish the dependence on $k$ at once; this would only
be applicable in a gauge invariant regularization).
The effective action reads
\parindent0.em

\parindent0.em
\begin{eqnarray}
\label{DF1}
\Gamma_{II} [e^{-1} k,0,0]\ &=&\ const.\ -
\ i\ln\ {\rm Det}_\Lambda\ \left( S^{-1}_I[e^{-1} k]\right)\\
\label{DF2}
&=&\ const.\ -\ 2i\ V_4 \int\limits_\Lambda {d^4p\over (2\pi)^4}\ \
h\left(-(p+k)^2\over m^2 \right)
\end{eqnarray}

where

\begin{equation}
\label{DF3}
h(s)\ =\ \ln\left[ \ s\ \tilde a_I^2 (s)\ +\ \tilde b_I^2 (s)\ \right]
\hspace{0.5cm},\hspace{0.5cm}
s\ \ =\ \ - {p^2\over m^2}\ \ =\ \ {p_E^2\over m^2}\ \ \ .
\end{equation}

The subscript $\Lambda$ in eqs.\ (\ref{DF1}), (\ref{DF2})
indicates that we apply a cut-off regularization with a (radial)
momentum space UV cut-off at $\Lambda$. The subscript $E$ in eq.\ (\ref{DF3})
refers to the (Wick rotated) Euclidean momentum variable.\\
\parindent1.5em

Now, let us further transform the integral appearing
in eq.\ (\ref{DF2}). First, we perform a Wick rotation
and then we expand the integrand in powers of $k$ (up to
$O(k^4)$; the notation is $h^\prime= d/ds\ h$).
\parindent0.em

\begin{eqnarray}
\label{DF4}
\hspace{-0.5cm}\int\limits_\Lambda d^4p_E\
h\left( {(p_E+k_E)^2\over m^2}\right)\ &=&
\ \int\limits_\Lambda d^4p\ \left\{ \ h(s)
\ +\ 2\ {pk\over m^2}\ h^\prime (s)
\ +\ {k^2\over m^2}\ h^\prime (s)\ +\right.\nonumber\\
\vspace{0.3cm}\nonumber\\
&&\hspace{-0.25cm}+\ 2\ {(pk)^2\over m^4}\ h^{\prime\prime} (s)\ +\
2\ {k^2\ pk\over m^4}\ h^{\prime\prime} (s)\ + \nonumber\\
\vspace{0.3cm}\nonumber\\
&&\hspace{-0.25cm}
+\ {4\over 3}\ {(pk)^3\over m^6}\ h^{\prime\prime\prime} (s)\ +\
{1\over 2}\ {(k^2)^2\over m^4}\ h^{\prime\prime} (s)\ +\nonumber\\
\vspace{0.3cm}\nonumber\\
&&\hspace{-0.25cm}
\left. +\ 2\ {k^2(pk)^2\over m^6}\ h^{\prime\prime\prime} (s)\ +\
{2\over 3}\ {(pk)^4\over m^8}\ h^{\prime\prime\prime\prime} (s)\ +\
\ldots\ \right\}
\end{eqnarray}

For convenience, we have omitted the index $E$ on the
r.h.s.. Deleting in the integrand terms
antisymmetric with respect to $p \longrightarrow -p$ and applying
following equivalences (valid under the 4D integral)

\begin{eqnarray}
\label{DF5}
(pk)^2\ &{\displaystyle\hat{=}}&\ {1\over 4}\ k^2\ p^2\\
\label{DF6}
(pk)^4\ &{\displaystyle\hat{=}}&\ {1\over 8}\ (k^2)^2\ (p^2)^2
\end{eqnarray}

we find after some manipulations

\begin{eqnarray}
\label{DF7}
\Gamma_{II} [e^{-1} k,0,0]\ &=&\ const.\ +\  {V_4\over 8\pi^2}\ m^4\
\left\{\ \int_0^{\Lambda^2\over m^2} ds\ s\ h(s)\ - \right.\nonumber \\
\vspace{0.3cm}\nonumber\\
&&\hspace{0.2cm}-\ {1\over 2}\ {k^2\over m^2}\ \left[ \ s^2\ h^\prime (s)\
\right]_0^{\Lambda^2\over m^2}\ +\ \nonumber \\
\vspace{0.3cm}\nonumber\\
&&\hspace{0.2cm}\left. +\ {1\over 12}\ {(k^2)^2\over m^4}\
\left[ \ 3\ s^2\  h^{\prime\prime} (s)\ +\ s^3\
h^{\prime\prime\prime} (s)\ \right]_0^{\Lambda^2\over m^2}\
+\ \ldots\ \right\}
\end{eqnarray}

where $k_\mu$ denotes the constant (Minkowski space) gauge
potential.\\
\parindent1.5em

 From the term proportional to $k^2$ in
eq.\ (\ref{DF7}) we see that the
requirement of gauge invariance (i.e., vanishing of
any mass term) yields that $h(s)$ should behave for
$s \longrightarrow \infty $ like
\parindent0.em

\begin{equation}
\label{DF8}
h(s)\ \ \stackrel{s \longrightarrow\infty }{\sim }
\ \ const.\ +\ O\left(s^\kappa\right)
\hspace{0.5cm},\hspace{0.5cm}
\kappa\ < \ -1 \ \ .
\end{equation}

Above condition obviously is also sufficient in order
to make vanish all higher (in powers of $k$) gauge non-invariant structures.
By translating information contained in (\ref{DF8})
one finds following conditions to obey it
\footnote{\label{foot1}We disregard here the somewhat weaker condition
\begin{eqnarray*}
\tilde a(s)\ \ \stackrel{s \longrightarrow \infty}{\sim}
\ \ s^{-1/2}\ +\ O\left(s^\kappa\right)
\hspace{0.5cm},\hspace{0.5cm}\kappa\ <\ -{3\over 2}\ \ \ \ ,
\end{eqnarray*}
and all other variants requiring some fine tuning between $\tilde a$
and $\tilde b$.}.

\begin{eqnarray}
\label{DF9}
&&\tilde a_I(s)\ \ \stackrel{s \longrightarrow \infty}{\sim}
\ \ O\left(s^\kappa\right)
\hspace{0.5cm},\hspace{0.5cm}\kappa\ <\ -1\ \ \ \ ,\\
\vspace{0.4cm}\nonumber\\ \label{DF10}
&&\tilde b_I(s)\ \ \stackrel{s \longrightarrow \infty}{\sim}
\ \  const.\ +\ O\left(s^\kappa\right)
\hspace{0.5cm},\hspace{0.5cm}const. \not= 0\ \ ,\ \ \kappa\ <\ -1
\ \ \ .\
\end{eqnarray}

 From these relations one recognizes that $\tilde a_I$ and
$\tilde b_I$ should behave differently for
$s \longrightarrow \infty$, i.e., they cannot be identical.
This requirement is in line with results for the fermion self-energy calculated
in lowest order of standard QED perturbation theory where
$\tilde a$ and $\tilde b$ already differ (see, e.g., \cite{e}).\\

\parindent1.5em
Finally, let us come back to the purpose of this subsection.
Although, with (\ref{DF9}), (\ref{DF10}) we have
found certain expectations for the UV behaviour of
$a_I$ and $b_I$ this result does not seem to improve
our situation. Even worse, it indicates that $a_I$
and $b_I$ cannot adequately be approximated by any
local operator Ansatz because it would exhibit an
unacceptable UV behaviour. So, from this analysis we conclude
that for the moment $a_I$ and $b_I$ should indeed be
kept arbitrary and the hope for simplifying our Ansatz
is exclusively placed on the kernel of the gauge field
action $\Gamma_I^G$ which we will discuss now.\\

\subsubsection{\label{DS22}Requirements on the Kernel of
the Gauge Field Action $\Gamma_I^G$}

The requirements on the kernel of the gauge field
action to be given below will not be made obvious
immediately in this subsection but will be commented
at the appropriate place in the course of the further
calculation. Here we simply mention them in order to
explain the approximation strategy and the reader is
asked to find justification for them later on only.\\

The first requirement (cf.\ subsection 4.3.1, eq.\ (\ref{DI8}))
is that we expect the (time integrated) self-energy
($D^{\mu\nu}_I$ is the photon propagator
derived from the action $\Gamma^G_I + \Gamma_{gf}$.)

\begin{eqnarray}
\label{DG1}
{1\over2}\ \int d^4 y\ d^4 y^\prime\ \ \ \bar{J}_\mu (x,x^\prime;y)\
D^{\mu\nu}_I(y - y^\prime)\ \bar{J}_\nu (x,x^\prime;y^\prime)\
\end{eqnarray}

\parindent0.em
of a charged point particle represented by the current

\begin{eqnarray}
\label{DG2}
\bar{J}_\mu (x,x^\prime;y)\ &=\ e\ {\displaystyle \int_0^1} d\tau\
\dot{z}_\mu\ \delta^{(4)}(z(\tau) - y)&\ \ \ ,\\
\vspace{0.3cm}\nonumber\\
&&\hspace{-0.5cm} z_\mu (\tau)\ = (x^\prime -x)_\mu\ \tau + x_\mu
\ \ \ \ ,\nonumber
\end{eqnarray}

and propagating over a finite time interval to be finite.
This is needed in order to properly define the map $f$.
Above requirement yields the condition

\begin{eqnarray}
\label{DG3}
&&\tilde d_I(s)\ \ \stackrel{s \longrightarrow \infty}{\sim}
\ \ O\left(s^\kappa\right)
\hspace{0.5cm},\hspace{0.5cm}\kappa\ >\ {1\over 2}\ \ \ .
\end{eqnarray}\\

\parindent1.5em
To take into account condition (\ref{DG3}) is sufficient for most
part of the explicit calculation we are attempting.
However, it turns out that in finally imposing our
approximation to the fixed point condition for the map
$f$ and then searching for a solution to it we need to
request more in order to find some \footnote{More precisely, this
concerns the integral equation for the kernel of the
fermion action to be studied further below (see
subsection 4.3.2.1).}. Specifically,
a solution correct in the asymptotic UV region can only
be found if the photon propagator $D^{\mu\nu}_I(x)$ is
finite in the coincidence limit $x \rightarrow 0$. This
entails for the kernel of the gauge field action the
stronger requirement

\begin{eqnarray}
\label{DG4}
&&\tilde d_I(s)\ \ \stackrel{s \longrightarrow \infty}{\sim}
\ \ O\left(s^\kappa\right)
\hspace{0.5cm},\hspace{0.5cm}\kappa\ >\ 1\ \ \ .
\end{eqnarray}

We see that $d_I$ characterizing the kernel of
the gauge field action should behave qualitatively
quite different than $a_I$ and $b_I$ defining the
kernel of the fermion action do. Conditions (\ref{DG3}),
(\ref{DG4}) induce justified hope that $d_I$ can
indeed be modelled by a local operator. Inasmuch as to
respect condition (\ref{DG3}) is sufficient for most of
the further explicit calculation (i.e., in particular for the
analysis of the asymptotic IR region) we choose the Ansatz

\begin{eqnarray}
\label{DG5}
d_I(x)\ &=&\
\left[\ 1 + \beta\ {\Box\over m^2}\ \right]\ \delta^{(4)}(x)
\ \ \ ,
\end{eqnarray}

\parindent0.em
where $\beta$ is an arbitrary real (positive) constant
parameterizing the Ansatz
\footnote{We have immediately normalized the first term to 1 hereby
freezing the arbitrariness against (finite) gauge field
renormalizations the formalism admits.}. The analysis of the
asymptotic IR region will be merely independent of
further terms to be introduced in (\ref{DG5}) to
satisfy (\ref{DG4}) and therefore they are ignored
in the present Ansatz for calculational simplicity.
Of course, the Ansatz introduces an additional
(spurious) pole at $p^2 = \beta^{-1} m^2$ in the
momentum space photon propagator representation.
However, we will not be worried
by this fact because we simply see eq.\ (\ref{DG5})
as a model representation of an unknown and possibly complicated
kernel of the gauge field action, and so it cannot be
expected to be free of perhaps unpleasant properties in any
respect. Also, eq.\ (\ref{DG5}) can be understood as
some low energy (i.e., IR) approximation that
however can safely be extended to arbitrary high
energies without severely misrepresenting the
required true UV behaviour. For a discussion of
some features and drawbacks of the particular model Ansatz
(\ref{DG5}) see \cite{pais},\cite{barc} and
references therein.
The analysis of the asymptotic UV region will
not demand any further explicit knowledge of the photon
propagator beyond condition (\ref{DG4}) so that the
Ansatz (\ref{DG5}) can be used for most of the
further calculation (that focuses on the IR analysis)
and needs not to be supplemented by any specific UV Ansatz.\\

\subsubsection{\label{DS23}\label{SH3}The Approximation Strategy in Ideal,
and in Practice}

After above considerations we are ready to define the
approximation strategy to be followed in the explicit calculation.
For reducing the calculational complexity we
will make use of the map $\tilde f$ (i.e., source terms are
given by $\Gamma_I$ and not by $\Gamma_{II}$) instead
of the map $f$. In practice, $\tilde f$ will
be slightly modified still further as we will explain in
section \ref{SH1} below.\\

\parindent1.5em
The local operator Ansatz (\ref{DG5}) for the
kernel of the gauge field action admits following
procedure for applying the map $\tilde f$.
First, starting from $\Gamma_I$ with eq.\ (\ref{DG5})
inserted we will perform the functional
integration over the gauge potentials. This can be
done exactly, independently of the Ansatz (\ref{DG5}).
Then, we perform the integration over the fermion
fields, and consequently we impose the fixed point
approximation $a_I = a_{II}$, $b_I = b_{II}$.
These integral equations are to be solved. In practice,
solution of these coupled integral equations can be
attempted in a certain approximation only.
Specifically, we will explicitly solve them in
the asymptotic UV region and in the asymptotic
IR region respectively. Solutions $a$, $b$ of these integral
equations are still parameterized by $\alpha$
($\alpha = e^2/4\pi$) \footnote{Let us
assume that there is an unique solution $a$, $b$ only
what is supported (in practice) by explicit
calculation to be discussed further below.} while
we find that the parameter $\beta$
(of the kernel of the gauge field action) has to be considered as
function of $\alpha$ in order to find any
consistent solution at all.
However, it remains to impose the third
condition $d_I = d_{II}$ yet.\\

The fermionic integration finally has
induced a contribution $\Delta\Gamma^G_I$ to the gauge
field action as follows \footnote{Gauge non-invariant structures
do not occur because the solutions $a$ and $b$
exhibit an UV behaviour as will be shown
preventing those from occurring even in a
gauge non-invariant regularization (at removing
the cut-off).}.

\begin{eqnarray}
\label{DP1}
\hspace{-0.5cm}\Delta\Gamma^G_I [A]\ &=&\nonumber\\
\vspace{0.3cm}\nonumber\\
\hspace{-0.7cm}
= \ {\alpha\over 4\pi} &\displaystyle{\int} d^4x &A^\mu(x)\
\left[ g_{\mu\nu} \Box\ -\ \partial_\mu \partial_\nu\right]\
\left[\ C_{1a}\ +\ C_{2a}\ {\Box\over m^2}\ +\ \ldots\ \right]
\ A^\nu(x)
\end{eqnarray}

\parindent0.em
$C_{1a}$, $C_{2a}$ are functionals of the distributions $a$ and $b$.
Therefore, they can also be viewed as certain functions of $\alpha$
and of the parameter $\beta(\alpha)$.
For the moment let us vary the parameter $\beta$ independently
of $\alpha$ although we believe that the necessity to consider
the parameter $\beta$ as a function of $\alpha$ in course of solving the
integral equation for the quadratic kernel of the fermion action is not
bound to the particular method we will apply.
The condition $d_I = d_{II}$ then reads

\begin{eqnarray}
\label{DP2}
C_{1a}(\alpha,\beta)\ &=&\ 0\ \ \,\\
\vspace{-0.1cm}\nonumber\\
\label{DP3}
C_{2a}(\alpha,\beta)\ &=&\ 0
\end{eqnarray}

and both these equations define an implicit
function $\alpha(\beta)$ (or $\beta(\alpha)$),
i.e., certain curves in the $\alpha$-$\beta$-plane.
The crossing points of these curves correspond to the
set of allowed values ($\alpha ,\beta$).
So far, the functional $C_{1a}$
has been explicitly calculated (see Appendix A) with
considerable effort in 1-loop approximation only
(i.e., taking into account the quadratic
kernel of the fermion action in the presence
of an arbitrary gauge potential).
To determine $C_{2a}$ in 1-loop approximation along
the same lines is a trivial but extremely
laborious task reserved to be undertaken in the
future. But, if as mentioned the parameter $\beta$ has to be viewed as a
function of $\alpha$ in advance of imposing
$d_I = d_{II}$ eqs.\ (\ref{DP2}), (\ref{DP3})
cannot be satisfied simultaneously anyway (To
expect that they are degenerate seems not to
be very realistic.). Requiring that at least
in the asymptotic IR (long distance, long
wavelength) region the fixed point condition
should be fulfilled we choose eq.\ (\ref{DP2})
as condition to be respected.
So, in principle the equation

\begin{eqnarray}
\label{DP4}
C_{1a}(\alpha,\beta(\alpha))\ =\ 0
\end{eqnarray}

admits us to determine the QED coupling constant
$\alpha$ within the present approximative approach.
It is clear that the above method can easily be accommodated
to the inclusion of additional terms in the Ansatz
(\ref{DG5}).\\

\parindent1.5em
So, at the end of this section we are equipped
with a plan for the explicit calculations,
and we will now proceed along the lines just discussed.\\

\subsection{\label{DS3}Bringing the Approximation Strategy to Work:
\hfill\break Explicit Calculation}

\subsubsection{\label{DS31}\label{SH1}Performing the Functional Integration}

\parindent0.em
According to our approximation strategy first we have
to calculate the functional integral
(cf.\ eqs.\ (\ref{BC3}), (\ref{C11}))

\begin{eqnarray}
&&\hspace{-0.7cm}{\rm e}^{\displaystyle\ i \Gamma_{II} [A,\Psi,\bar\Psi]}\ \
=\hfill\nonumber\\ \vspace{0.3cm}\nonumber\\
&&=\ C\ \int D\left[a_{\mu}\right]\ D\psi D\bar\psi\
\ {\rm e}^{\displaystyle\
i\Gamma_{I} [a + A,\psi + \Psi,\bar\psi + \bar\Psi]}\ \
\cdot \nonumber\\
\label{DI1}
&&\hspace{1.2cm}\cdot\ {\rm e}^{\displaystyle\ i\Gamma_{gf}[a]
+\ i \int d^4x \left[ J_{I\mu}(x) a^{\mu} (x)
+ \bar\eta_I (x) \psi (x) +  \bar\psi (x) \eta_I (x)\right]}
\end{eqnarray}

with

\begin{eqnarray}
\label{DI2}
{\displaystyle{\delta \Gamma_I [A,\Psi,\bar\Psi]
\over \delta\ A^{\mu} (x)}}\ &=&\ -\ J_{I\mu}(x)
\hspace{0.5cm},\hspace{0.5cm}\\
\vspace{0.2cm}\nonumber\\
\label{DI3}
{\displaystyle{\delta \Gamma_I [A,\Psi,\bar\Psi]
\over \delta\ \Psi (x)}}\ &=&\ \bar\eta_I (x)
\hspace{0.5cm},\hspace{0.5cm}\\
\vspace{0.2cm}\nonumber\\
\label{DI4}
{\displaystyle{\delta \Gamma_I [A,\Psi,\bar\Psi]
\over \delta\ \bar\Psi (x)}}\
&=&\ -\ \eta_I (x)
\end{eqnarray}

inserted. In calculating $J_{I\mu}$ we may neglect
the term stemming from $\Gamma^F_I$ because in
$\Gamma_{II}$ it gives rise to fermion interactions only
\footnote{Incidentally, it should be noted that reasoning
leading to this fact also makes use of Furry's theorem
(i.e., an appropriate generalization of it)
which applies to our situation. It excludes a closed
fermion loop tadpole contribution.}. Furthermore,
by using a partial integration we rewrite eq.\ (\ref{D3})
in the following manner (for the definition of $\bar{J}$ see
eq.\ (\ref{DG2}))

\begin{eqnarray}
\label{DI5}
\Gamma_I^{F} [A,\Psi,\bar\Psi]\ &=&\ \int d^4x\ d^4x^\prime\ \ \bar\Psi (x)
\ \ {\rm e}^{\displaystyle\ i \int d^4y\
\bar{J}_\mu (x,x^\prime;y) \ A^\mu(y)}\ \
\cdot\nonumber\\
\vspace{0.3cm}\nonumber\\
&&\hspace{1.cm}\cdot\ \left[
i\not\hspace{-0.07cm}\partial_x\ a_I\left(x-x^\prime\right)
\ -\ m\ b_I\left(x-x^\prime\right) \right]\Psi (x^\prime)\ \ \ .
\end{eqnarray}

This will admit us to represent the result of the gauge
field integration to be treated first
in a very convenient way. To perform
the gauge field integration we temporarily
expand in eq.\ (\ref{DI1}) the term
$\exp\{i \Gamma^F_I\}$ in a power series,

\begin{eqnarray}
\label{DI6}
{\rm e}^{\displaystyle\ i \Gamma^F_I}\ =\ 1\ +\  i\ \Gamma^F_I\
-\ {1\over 2}\ \left( \Gamma^F_I \right)^2\ +\ \ldots\hspace{1.cm} ,
\end{eqnarray}

what is a very natural procedure in view of the Grassmann
integration. This way it turns out that the result
of the gauge field integration can be given as an
infinite sum of Gaussian integrals. Each term of
this sum corresponds to a certain power $n$ of $\Gamma^F_I$
and contains the expression

\begin{eqnarray}
\label{DI7}
\int D\left[a_{\mu}\right]&&
{\rm e}^{\displaystyle\ \ i\Gamma^G_{I} [a]\ +\ i\Gamma_{gf}[a]\
+\ i\sum_{k=1}^n\
\int d^4y\ \bar{J}_\mu (x_k,x_k^\prime;y)\ a^\mu (y)}
\end{eqnarray}

where the arguments $\{x_k,x_k^\prime\}$ refer to
the integration variables in the $k$-th copy of
$\Gamma^F_I$. Having performed the Gaussian
integration eq.\ (\ref{DI7}) reads

\begin{eqnarray}
\label{DI8}
C\ \ {\rm e}^{\displaystyle\ -{i\over 2}\sum_{k=1}^n \sum_{l=1}^n\
\int d^4y\ d^4y^\prime\ \
\bar{J}_\mu (x_k,x_k^\prime;y)\ D_I^{\mu\nu}(y - y^\prime)\
\bar{J}_\nu (x_l,x_l^\prime;y^\prime)}\ \ \ .
\end{eqnarray}

Terms with $k=l$ are self-energy contributions while
off-diagonal terms of the double sum in the exponent
generate fermion interactions. We see that the requirement
(\ref{DG3}) arises naturally in the course of the
functional integration. Let us define following
function from the self-energy term.

\begin{eqnarray}
\label{DI9}
g(x - x^\prime)\ =\
{\rm e}^{\displaystyle\ -{i\over 2}\int d^4y\ d^4y^\prime\
\bar{J}_\mu (x,x^\prime;y)\ D_I^{\mu\nu}(y - y^\prime)\
\bar{J}_\nu (x,x^\prime;y^\prime)}
\end{eqnarray}

$g$ can be calculated explicitly, and for the Ansatz
(\ref{DG5}) this is done in Appendix B. Using $g$ we
introduce the new functions $a_{Ig}$, $b_{Ig}$ by
defining a map ${\bf g}: a \longrightarrow a_g,\
b \longrightarrow b_g$ specified by the prescriptions

\begin{eqnarray}
\label{DI10a}
a^\prime_{Ig}(x)\ &=&\ g(x)\ a^\prime_I(x)\ \ \ ,\\
\vspace{-0.1cm}\nonumber\\
\label{DI10b}
b_{Ig}(x)\ &=&\ g(x)\ b_I(x)\ \ \ .
\end{eqnarray}

Here, the notation $a^\prime (x)\ =\ d/dr\ a(r)$, $r = - m^2 x^2$
is used. The uncertainty in $a_{Ig}$ due to the free
integration constant is removed by noting that
$g(0) = 1$ (this follows from condition (\ref{DG3}))
and consequently requiring the same behaviour for
$a_{Ig}(x)$ and $a_I(x)$ at $x \rightarrow 0$.\\

\parindent1.5em
Now, we may reverse the procedure indicated in
eq.\ (\ref{DI6}) and re-exponentiate the terms of
the infinite sum under the remaining fermionic
integration what however cannot be done in a
closed form. Proceeding this way we obtain

\begin{eqnarray}
\label{DI11}
&&\hspace{-0.7cm}{\rm e}^{\displaystyle\ i \Gamma_{II} [A,\Psi,\bar\Psi]}\ \
=\hfill\nonumber\\ \vspace{0.3cm}\nonumber\\
&&\hspace{-1.cm}=\ C\ \ {\rm e}^{\displaystyle\ i\Gamma^G_{I} [A]}\
\ \int D\psi D\bar\psi\ \ {\rm e}^{\displaystyle\ i \int d^4x
\left[ \bar\eta_I (x) \psi (x) +  \bar\psi (x) \eta_I (x)\right]}\
\cdot\nonumber\\ \vspace{0.3cm}\nonumber\\
&&\hspace{-0.9cm}\ \ \ \cdot\ \exp\left\{ i\int d^4x\ d^4x^\prime\ \
\left( \bar\psi (x)\ +\ \bar\Psi (x)\right)
\ \ {\rm e}^{\displaystyle\ ie \int^{x^\prime}_x dy_\mu\ A^\mu(y)}\ \
\right. \cdot\nonumber\\ \vspace{0.3cm}\nonumber\\
&&\ \ \ \cdot\ \left[ a_{Ig}\left(x-x^\prime\right)
\ \left( i\not\hspace{-0.07cm}\partial_{x^\prime}
- e \not\hspace{-0.13cm}A(x^\prime) \right)\
-\ m\ b_{Ig}\left(x-x^\prime\right) \right]
\left( \psi (x^\prime)\ +\ \Psi (x^\prime)\right)\ -\nonumber\\
\vspace{0.3cm}\nonumber\\
&&\hspace{-0.6cm}-\ {1\over 2}\int d^4x\ d^4x^\prime\
d^4z\ d^4z^\prime\ \cdot\nonumber\\
\vspace{0.3cm}\nonumber\\
&&\hspace{-0.6cm}\cdot\ \Bigg[\
\left( \bar\psi (x)\ +\ \bar\Psi (x)\right)\ \left[
i\not\hspace{-0.07cm}\partial_x\ a_{Ig}\left(x-x^\prime\right)
\ -\ m\ b_{Ig}\left(x-x^\prime\right) \right]
\left( \psi (x^\prime)\ +\ \Psi (x^\prime)\right)
\ \cdot\nonumber\\
\vspace{0.3cm}\nonumber\\
&&\hspace{-0.6cm}\cdot\
\left(\bar\psi (z)\ +\ \bar\Psi (z)\right)\ \left[
i\not\hspace{-0.07cm}\partial_z\ a_{Ig}\left(z-z^\prime\right)
\ -\ m\ b_{Ig}\left(z-z^\prime\right) \right]
\left( \psi (z^\prime)\ +\ \Psi (z^\prime)\right)\
\ \cdot\nonumber\\
\vspace{0.3cm}\nonumber\\
&&\hspace{-0.6cm}\cdot\ \left.
\left. \left( {\rm e}^{\displaystyle\ -i \int d^4y\ d^4y^\prime\ \
\bar{J}_\mu (x,x^\prime;y)\ D_I^{\mu\nu}(y - y^\prime)\
\bar{J}_\nu (z,z^\prime;y^\prime)}\ -\ 1\ \right)\right]\
+\ \ldots\ \right\} .\ \
\end{eqnarray}

\parindent0.em
In the last term of eq.\ (\ref{DI11}) we have already
put $A_\mu = 0$ because we will consider 1-loop
contributions (i.e., those stemming in eq.\ (\ref{DI11})
from the quadratic kernel of the fermion action in the
presence of the arbitrary gauge potential $A_\mu$)
to the quadratic kernel of
the gauge field action $\Gamma^G_{II}$ only \footnote{As long as
$\alpha$ is sufficiently small higher loop
contributions will only lead to small quantitative
changes.}.
In eq.\ (\ref{DI11}) the remaining fermionic
integration is now done (in the sense of perturbation
theory, and which after integration is formally summed up
again). In performing the Gaussian integration (i.e.,
treating the last term (and all further terms) in
eq.\ (\ref{DI11}) as perturbation) for calculational
simplicity we neglect the source terms (linear
in $\Psi$, $\bar\Psi$; others are not of our
present interest) that contain
$[ g(x) - 1 ]$ factors. For our envisaged study due
to $g(0) = 1$ these source terms are irrelevant in
the asymptotic UV region and in the asymptotic IR
region they will lead to certain changes which are
apparently small however as long as $\alpha$ is sufficiently
small. Now, without appealing to the eventual
range of $\alpha$ we simply understand this neglect
as a certain further modification of the map
$\tilde f$ but which preserves all important
features (In particular, it does not lead to any
change in the asymptotic UV region.). So we obtain
for eq.\ (\ref{DI11}) the following result \footnote{We
display only non-interaction terms of $\Gamma_{II}$ which we are
exclusively interested in. Furthermore, on the r.h.s.\ only the
term containing one photon propagator is shown.}.

\begin{eqnarray}
\label{DI12}
&&\hspace{-1.2cm}{\rm e}^{\displaystyle\ i \Gamma_{II} [A,\Psi,\bar\Psi]}\ \
=\hfill\nonumber\\ \vspace{0.3cm}\nonumber\\
&&\hspace{-1.cm}=\ C\ \ {\rm e}^{\displaystyle\ i\Gamma^G_{I} [A]\ +\
i \Delta\Gamma^G_I [A]}\ \ \cdot\nonumber\\
\vspace{0.3cm}\nonumber\\
&&\hspace{-0.9cm}\ \ \ \cdot\ \exp\left\{\ i\int d^4x\ d^4x^\prime\
\bar\Psi (x)\ \left[
i\not\hspace{-0.07cm}\partial_x\ a_{Ig}\left(x-x^\prime\right)
\ -\ m\ b_{Ig}\left(x-x^\prime\right) \right] \Psi (x^\prime)
\right. -\nonumber\\
\vspace{0.3cm}\nonumber\\
&&\hspace{0.5cm} -\
\int d^4x\ d^4x^\prime\ d^4z\ d^4z^\prime\ \cdot\nonumber\\
\vspace{0.3cm}\nonumber\\
&&\hspace{0.7cm} \ \ \cdot\ \bar\Psi (x)\ \left[
i\not\hspace{-0.07cm}\partial_x\ a_{Ig}\left(x-x^\prime\right)
\ -\ m\ b_{Ig}\left(x-x^\prime\right) \right]\ \cdot\nonumber\\
\vspace{0.3cm}\nonumber\\
&&\hspace{0.7cm} \ \ \cdot\ S_{I(g)} (x^\prime - z)\ \left[
i\not\hspace{-0.07cm}\partial_z\ a_{Ig}\left(z-z^\prime\right)
\ -\ m\ b_{Ig}\left(z-z^\prime\right) \right]\ \Psi (z^\prime)
\ \cdot\nonumber\\
\vspace{0.3cm}\nonumber\\
&&\hspace{0.7cm} \ \ \cdot\ \left. \int d^4y\ d^4y^\prime\ \
\bar{J}_\mu (x,x^\prime;y)\ D_I^{\mu\nu}(y - y^\prime)\
\bar{J}_\nu (z,z^\prime;y^\prime)\ +\ \ldots\ \ \right\}
\end{eqnarray}

Here, $\Delta\Gamma^G_I$ is defined by eq.\ (\ref{ZA1}).
However, as is clear from eq.\ (\ref{DI11}) for the
present purpose in eq.\ (\ref{ZA2}) $a_I$, $b_I$ have
to be replaced by $a_{Ig}$, $b_{Ig}$ respectively.
Accordingly, the fermion propagator $S_{I(g)} (x)$
used here reads

\begin{eqnarray}
\label{DI13}
S_{I(g)} (x)\ =\ -\ \int {d^4p\over (2\pi)^4}\ {\rm e}^{\ ipx}\ \
{\not\hspace{-0.07cm}p\ \tilde a_{Ig}(p)\ -\ m\ \tilde b_{Ig}(p) \over
p^2\ \tilde a_{Ig}^2 (p)\ -\ m^2\ \tilde b_{Ig}^2 (p)\ +\ i\epsilon}
\hspace{1.cm} .
\end{eqnarray}

Eq.\ (\ref{DI12}) provides us with those terms of
the image $\Gamma_{II}$ of $\Gamma_I$ we need to
know for our further investigation. So we may now
proceed to apply the fixed point condition to
the kernel of the fermion action.\\

\subsubsection{\label{DS32}\label{SH4}The Integral Equation for the Kernel
of the Fermion Action}

Considering $\Gamma_{II} [0,\Psi,\bar\Psi] =
\Gamma^F_{II} [0,\Psi,\bar\Psi]$ and writing
the quadratic terms as

\begin{eqnarray}
\label{DK1a}
&&\hspace{-2.cm}\Gamma^F_{II} [0,\Psi,\bar\Psi]\ =\nonumber\\
\vspace{0.3cm}\nonumber\\
&&\hspace{-1.cm}=\ \int d^4x\ d^4x^\prime\ \ \bar\Psi (x)
\ \left[ i\ a_{II}\left(x-x^\prime\right)
\ \not\hspace{-0.07cm}\partial_{x^\prime}\
-\ m\ b_{II}\left(x-x^\prime\right) \right]\Psi (x^\prime)
\end{eqnarray}

eq.\ (\ref{DI12}) provides us with expressions for $a_{II}$, $b_{II}$.
Consequently, we may explicitly write down the
fixed point condition $a_I = a_{II}$, $b_I = b_{II}$.
For convenience, we do it in terms of $a_g$, $b_g$,
but any information obtained for these quantities
can be translated into terms of $a$, $b$ by means of relations
(\ref{DI10a}), (\ref{DI10b}). The integral equation
reads\footnote{Note, that the fixed point condition has been
multiplied by $g(x-z^\prime)$ yet.}

\begin{eqnarray}
\label{DK1}
&&\hspace{-1.5cm}\left[ g(x-z^\prime)\ -\ 1 \right] \left[
i\not\hspace{-0.07cm}\partial_x\ a_g\left(x-z^\prime\right)
\ -\ m\ b_g\left(x-z^\prime\right) \right]\ = \nonumber\\
\vspace{0.3cm}\nonumber\\
&&\hspace{-1.cm}=\ -i\ g(x-z^\prime)\
\left\{\ \ \int d^4x^\prime\ d^4z\ \left[
i\not\hspace{-0.07cm}\partial_x\ a_g\left(x-x^\prime\right)
\ -\ m\ b_g\left(x-x^\prime\right) \right]\right. \ \cdot\nonumber\\
\vspace{0.3cm}\nonumber\\
&&\hspace{0.5cm}\cdot\ S_{(g)} (x^\prime - z)\ \left[
i\not\hspace{-0.07cm}\partial_z\ a_g\left(z-z^\prime\right)
\ -\ m\ b_g\left(z-z^\prime\right) \right]\ \cdot\nonumber\\
\vspace{0.3cm}\nonumber\\
&&\hspace{0.5cm}\cdot\ \left. \int d^4y\ d^4y^\prime\ \
\bar{J}_\mu (x,x^\prime;y)\ D_I^{\mu\nu}(y - y^\prime)\
\bar{J}_\nu (z,z^\prime;y^\prime)\ +\ \ldots\ \ \right\}\ \ \ .
\end{eqnarray}

Eq.\ (\ref{DK1}) represents two coupled integral
equations for $a_g$, $b_g$ and needs now
to be solved. In general, this is a complicated
task and we will restrict ourselves to the solution
of eq.\ (\ref{DK1}) in the asymptotic UV (i.e.,
$-m^2 (x-z^\prime)^2 \rightarrow 0$) and IR (i.e.,
$-m^2 (x-z^\prime)^2 \rightarrow \infty $) regions
respectively \footnote{We always have the Euclidean region
in mind, of course.}. Before studying these cases let us
mention that eq.\ (\ref{DK1}) has an exact but trivial
solution, namely

\begin{eqnarray}
\label{DK2}
a_g(x)\ =\ a(x)\ \equiv & \ 0\ &\ \ \ \ , \\
\vspace{0.3cm}\nonumber\\
\label{DK3}
b_g(x)\ =\ b(x)\ =&\ \tilde b(\infty)\ \delta^{(4)}(x)&\ \ \ \ ,
\end{eqnarray}

where $\tilde b(\infty)$ is some arbitrary real
constant. Of course, this solution corresponds
to the non-interacting case where the gauge and
fermion sectors are decoupled and it is not very
interesting therefore. However, in the following
we will search the interacting solution of
eq.\ (\ref{DK1}) as sum of the trivial solution
(\ref{DK2}), (\ref{DK3}) and
some additional nontrivial contribution. As
already mentioned it seems to be rather
complicated to find a nontrivial and exact
solution of eq.\ (\ref{DK1}), but it appears possible
to analyze it merely exactly at least in the asymptotic UV region and
for small $\alpha$ to leading order in the IR region
solely based on those terms explicitly
displayed in eq.\ (\ref{DK1}). First we will turn
to the asymptotic UV region.\\

\paragraph{\label{DS321}Solving the Integral Equation in the
Asymptotic UV Region}

\hspace{4.cm}\\

Playing around with eq.\ (\ref{DK1}) one soon
recognizes, that to find a solution correct in
the asymptotic UV region one needs to assume
that the photon propagator $D_I^{\mu\nu}(x)$
is finite in the coincidence limit
$x \rightarrow 0$. Consequently, the photon
propagator written as
\footnote{An appropriate gauge fixing term $\Gamma_{gf}$
has been added to the gauge field action $\Gamma^G_I$,
i.e., it has been chosen
$\tilde n_\mu(p) = i p_\mu\ \tilde d(p)^{1/2}$.}

\begin{eqnarray}
\label{DV1}
D_I^{\mu\nu}(x)\ =\ -\ \int {d^4p\over (2\pi)^4}\
{{\rm e}^{\ ipx}\over p^2\ +\ i\epsilon}\ {1\over \tilde d(p)}\
\left[\ g^{\mu\nu}\
-\ (1-\lambda)\ {p^\mu p^\nu\over p^2\ +\ i\epsilon}\ \right]
\end{eqnarray}

reads in the coincidence limit

\begin{eqnarray}
\label{DV2}
D_I^{\mu\nu}(0)\ &=&\ i\ g^{\mu\nu}\ {3+\lambda\over 4}\
K_A\ m^2\ \ \ \ ,\\
\vspace{0.3cm}\nonumber\\
&&\hspace{1.7cm}K_A\ =\ {1\over 4\pi^2}\ \int_0^\infty\ ds\
{1\over \tilde d(s)}\ \ \ \ ,\nonumber
\end{eqnarray}

where $K_A$ is some finite, real constant.\\

\parindent1.5em
The analysis of the integral equation (\ref{DK1})
in the asymptotic UV region now starts by replacing
the photon propagator (\ref{DV1}) by its leading
short distance term (\ref{DV2}). Consequently, the
current-current interaction then reads in the short
distance limit

\begin{eqnarray}
\label{DV3}
&&\hspace{-2.5cm}\int d^4y\ d^4y^\prime\ \
\bar{J}_\mu (x,x^\prime;y)\ D_I^{\mu\nu}(y - y^\prime)\
\bar{J}_\nu (z,z^\prime;y^\prime)\ =\nonumber\\
\vspace{0.3cm}\nonumber\\
&&\ \ =\ i\ \alpha\pi\ (3+\lambda)\ K_A\
m^2\ (x-x^\prime)(z-z^\prime)\ +\ \ldots\hspace{1.cm},
\end{eqnarray}

\parindent0.em
and the function $g$ has the short distance behaviour

\begin{eqnarray}
\label{DV4}
g(x)\ =\ 1\ +\ {\alpha\pi\over 2}\ (3+\lambda)\
K_A\ m^2 x^2\ +\ \ldots\hspace{1.cm}.
\end{eqnarray}

The leading short distance terms (\ref{DV3}), (\ref{DV4})
have to be inserted into the integral equation (\ref{DK1})
yielding

\begin{eqnarray}
\label{DV5}
&&\hspace{-1.5cm}{1\over 2}\ (x-z^\prime)^2\ \left[
i\not\hspace{-0.07cm}\partial_x\ a_g\left(x-z^\prime\right)
\ -\ m\ b_g\left(x-z^\prime\right) \right]\ = \nonumber\\
\vspace{0.3cm}\nonumber\\
&&\hspace{-1.cm}=\ \int d^4x^\prime\ d^4z\ \left[
i\not\hspace{-0.07cm}\partial_x\ a_g\left(x-x^\prime\right)
\ -\ m\ b_g\left(x-x^\prime\right) \right]\ S_{(g)} (x^\prime - z)\ \
\cdot\nonumber\\
\vspace{0.3cm}\nonumber\\
&&\hspace{0.5cm}\cdot\ \left[
i\not\hspace{-0.07cm}\partial_z\ a_{g}\left(z-z^\prime\right)
\ -\ m\ b_{g}\left(z-z^\prime\right) \right]\
(x-x^\prime)(z-z^\prime)\ \ +\ \ldots\ \ \ \ \ .
\end{eqnarray}

Here, certain constants have been divided out.
Conveniently, we will now further consider
above integral equation in momentum space.
For this purpose we translate coordinate
difference factors occurring (i.e., $(x-z^\prime)^2$,
$(x-x^\prime)(z-z^\prime)$) into momentum
space derivatives. Having this in mind one may convince
oneself that to leading order terms
(indicated by dots $\ldots$) containing more
than just one photon propagator and which are not all
coupled to a closed fermion loop do not
contribute because they are related to a
higher number of derivatives in momentum
space (and those terms then are falling
faster off in the UV (i.e, high momentum) region).
Effectively, to those terms shown in
eq.\ (\ref{DV5}) only diagrams additionally
contribute where all photon propagators
are coupled to closed fermion loops. However,
these closed fermion loops can always be summed up to give
an effective (modified) photon propagator.
As long as its coincidence limit remains
finite eq.\ (\ref{DV5}) stays in effect.
So, the UV analysis can be done merely exactly.
In addition, already from eq.\ (\ref{DV5})
we recognize that the leading UV term of
the solution we are in search of is independent
of the coupling constant $\alpha$ as well as
of the structure of the gauge field action
(beyond condition (\ref{DG4})) determining
the constant $K_A$ what will give the UV
behaviour a kind of universal character.\\

\parindent1.5em
Eq.\ (\ref{DV5}) now reads in momentum space (the
subscript $_g$ is omitted for the moment)

\begin{eqnarray}
\label{DV6}
&&\hspace{-0.6cm}\not\hspace{-0.07cm}p\
\left[\ s\ \tilde a^{\prime\prime}\ +\ 3\ \tilde a^\prime\ \right]\ +\
m\ \left[\ s\ \tilde b^{\prime\prime}\ +\ 2\ \tilde b^\prime\ \right]\ =
\nonumber\\
\vspace{0.3cm}\nonumber\\
&&\hspace{-0.3cm}=\ {2\over s \tilde a^2 + \tilde b^2}\ \cdot\nonumber\\
\vspace{0.3cm}\nonumber\\
&&\ \cdot\left\{\ \not\hspace{-0.07cm}p\ \left[\
s^2\ \tilde a \left( \tilde a^\prime\right)^2\
+\ s\ \tilde a^2 \tilde a^\prime\
+\ 2\ s\ \tilde a^\prime \tilde b \tilde b^\prime\
-\ s\ \tilde a \left( \tilde b^\prime\right)^2\
-\ {1\over 2}\ \tilde a^3\
+\ \tilde a \tilde b \tilde b^\prime\
\right]\right.\ +\nonumber\\
\vspace{0.3cm}\nonumber\\
&&\ +\ m\ \left.\left[\
2\ s^2\ \tilde a \tilde a^\prime \tilde b^\prime\
-\ s^2\ \left( \tilde a^\prime\right)^2 \tilde b\
+\ s\ \tilde a^2 \tilde b^\prime\
-\ s\ \tilde a \tilde a^\prime \tilde b\
+\ s\ \tilde b \left( \tilde b^\prime\right)^2\
-\ \tilde a^2 \tilde b\
\right]\ \right\}\ +\nonumber\\
\vspace{0.3cm}\nonumber\\
&&\ +\ \ldots\ \
\end{eqnarray}

\parindent0.em
Here, the notation is $\tilde a = \tilde a (s)$,
$\tilde a^\prime = d/ds\ \tilde a$, $s = -p^2/m^2$.
We will now solve the two coupled differential
equations represented by eq.\ (\ref{DV6}) in the
asymptotic UV region $s\rightarrow\infty$. Our
Ansatz in accordance with conditions (\ref{DF9}), (\ref{DF10})
will be $\tilde a = \tilde a_s$,
$\tilde b = \tilde b(\infty) + \tilde b_s$,
where $\tilde a_s$, $\tilde b_s$ are assumed
to vanish power-like in leading order for
$s\rightarrow\infty$. Neglecting all clearly
nonleading terms the two coupled differential
equations yielded by eq.\ (\ref{DV6}) then read
\footnote{Note, that also a temporary transition
$\tilde a_s\rightarrow \tilde b(\infty)\ \tilde a_s$,
$\tilde b_s\rightarrow \tilde b(\infty)\ \tilde b_s$
has been applied and then the factor
$2\ \tilde b(\infty)$ has been divided out of
the equations below.}

\begin{eqnarray}
\label{DV7}
&&\hspace{-1.5cm}
{1\over 2}\ s\ \tilde a^{\prime\prime}_s\
+\ {3\over 2}\ \tilde a^\prime_s\ \ \
\stackrel{s \longrightarrow \infty}{=}\nonumber\\
\vspace{0.2cm}\nonumber\\
&&=\
s^2\ \tilde a_s \left(\tilde a^\prime_s\right)^2\
+\ s\ \tilde a^2_s \tilde a^\prime_s\
-{1\over 2}\ \tilde a^3_s\
+\ \left[\ 2\ s\ \tilde a^\prime_s\ +\ \tilde a_s\ \right]\
\tilde b^\prime_s\ +\ \ldots\ \ \ \ ,\\
\vspace{0.6cm}\nonumber\\
\label{DV8}
&&{1\over 2}\ s\ \tilde b^{\prime\prime}_s\
+\ \tilde b^\prime_s\ \ \
\stackrel{s \longrightarrow \infty}{=}\ \ \
-\ s^2\ \left(\tilde a^\prime_s\right)^2\
-\ s\ \tilde a_s \tilde a^\prime_s\
-\ \tilde a^2_s\
+\ \ldots\ \ \ \ .
\end{eqnarray}

Let us first discuss eq.\ (\ref{DV7}) and its
consequences on the asymptotic UV behaviour of
$\tilde b_s$. As long as the term on the
l.h.s.\ of eq.\ (\ref{DV7}) does not vanish
to leading order we are forced to conclude that
$\tilde b^\prime_s \stackrel{s \longrightarrow
\infty}{\sim}\ 1/s $,
i.e., $\tilde b_s \stackrel{s \longrightarrow
\infty}{\sim}\ \ln s$
\footnote{Of course, one could also try the assumption that the term
in front of $\tilde b^\prime_s$ vanishes (i.e.,
$\tilde a_s \stackrel{s \longrightarrow \infty}{\sim}\
 s^{-1/2}$), however eq.\
(\ref{DV8}) immediately leads to the same result
then.}. But, such a behaviour is in conflict with
gauge invariance because it is not in line with
condition (\ref{DF10}). So, we are forced to
conclude that the l.h.s.\ of eq.\
(\ref{DV7}) should vanish to leading order,
consequently it must hold ($C_{\tilde a}$
is some constant)

\begin{eqnarray}
\label{DV9}
\tilde a_s\ &\stackrel{s \longrightarrow \infty}{=}
&\ {C_{\tilde a}\over s^2}\ +\ \ldots \ \ \ .
\end{eqnarray}

This information is sufficient to determine
the leading behaviour of $\tilde b_s$ from
eq.\ (\ref{DV8}), and we find

\begin{eqnarray}
\label{DV10}
\tilde b_s\ &\stackrel{s \longrightarrow \infty}{=}
&\ -\ {C^2_{\tilde a}\over s^3}\ +\ \ldots\ \ \ .
\end{eqnarray}

We may now come back to eq.\ (\ref{DV7}) and determine
the next-to-leading term of $\tilde a_s$. Writing
$\tilde a_s$ without any loss of generality as

\begin{eqnarray}
\label{DV11}
\tilde a_s &=&{C_{\tilde a}\over s^2}\ \ \tilde v(s) \hspace{2.cm},\\
\vspace{0.3cm}\nonumber\\
&&\hspace{4.5cm}\tilde v(\infty)\ =\ 1\ \ \ \ ,\nonumber
\end{eqnarray}

and taking into account (\ref{DV9}), (\ref{DV10})
eq.\ (\ref{DV7}) then reads
\footnote{To be more precise, the vanishing of
the leading term on the l.h.s.\ of eq.\ (\ref{DV12})
(eq.\ (\ref{DV11}) inserted) rests on the relation
($\hat{p} = (-p_0,{\bf p})$)

\begin{eqnarray}
_p\Box\ {\not\hspace{-0.07cm}p\over\left[ p^2\right]^2}&=&
i\ 2\pi^2\ \not\hspace{-0.07cm}\partial_{\hat{p}}\
\delta^{(4)}(p)\ \ \ ,\nonumber
\end{eqnarray}

\parindent0.em
accompanied by certain reasonable assumptions about
$\tilde a_s(s\rightarrow 0)$ (i.e.,
$\tilde v \sim s^2, s\rightarrow 0$; or even some
weaker condition).}

\begin{eqnarray}
\label{DV12}
{C_{\tilde a}\over 2}\ {1\over s^2}\
\left[\ s\ \tilde v^{\prime\prime}\ -\ \tilde v^\prime\ \right]\ \
&\stackrel{s \longrightarrow \infty}{=}
&\ -\ {15\over 2}\ {C^3_{\tilde a}\over s^6}\ +\ \ldots\ \ \ \ .
\end{eqnarray}

And we find

\begin{eqnarray}
\label{DV13}
\tilde v(s)\ &\stackrel{s \longrightarrow \infty}{=}
&\ 1\ -\ {C^2_{\tilde a}\over s^3}\ +\ \ldots\ \ \ \ .
\end{eqnarray}

\parindent1.5em
Summarizing above results, one can say that eq.\
(\ref{DK1}) admits a (unique) solution respecting conditions
(\ref{DF9}), (\ref{DF10}). It behaves in the
asymptotic UV region as follows:

\begin{eqnarray}
\label{DV14}
\tilde a_g(s)\ &\stackrel{s \longrightarrow \infty}{=}
&\ {C_{\tilde a}\over s^2}\ \ \tilde b(\infty)\
\left[\ 1\ -\ {C^2_{\tilde a}\over s^3}\ +\ \ldots\ \right]\\
\vspace{0.3cm}\nonumber\\
\label{DV15}
\tilde b_g(s)\ &\stackrel{s \longrightarrow \infty}{=}
&\ \tilde b(\infty)\
\left[\ 1\ -\ {C^2_{\tilde a}\over s^3}\ +\ \ldots\ \right]
\end{eqnarray}

\parindent0.em
Most important, in qualitative respect
this asymptotic UV behaviour is
independent of the coupling constant $\alpha$ and
of any specific details of the photon propagator structure
beyond condition (\ref{DG4}). Furthermore, due to
$g(0) = 1\ $ (cf.\ eq.\ (\ref{DI9}))
$\tilde a$, $\tilde b$ exhibit the
same leading UV behaviour like $\tilde a_g$,
$\tilde b_g$. We will discuss
consequences of above results further below (see
subsection 4.3.3 and chapter 5). In the
next subsection we will study eq.\ (\ref{DK1}) in
the asymptotic IR region.\\

\paragraph{\label{DS322}Solving the Integral Equation in the
Asymptotic IR Region}

\hspace{3.cm}\\

For the IR analysis of the integral equation
(\ref{DK1}) we need to apply our Ansatz
(\ref{DG5}) to the photon propagator
\footnote{To obtain this propagator a gauge fixing term $\Gamma_{gf}$
with $\tilde n_\mu = i p_\mu\ \tilde d(p)^{1/2}$ has been
added to the gauge field action $\Gamma^G_I$.}.
Consequently, the current-current interaction
reads in the long distance limit to leading
order \footnote{Of course, it is not specifically
related to the Ansatz (\ref{DG5}), only
next-to-leading terms will be influenced.}

\begin{eqnarray}
\label{DR1}
&&\hspace{-1.5cm}\int d^4y\ d^4y^\prime\ \
\bar{J}_\mu (x,x^\prime;y)\ D_I^{\mu\nu}(y - y^\prime)\
\bar{J}_\nu (z,z^\prime;y^\prime)\ =\nonumber\\
\vspace{0.3cm}\nonumber\\
&&=\ i\ {\alpha\over\pi}\ \left\{\ {(1+\lambda)\over 2}\
{(x-x^\prime)(z-z^\prime)\over (x-z^\prime)^2}\ +\right.\nonumber\\
\vspace{0.3cm}\nonumber\\
&&\hspace{1.cm}+\ \left. (1-\lambda)\
{(x-x^\prime)(x-z^\prime)\ (x-z^\prime)(z-z^\prime)\
\over \left[\ (x-z^\prime)^2\ \right]^2}\ \right\}\ +\
\ldots\hspace{1.cm},
\end{eqnarray}

Here, $(x-x^\prime)^2$, $(z-z^\prime)^2$ are
understood to be small compared with $(x-z^\prime)^2$
\footnote{We always have in mind the region
$-m^2(x-z^\prime)^2 \rightarrow\infty $. More precisely,
for any large but fixed value of $(x-z^\prime)^2$
contributions from integration regions
in the integral equation (\ref{DK1}) where
$(x-x^\prime)^2$, $(z-z^\prime)^2$ are not small
compared to $(x-x^\prime)^2$ can be expected to be small due to
the expected decay of $a_g$, $b_g$ there. Furthermore, terms containing
higher powers of $1/(x-z^\prime)^2$ are suppressed in the asymptotic IR region
whatever their coefficient numerically might be.}.
The function $g$ has the long distance
behaviour (We give it here right for Euclidean space.
For the full expression and its derivation see Appendix B.)

\begin{eqnarray}
\label{DR2}
g(x_E)&=&C_g\ \left(m^2 x_E^2\right)^{\displaystyle\
\alpha (3-\lambda)/4\pi}\ \
{\rm e}^{\displaystyle\ -\ {\alpha\over 2 \sqrt{\beta}}
\ m \vert x_E \vert}\ \ \Big[\ 1\ + \ \ldots\ \Big]\ \ \ \ ,\ \ \ \\
\vspace{0.3cm}\nonumber\\
C_g&=&\left( 4\beta\right)^{\displaystyle\
- \alpha (3-\lambda)/4\pi}\ \
\exp\left\{\ {\alpha\over 4\pi}\left[\ (3+\lambda)\ +\
2\ (3-\lambda)\ \gamma\ \right]\ \right\}\ \ .\nonumber
\end{eqnarray}

Please note, that eq.\ (\ref{DR2}) contains the
Bloch-Nordsieck contribution
(cf.\ \cite{h},\cite{i} and references therein) exhibiting a
power-like behaviour with the
well-known exponent $\alpha (3-\lambda)/ 4\pi$.
It appears justified to assume that the leading
IR behaviour displayed in eqs.\ (\ref{DR1}),
(\ref{DR2}) will depend on additional terms to
be introduced in the Ansatz (\ref{DG5}) in order
also to satisfy condition (\ref{DG4}) very weakly
only. For the purpose of calculational simplicity
those terms can be safely disregarded therefore.\\

\parindent1.5em
We may now insert eqs.\ (\ref{DR1}), (\ref{DR2}) into the
integral equation (\ref{DK1}). Having in mind IR
analysis in Euclidean space on the l.h.s.\ of eq.\
(\ref{DK1}) we replace the factor
$[ 1 - g(x-z^\prime) ]$ simply by $1$ because this
is the leading contribution due to the exponential
decay (i.e., oscillation in Minkowski space) of
$g(x_E)$ for $m^2 x^2_E\rightarrow\infty$. Furthermore,
coordinate difference factors (i.e.,
$(x-x^\prime)_\mu$, $(z-z^\prime)_\nu$) occurring
on the r.h.s.\ of eq.\ (\ref{DR1}) are translated
into momentum space derivatives acting on the
Fourier transform of the kernel $S^{-1}$ of
the fermion action. So, eq.\ (\ref{DK1}) reads now

\begin{eqnarray}
\label{DR3}
&&\hspace{-1.5cm}\left[
i\not\hspace{-0.07cm}\partial_x\ a_g\left(x-z^\prime\right)
\ -\ m\ b_g\left(x-z^\prime\right) \right]\ = \nonumber\\
\vspace{0.3cm}\nonumber\\
&&\hspace{-1.cm}=\ {\alpha\over\pi}\ \ g(x-z^\prime)\ \cdot\nonumber\\
\vspace{0.3cm}\nonumber\\
&&\cdot\ \left[\ {(1+\lambda)\over 2}\
{g^{\mu\nu}\over (x-z^\prime)^2}\ +\ (1-\lambda)\
{(x-z^\prime)^\mu\ (x-z^\prime)^\nu \
\over \left[\ (x-z^\prime)^2\ \right]^2}\ \right]\ \cdot\nonumber\\
\vspace{0.3cm}\nonumber\\
&&\cdot\ \int {d^4p\over (2\pi)^4}\ \ {\rm e}^{\ ip(x-z^\prime)}\
\left[\ _p\partial_\mu\
\left(\ \not\hspace{-0.07cm}p\ \tilde a_g(p)\ +\ m\
\tilde b_g(p)\ \right)\ \right]\ \cdot\nonumber\\
\vspace{0.3cm}\nonumber\\
&&\cdot\ \
{\not\hspace{-0.07cm}p\ \tilde a_g(p)\ -\ m\ \tilde b_g(p) \over
p^2\ \tilde a_g^2 (p)\ -\ m^2\ \tilde b_g^2 (p)\ +\ i\epsilon}\
\left[\ _p\partial_\nu\
\left(\ \not\hspace{-0.07cm}p\ \tilde a_g(p)\ +\ m\
\tilde b_g(p)\ \right)\ \right]\ +\nonumber\\
\vspace{0.3cm}\nonumber\\
&&+\ \ldots\hspace{8.cm}.
\end{eqnarray}

\parindent0.em
What concerns the contribution of terms containing
more than just one photon propagator (indicated by
dots $\ldots $) following comments are due. Most of
those terms will finally yield higher powers of
$1/(x-z^\prime)^2$ at least and these can therefore
be neglected in the asymptotic IR region. However,
one should expect that also terms are occurring which
are of the same order as the 1-loop term given
above. However, such terms should be expected to only weakly
contribute numerically as long as $\alpha$ is
sufficiently small because each additional photon
propagator is accompanied by an additional
factor of $\alpha$. This argument is what is
left within the present approximative approach
of the line of reasoning applied in standard QED
perturbation theory. Of course, the belief based
on this reasoning may turn out wrong by nonperturbative
mechanisms which are not easily seen at the present
stage of the investigation. Anyway, in the region
where $\alpha$ is of order 1 terms containing more
than just one photon propagator cannot be neglected
anymore in principle. But for the purpose of the
present model calculation (without appealing to
the eventual range of $\alpha$) we simply ignore all
terms containing more than just one photon
propagator also in the region where $\alpha$
is not small.\\

\parindent1.5em
To determine the IR tail of $a_g$,
$b_g$ (i.e., the l.h.s.\ of eq.\ (\ref{DR3}))
it remains to find the leading long distance contribution
of the Fourier integral on the r.h.s.\ of eq.\ (\ref{DR3}).
To further proceed we would preferably need to
know the analytic structure of the integrand,
in particular that of the denominator.
We do not have any reliable information on this,
but as appears reasonable we will assume that
the integrand has a simple pole at some
$p_0 = \pm \sqrt{{\bf p}^2 - s_0\ m^2}$ with

\begin{eqnarray}
\label{DR4}
s_0 &=& -\ {\tilde b^2_g(s_0)\over\tilde a^2_g(s_0)}\ \ \ \ \ \ ,\ \
(\ s_0\ <\ 0\ )\ \ \ \ ,
\end{eqnarray}

\parindent0.em
and that just this pole determines the leading
long distance behaviour of the Fourier integral. Consequently,
we may exploit the residue of this pole and the
leading long distance contribution of the Fourier integral
is simply given by the product of the nominator
of its integrand (appropriately treated by
considering $p_\kappa$ factors occurring as configuration space
derivatives acting on eq.\ (\ref{DR5})) taken at $p^2/m^2 = - s_0$ and
the leading long distance term of

\begin{eqnarray}
\label{DR5}
{1\over \tilde a_g^2 (s_0)}\
\int {d^4p\over (2\pi)^4}\ {{\rm e}^{\ ip(x-z^\prime)}
\over p^2\ +\ s_0\ m^2\ +\ i\epsilon}\hspace{2.cm}.\
\end{eqnarray}

\parindent1.5em
The explicit calculation is now straightforward
but somewhat tedious and we will
comment few points only. So, as one intermediate step
one calculates the following useful
relation ($\tilde a^\prime = d/ds\ \tilde a(s)$, $s = - p^2/m^2$).

\begin{eqnarray}
\label{DR6}
&&\hspace{-1.6cm}\left[\ _p\partial_\mu\
\left(\ \not\hspace{-0.07cm}p\ \tilde a(p)\ +\ m\
\tilde b(p)\ \right)\ \right]\
\left[\ \not\hspace{-0.07cm}p\ \tilde a(p)\ -\ m\ \tilde b(p)\ \right]\
\cdot\nonumber\\
\vspace{0.3cm}\nonumber\\
&&\hspace{-1.4cm}\cdot\
\left[\ _p\partial_\nu\
\left(\ \not\hspace{-0.07cm}p\ \tilde a(p)\ +\ m\
\tilde b(p)\ \right)\ \right]\ =\nonumber\\
\vspace{0.3cm}\nonumber\\
&&\hspace{-0.8cm}=\
\left[\ \not\hspace{-0.07cm}p\ \tilde a\ -\ m\ \tilde b\ \right]\
\left[\
\gamma_\mu \gamma_\nu\ \tilde a^2\
+\ 4\ {p_\mu p_\nu\over m^2}\ {p^2\over m^2}\ \left(\tilde a^\prime\right)^2\
+\ 4\ {p_\mu p_\nu\over m^2}\ \left(\tilde b^\prime\right)^2\
\right. -\nonumber\\
\vspace{0.3cm}\nonumber\\
&&
-\ {2\over m^2}\ \left(\gamma_\mu p_\nu\ +\ \gamma_\nu p_\mu\right)\
 \not\hspace{-0.07cm}p\ \tilde a \tilde a^\prime\
-\ {2\over m^2}\ \left(\gamma_\mu p_\nu\ +\ \gamma_\nu p_\mu\right)\
\tilde a \tilde b^\prime\ + \nonumber\\
\vspace{0.3cm}\nonumber\\
&&\left.
+\ 8\ \not\hspace{-0.07cm}p\ {p_\mu p_\nu\over m^3}\
 \tilde a^\prime \tilde b^\prime\ \right]\
+\ \left(\gamma_\mu \not\hspace{-0.07cm}p \gamma_\nu\
-\ \not\hspace{-0.07cm}p \gamma_\mu \gamma_\nu\right)\ \tilde a^3\
-\ 2\ \gamma_\mu p_\nu\ {p^2\over m^2}\ \tilde a^2\tilde a^\prime\
+\nonumber\\
\vspace{0.3cm}\nonumber\\
&&
+\ 2\ \not\hspace{-0.07cm}p \gamma_\mu \not\hspace{-0.07cm}p\
 {p_\nu\over m^2}\ \tilde a^2 \tilde a^\prime\
-\ 2\ \left(\gamma_\mu \not\hspace{-0.07cm}p\
 -\ \not\hspace{-0.07cm}p \gamma_\mu\right)\ {p_\nu\over m}\
 \tilde a^2 \tilde b^\prime\
\end{eqnarray}

\parindent0.em
In performing the calculation we always
keep track of those terms contributing
in the long distance region to leading order only.
In particular, the leading long distance term
of eq.\ (\ref{DR5}) is read off from the
relation (written for Euclidean space here)

\begin{eqnarray}
\label{DR7}\hspace{-0.4cm}
\int {d^4p_E\over (2\pi)^4}\ {{\rm e}^{\ ip_E x_E}
\over p^2_E\ +\ m^2 }\ &=&\
{m\over 4\pi^2\ \vert x_E\vert}\ K_1\left( m\vert x_E\vert\right)
\nonumber\\
\vspace{0.8cm}\nonumber\\
&\stackrel{m^2 x^2_E \gg 1}{=}&\
{\sqrt{m}\over 2\ (2\pi)^{3/2}\ \vert x_E\vert^{3/2}}\ \
{\rm e}^{\displaystyle\ -\ m \vert x_E\vert}\ \
\Big[\ 1\ +\ \ldots\ \Big]\ .\ \ \
\end{eqnarray}

The result obtained this way for the IR tail of $a_g$, $b_g$
then is (We give this and all further results for Euclidean space.)

\begin{eqnarray}
\label{DR8}
a_g(x_E)&\stackrel{m^2 x^2_E \rightarrow \infty}{=}&
m^4\ {\alpha\ C_g\ G\over (2\pi)^{5/2}}\ \
{(-s_0)^{3/4}\ \tilde a_g(s_0)\over \sqrt{-s_0}\ +\ \alpha/2\sqrt{\beta}}\ \
(m \vert x_E\vert)^{-7/2\ +\ \alpha (3-\lambda)/2\pi}
\ \cdot\nonumber\\
\vspace{0.3cm}\nonumber\\
&&\ \ \ \cdot\
{\rm e}^{\displaystyle\ -\ (\ \sqrt{-s_0}\ +\ \alpha/2\sqrt{\beta}\ )\
m \vert x_E\vert}\ \
\Big[\ 1\ +\ \ldots\ \Big]\\
\vspace{0.3cm}\nonumber\\
\label{DR9}
b_g(x_E)&\stackrel{m^2 x^2_E \rightarrow \infty}{=}&
m^4\ {\alpha\ C_g\ H\over (2\pi)^{5/2}}\ \
(-s_0)^{3/4}\ \tilde a_g(s_0)\ \
(m \vert x_E\vert)^{-7/2\ +\ \alpha (3-\lambda)/2\pi}
\ \cdot\nonumber\\
\vspace{0.3cm}\nonumber\\
&&\ \ \ \cdot\
{\rm e}^{\displaystyle\ -\ (\ \sqrt{-s_0}\ +\ \alpha/2\sqrt{\beta}\ )\
m \vert x_E\vert}\ \
\Big[\ 1\ +\ \ldots\ \Big]\\
\vspace{0.3cm}\nonumber\\
\label{DR10}
G&=&-{3\over 2}\ (1+\lambda)\ + \nonumber\\
\vspace{0.3cm}\nonumber\\
&&\ \ \ +\ 2\ (3-\lambda)\
\left[\ s_0\ \ {\tilde a^\prime_g(s_0)\over \tilde a_g(s_0)}\
+\ \sqrt{-s_0}\ \ {\tilde b^\prime_g(s_0)\over \tilde a_g(s_0)}\
+\ {1\over 2}\ \right]^2 \\
\vspace{0.3cm}\nonumber\\
\label{DR11}
H&=&-3\ (1+\lambda)\ -\ G
\end{eqnarray}

It should find mention that eq.\ (\ref{DR10}) provides
us with an implicit expression for $G$ only
because in view of eqs.\ (\ref{DR9}), (\ref{DR11})
its r.h.s.\ also depends on $G$ via the term
$\tilde b^\prime_g(s_0)/\tilde a_g(s_0)$. Therefore,
eq.\ (\ref{DR10}) represents a cubic equation for the
value of $G$ which has always at least one
(real) solution. From eq.\ (\ref{DR10}) one recognizes
that $G$ is a RG invariant quantity, i.e.,
it is invariant against (finite) mass and (fermion) wave
function renormalizations (We will discuss the
normalization issue further below.).\\

\parindent1.5em
Taking into account the definitions
(\ref{DI10a}), (\ref{DI10b})  we find from
eqs.\ (\ref{DR8}), (\ref{DR9}) the IR tail of $a$, $b$.

\begin{eqnarray}
\label{DR12}
a(x_E)&\stackrel{m^2 x^2_E \rightarrow \infty}{=}&
m^4\ {\alpha\ G\over (2\pi)^{5/2}}\ \ (-s_0)^{1/4}\
\tilde a_g(s_0)\ \ \
(m \vert x_E\vert)^{-7/2}\ \cdot\nonumber\\
\vspace{0.3cm}\nonumber\\
&&\ \ \ \cdot\
{\rm e}^{\displaystyle\ -\ \sqrt{-s_0}\ m \vert x_E\vert}\ \
\Big[\ 1\ +\ \ldots\ \Big]\\
\vspace{0.3cm}\nonumber\\
\label{DR13}
b(x_E)&\stackrel{m^2 x^2_E \rightarrow \infty}{=}&
m^4\ {\alpha\ H\over (2\pi)^{5/2}}\ \ (-s_0)^{3/4}\
\tilde a_g(s_0)\ \ \
(m \vert x_E\vert)^{-7/2}\ \cdot\nonumber\\
\vspace{0.3cm}\nonumber\\
&&\ \ \ \cdot\
{\rm e}^{\displaystyle\ -\ \sqrt{-s_0}\ m \vert x_E\vert}\ \
\Big[\ 1\ +\ \ldots\ \Big]
\end{eqnarray}

\parindent0.em
 From above equations we see that the
IR tails of $a$, $b$ agree qualitatively
(The same is true for $a_g$, $b_g$.).\\

\parindent1.5em
After having obtained the functional dependence
of the kernel of the fermion action in the
asymptotic IR region we still need to fix the
arbitrary constants involved (In particular,
this will require to discuss the normalization
issue not touched so far.). For this
purpose we have to calculate the Fourier
transforms of $a_g$, $b_g$, and that of
$a$, $b$ the latter of which are determined by
the solution of the integral equation (\ref{DK1})
via eqs.\ (\ref{DI10a}), (\ref{DI10b}). It appears
reasonable to represent these Fourier transforms
in the low $s$ region
\footnote{In the following we will
deliberately leave open the precise meaning of this term
and we will return to the issue in section \ref{SH2} only.}
appropriate for the normalization purposes we are aiming at
by the sum of the Fourier
transforms of the trivial solution (\ref{DK2}),
(\ref{DK3}) and the Fourier transforms $\tilde a_{sg}$,
$\tilde b_{sg}$, $\tilde a_s$, $\tilde b_s$ of the
IR tails of $a_g$, $b_g$ and $a$, $b$ given
in eqs.\ (\ref{DR8}), (\ref{DR9}) and (\ref{DR12}),
(\ref{DR13}) respectively. So, we simply extend the long
distance representations (\ref{DR8}), (\ref{DR9}),
(\ref{DR12}), (\ref{DR13})
to the whole configuration space and expect that this
procedure will give reasonable results in the
low $s$ region at least.\\

To the calculation of the Fourier transforms following
formula applies \cite{g}.

\begin{eqnarray}
\label{DR14}
&&\hspace{-1.5cm}\int d^4x_E\ \ {\rm e}^{\ -ip_E x_E}\ \
\left( x^2_E\right)^\kappa\ \
{\rm e}^{\displaystyle\ - \rho \vert x_E\vert}\ \ \ =\nonumber\\
\vspace{0.3cm}\nonumber\\
&&=\ -\ {4\pi^2\ \Gamma\left(4+2\kappa\right)\over
\vert p_E\vert\ \left(\rho^2+p^2_E\right)^{3/2\ +\ \kappa}}\ \
P^{-1}_{2(1+\kappa)}\left({\rho\over\sqrt{ \rho^2+p^2_E}}\right)\nonumber\\
\vspace{0.5cm}\nonumber\\
&&=\ \ \ {4\pi^2\ \Gamma(3+2\kappa)\ \ \over p^2_E\
\left(\rho^2+p^2_E\right)^{3/2\ +\ \kappa} }\ \cdot\nonumber\\
\vspace{0.3cm}\nonumber\\
&&\ \ \ \ \ \ \cdot\
\left[\ \sqrt{ \rho^2+p^2_E}\
P_{1+2\kappa}\left({\rho\over\sqrt{ \rho^2+p^2_E}}\right)\ -\
\rho\ P_{2(1+\kappa)}\left({\rho\over\sqrt{ \rho^2+p^2_E}}\right)\
\right]\\
\vspace{0.5cm}\nonumber\\
&&\hspace{7.cm}Re\ \rho\ >\ 0\ ,\ \
Re\ \kappa\ > -2\nonumber
\end{eqnarray}

\parindent0.em
Having in mind continuation to Minkowski space,
please note that (more precisely) the condition
$\left\vert Im\ \vert p_E \vert \right\vert\ <\ Re\ \rho$
is to be respected.
Although it is less compact in the following
we will always exploit the lower representation
of eq.\ (\ref{DR14}) because we find it more convenient for
an eventual transition back to Minkowski space.\\

\parindent1.5em
For $\tilde a_g$ given in the
low $s$ region as Fourier transform of
eq.\ (\ref{DR8}) we obtain the following result.

\begin{eqnarray}
\label{DR15}
\hspace{-0.5cm}
\tilde a_g(s)&=&{\alpha\ C_g\ G\over \sqrt{2\pi}}\ \
\Gamma\left(-{1\over 2}\ +\ \alpha\ {(3-\lambda)\over 2\pi}\ \right)\
(-s_0)^{3/4}\ \tilde a_g(s_0)\ \cdot\nonumber\\
\vspace{0.3cm}\nonumber\\
&&\ \ \ \cdot\ {1\over s}\
\left[\ (\sqrt{-s_0}\ +\ \alpha/2\sqrt{\beta})^2\ + s\ \right]^{1/4\
-\ \alpha (3-\lambda)/4\pi}\ \cdot\nonumber\\
\vspace{0.3cm}\nonumber\\
&&\ \ \ \cdot\ \left[\ \sqrt{\ 1\ +\
{s\over (\sqrt{-s_0}\ +\ \alpha/2\sqrt{\beta})^2}}\
\right.\cdot\nonumber\\
\vspace{0.3cm}\nonumber\\
&&\ \ \ \ \ \ \cdot\ P_{-5/2\ +\ \alpha (3-\lambda)/2\pi}\left(\left(
1\ +\ {s\over (\sqrt{-s_0}\ +\ \alpha/2\sqrt{\beta})^2}\right)^{-1/2}
\right) \ -\nonumber\\
\vspace{0.3cm}\nonumber\\
&&\ \ \ \ \ \ -\ \left. P_{-3/2\ +\ \alpha (3-\lambda)/2\pi}\left(\left(
1\ +\ {s\over (\sqrt{-s_0}\ +\ \alpha/2\sqrt{\beta})^2}\right)^{-1/2}
\right)\ \right]
\end{eqnarray}

\parindent0.em
By specifying $s = s_0$ (this corresponds
to an analytic continuation to Minkowski space) above equation
leads to a consistency equation (the value of $\tilde a_g(s_0)$ drops out)
yielding a first relation among the parameters of the IR solution. It reads

\begin{eqnarray}
\label{DR16}
1&&=\ -\ \alpha^{1\ -\ \alpha(3-\lambda)/2\pi}\ \ {G\over \sqrt{2\pi}}\ \
\Gamma\left(-{1\over 2}\ +\ \alpha\ {(3-\lambda)\over 2\pi}\ \right)\ \
\ \cdot\nonumber\\
\vspace{0.3cm}\nonumber\\
&&\ \ \ \ \ \cdot\
\exp\left\{\ {\alpha\over 4\pi}\left[\ (3+\lambda)\ +\
2\ (3-\lambda)\ \gamma\ \right]\ \right\}\
w^{-1/2}\ \ (1 + 2 w)^{1/4\ -\ \alpha (3-\lambda)/4\pi}\ \cdot\nonumber\\
\vspace{0.3cm}\nonumber\\
&&\ \ \ \ \ \cdot\ \left[\ {\sqrt{1 + 2 w}\over 1 + w}\
P_{-5/2\ +\ \alpha (3-\lambda)/2\pi}\left(\
{1 + w\over\sqrt{1 + 2 w} }\ \right) \ - \right.\nonumber\\
\vspace{0.3cm}\nonumber\\
&&\hspace{3.5cm} -\ \left. P_{-3/2\ +\ \alpha (3-\lambda)/2\pi}\left(\
{1 + w\over\sqrt{1 + 2 w} }\ \right)\ \right]\ \ \ \ ,\ \\
\vspace{1.3cm}\nonumber\\
\label{DR17}
&&\hspace{7.5cm}w\ =\ {2\over\alpha}\ \sqrt{-s_0\ \beta}\ \ \ \ .
\end{eqnarray}

Here, $G$ is understood as a function of $w$ and $\alpha$
(and $\lambda$). It is given as
solution of the following cubic equation derived
from eq.\ (\ref{DR10}).

\begin{eqnarray}
\label{DR18}
&&\hspace{-1.5cm}
G^3\ +\ \left\{\ {3\over 2}\ (1 + \lambda)\ -\
2\ (3 - \lambda)\ \left[\ \left( 2 + {1\over w}\right)\
L(w,\alpha)\ +\ {1\over 2}\ \right]^2\ \right\}\ G^2\ -\nonumber\\
\vspace{0.3cm}\nonumber\\
&&\hspace{-1.cm}-\ 12\ (3 - \lambda)\
(1 + \lambda)\ \left( 1 + {1\over w}\right)\
\left[\ \left( 2 + {1\over w}\right)\ L(w,\alpha)\ +\ {1\over 2}\ \right]\
L(w)\ \ G -\nonumber\\
\vspace{0.3cm}\nonumber\\
&&\hspace{-1.cm}-\ 18\ (3 - \lambda)\ (1 + \lambda)^2\
\left(1 + {1\over w}\right)^2\ L(w,\alpha)^2\ \ =\ \ 0\\
\vspace{0.5cm}\nonumber\\
&&\hspace{6.5cm}
L(w,\alpha)\ =\ s_0\ \ {\tilde a^\prime_g(s_0)\over \tilde a_g(s_0)}\nonumber
\end{eqnarray}

To obtain this cubic equation we have made use of the relation

\begin{eqnarray}
\label{DR19}
\tilde b_g(s)&=&-\ \sqrt{-s_0}\
\left(\ 1 + {1\over w}\ \right)\
\left[1\ +\ {3\ (1 + \lambda)\over G}\right]\
\tilde a_g(s)\ +\ \tilde b(\infty)
\end{eqnarray}

based on eqs.\ (\ref{DR8}), (\ref{DR9}) and therefore
valid in the low $s$ region only. We see that
solutions $G$ of equation (\ref{DR18}) are functions
of $w$, $\alpha$ while solutions $w$ of eq.\
(\ref{DR16}) exclusively depend on $\alpha$ (and on $\lambda$,
in principle, if for conceptual reasons
we were not to set it to zero as outlined in chapter 3).
Clearly, they do not depend on $\tilde b(\infty)$.
Although numerically the discriminant of eq.\ (\ref{DR18})
turns always out to be negative in the relevant domain,
only one of the three real solutions of eq.\ (\ref{DR18})
then proves appropriate to find a solution
of eq.\ (\ref{DR16}) furthermore.
In general, solutions $G$, $w(\alpha)$ of
above equations can be found numerically only (For a
plot of numerical results see figs.\ 1, 2.). But,
for sufficiently small $\alpha$ ($\alpha \ll 1$)
$w(\alpha)$ turns out to be large ($w(\alpha) \gg 1$)
and eq.\ (\ref{DR16}) admits an analytical solution in
this region. This asymptotic solution will be studied now.\\

\parindent1.5em
We investigate the case $\alpha \ll 1$ (We
assume that the solution $w(\alpha)$ in this region
will be much larger than one.). Let us start with the following
asymptotic representation \cite{j}.

\begin{eqnarray}
\label{DR20}
&&\hspace{-1.5cm}z^{-1/2\ +\ \kappa}\
\left[\ z^{-1}\ P_{-5/2\ +\ \kappa}(z)\ -\ P_{-3/2\ +\ \kappa}(z)\ \right]\ =
\nonumber\\
\vspace{0.3cm}\nonumber\\
&&=\ \left( {1\over 2}\ -\ \kappa \right)\
{\Gamma\left(1\ -\ \kappa\right)\over
\Gamma\left({5\over2}\ -\ \kappa\right)}\
{2^{1/2\ -\ \kappa}\over \sqrt{\pi}}\
\cdot\nonumber\\
\vspace{0.3cm}\nonumber\\
&&\ \ \ \ \cdot\ \left[\ 1\ -\ {1\over 4 \kappa\ z^2}\
\left( {1\over 2}\ -\ \kappa \right)\
\left( {3\over 2}\ -\ \kappa \right)\ \ \right.
\cdot \nonumber\\
\vspace{0.3cm}\nonumber\\
&&\ \ \ \ \cdot\ \left.\left(\ {(2z)^{2\kappa}\over 1 - \kappa}\
{\Gamma\left({1\over2}\ -\ \kappa\right)\over
\Gamma\left({1\over2}\ +\ \kappa\right)}\
{\Gamma\left(1\ +\ \kappa\right)\over
\Gamma\left(1\ -\ \kappa\right)}\ -\ 1\ \right)\ +\
O\left( z^{-2(2\ -\ \kappa)}\right)\ \right]\\
\vspace{0.3cm}\nonumber\\
&&\hspace{7.5cm}\kappa\ >\ 0\ \ ,\ \ \vert z\vert \ \gg\ 1\nonumber\\
\vspace{0.3cm}\nonumber\\
\label{DR21}
&&=\ {2\ \sqrt{2}\over 3\pi}\
\left[\ 1\ -\ {3\over 16}\ z^{-2}\
\left[\ 2\ \ln 8z\ +\ 1\ \right]\ +\ O(z^{-4}\ln z)\ \right]\ \ ,\\
\vspace{0.5cm}\nonumber\\
&&\hspace{7.5cm}\kappa\ =\ 0\ \ ,\ \ \vert z\vert \ \gg\ 1\nonumber
\end{eqnarray}

\parindent0.em
Then, from eq.\ (\ref{DR16}) one finds (Here,
$\ln w(\alpha)$ is thought to grow for small $\alpha$
like $\alpha^{-1/2}$ at most.)

\begin{eqnarray}
\label{DR22}
G&=&{3\pi\over 4\alpha}\
\left\{\ 1\ -\ {\alpha\over 4\pi}\
 \left[\ (3+\lambda)\ +\ 2\ (3-\lambda)\
  \left({8\over 3}\ -\ \ln\left[{2^5 w(\alpha)\over\alpha}\right]\ \right)\
\right]\right.\ +\nonumber\\
\vspace{0.3cm}\nonumber\\
&&\hspace{1.5cm}+\left. {1\over 2}\
\left({\alpha (3 - \lambda)\over 2\pi}\right)^2\ \ln w(\alpha)\
\ln\left[\alpha^4 w(\alpha)\right]\ +\
O(\alpha^{3/2})\ \right\}\ \ .
\end{eqnarray}

Taking into account (\ref{DR22}) eq.\ (\ref{DR19}) can then be inserted on
the r.h.s.\ of eq.\ (\ref{DR10}) and eq.\ (\ref{DR22})
on its l.h.s.. The solution of the resulting
equation for $w(\alpha)$ is now straightforward.
One finds for small $\alpha$

\begin{eqnarray}
\label{DR23}
\hspace{-2.cm}w(\alpha)&=&{1\over 32}\
\exp\left\{\ {2\over 3}\ \sqrt{{2\pi\over \alpha\ (1-\lambda/3)}}\ +\
4\ +\ \sqrt{{\alpha\ (1-\lambda/3)\over 2\pi}}\ \ln \alpha\ -\
\right.\nonumber\\
\vspace{0.3cm}\nonumber\\
&&\hspace{1.5cm}-\left.
{1\over 6}\ \sqrt{{\alpha\ (1-\lambda/3)\over 2\pi}}\
\left[\ {59\over 3}\ +\ {38\lambda\over (3-\lambda)}\ \right]\
+\ O\left(\alpha\right)\ \right\}\ \ ,\\
\vspace{0.5cm}\nonumber\\
&&\hspace{8.5cm}\alpha\ \ll\ 1\ \ \ .\nonumber
\end{eqnarray}

Note, that higher loop contributions possibly to be taken
into account in the integral equation (\ref{DK1})
will influence above result via the
last term in the exponent only.
To see this simply replace in the
first term in the exponent $\alpha$ by
$\alpha [1 + O(\alpha)]$. Finally, using (\ref{DR23})
one finds from eq.\ (\ref{DR22}) following expression
for $G(\alpha)$.

\begin{eqnarray}
\label{DR24}
\hspace{-0.5cm}
G(\alpha)&=&{3\pi\over 4\alpha}\
\left\{\ 1\ +\ 2\ \sqrt{{\alpha\ (1 - \lambda/3)\over 2\pi}}\ +\
{\alpha\ (3 - \lambda)\over 2\pi}\ \ln \alpha\ +\right.\nonumber\\
\vspace{0.3cm}\nonumber\\
&&+\left. \ {\alpha\ (9 - 5 \lambda)\over 4\pi}\ +\
15\ \left({\alpha\ (1 - \lambda/3)\over 2\pi}\right)^{3/2}\ \ln \alpha\ +\
O\left(\alpha^{3/2}\right)\ \right]\\
\vspace{0.5cm}\nonumber\\
&&\hspace{8.5cm}\alpha\ \ll\ 1\ \ \ .\nonumber
\end{eqnarray}

\parindent1.5em
The next task is to find the solution $s_0$ of eq.\ (\ref{DR4}).
But, any solution $s_0$ can sensibly be related to physics only if the
mass normalization to be used is specified. So,
before attempting the task to find $s_0$
we discuss the normalization issue in somewhat
greater detail now.\\

Let us assume we had determined $s_0$. Then,
whatever normalization of $\tilde a_g(s_0)$ is
applied eq.\ (\ref{DR4}) yields the value of
$\tilde b_g(s_0)$, and in our specific
case the value of $\tilde b(\infty)$ because
eq.\ (\ref{DR9}) is not independent of eq.\ (\ref{DR8}).
Now, let a certain function $\hat{g}\ =\ \hat{g}(-m^2x^2)$
with $\hat{g}(0) = 1$ define a map
$\hat{{\bf g}}: a_g\longrightarrow a_{g\hat{g}},\
b_g\longrightarrow b_{g\hat{g}}$
by applying the prescriptions (\ref{DI10a}),
(\ref{DI10b}) to $\hat{g}$. Considering the equation

\begin{eqnarray}
\label{DR25}
s_1 &=& -\ {\tilde b^2_{g\hat{g}}(s_1)\over
\tilde a^2_{g\hat{g}}(s_1)}\ \ \ \ \ \ ,\ \
(\ s_1\ <\ 0\ )\ \ \ \ ,
\end{eqnarray}

\parindent0.em
the map $\hat{\bf g}$ obviously induces a map
$\hat{{\bf g}}_s: s_0 \longrightarrow s_1$. If
$\hat{g} \equiv 1$, $\hat{\bf g}$ and $\hat{{\bf g}}_s$ are the
identity maps. If we specifically choose
$\hat{g} = g^{-1}$, then $\hat{\bf g}$ is the inverse of
$\bf g$ and it holds $a_{g\hat{g}} = a$,
$b_{g\hat{g}} = b$ (cf.\ eqs.\ (\ref{DI10a}), (\ref{DI10b})).
However, $a$, $b$ are related to physics and we
would like to formulate normalization conditions
in their terms, i.e., we naturally prefer to impose standard
normalization conditions on $\tilde a$,
$\tilde b$ (i.e., mass shell normalization at
the physical electron mass $m$):

\begin{eqnarray}
\label{DR26}
\tilde a(s_1=-1)&=\ \pm\ \tilde b(s_1=-1)\ =&N^{-1}_2\ =\ 1\hspace{2.cm}.
\end{eqnarray}

By other words, we of course require that the fermion
propagator derived from the effective action we are in search of
has a pole related to the physical electron mass $m$.
In eq.\ (\ref{DR26}) $N_2$ is the (fermion) wave function normalization
constant \footnote{A (finite) wave function renormalization
corresponds to a change in $N_2$.}.
Note, that it is always possible to
choose $s_1 = -1$ because in our set-up there
exists a scaling symmetry $m \rightarrow \tau m$
($s \rightarrow s/\tau^2 $), $\beta \rightarrow \tau^2 \beta$,
$b \rightarrow b/\tau $ for any non-zero real parameter $\tau$
(RG invariance against (finite) mass renormalizations).
Consequently, we now apply the inverse map
$\ \hat{{\bf g}}^{-1}_s: s_1 \longrightarrow s_0$ to determine
$s_0$.\\

\parindent1.5em
Taking into account (cf.\ eqs.\ (\ref{DR12}), (\ref{DR13}))

\begin{eqnarray}
\label{DR27}
\tilde b(s)&=&-\ \sqrt{-s_0}\
\left[1\ +\ {3\ (1 + \lambda)\over G}\right]\
\tilde a(s)\ +\ \tilde b(\infty)
\end{eqnarray}

\parindent0.em
(valid in the low $s$ region) and the low $s$ result
for the Fourier transform of $a$

\begin{eqnarray}
\label{DR28}
&&\hspace{-1.cm}\tilde a(s)\ =\
\sqrt{2}\ \ \alpha\ G(\alpha)\ \tilde a(s_0)\ \ {s_0\over s}\
\left(1-{s\over s_0}\right)^{1/4}\ \cdot\nonumber\\
\vspace{0.3cm}\nonumber\\
&&\ \ \ \ \ \cdot\ \left[\ \sqrt{1-{s\over s_0}}\
P_{-5/2}\left(\left(1-{s\over s_0}\right)^{-1/2}\right)\ -\
P_{-3/2}\left(\left(1-{s\over s_0}\right)^{-1/2}\right)\ \right]
\end{eqnarray}

we conveniently calculate for the $s_1$-pole
via $\sqrt{-s_1}\ \tilde a(s_1) = \pm \tilde b(s_1)$
the value of the RG invariant quantity
$\tilde b(\infty)/(\sqrt{-s_0}\ \tilde a_g(s_0))$
(i.e., the value of the RG variant quantity
$\tilde b(\infty)$ expressed in terms of $s_0$
and $\tilde a_g(s_0)$). We find

\begin{eqnarray}
\label{DR29}
&&\hspace{-1.cm}{\tilde b(\infty)\over\sqrt{-s_0}\ \tilde a_g(s_0)}\ =
\nonumber\\
\vspace{0.3cm}\nonumber\\
&&\hspace{-0.5cm}
=\sqrt{2\ u}\ \alpha\ G(\alpha)\ \left\{\ \pm 1\ +\ \sqrt{u}\
\left[1\ +\ {3 (1 + \lambda)\over G(\alpha)}\right]\right\}
\left(1-u^{-1}\right)^{1/4}\ \cdot\nonumber\\
\vspace{0.3cm}\nonumber\\
&&\ \cdot\ \left[\ \sqrt{1-u^{-1}}\
P_{-5/2}\left(\left(1-u^{-1}\right)^{-1/2}\right)\ -\
P_{-3/2}\left(\left(1-u^{-1}\right)^{-1/2}\right)\ \right]\ ,\ \\
\vspace{0.5cm}\nonumber\\
&&\hspace{9.5cm}u\ =\ {s_0\over s_1}\hspace{1.5cm} .\nonumber
\end{eqnarray}

The same quantity can now be found from
the $s_0$-pole via
$\sqrt{-s_0}\ \tilde a_g(s_0) = \tilde b_g(s_0)$
\footnote{We omit the other root
$\sqrt{-s_0}\ \tilde a_g(s_0) = - \tilde b_g(s_0)$
because one does not find any solution $s_0$ in this case.}
and both values have to agree,
of course, what provides us with an equation
for $s_0$ measured in units
of $s_1$, which is in our case ($s_1 = -1$) related to the
physical electron mass $m$. The equation reads

\begin{eqnarray}
\label{DR30}
&&\hspace{-1.cm}
1\ +\ \left(1\ +\ {1\over w(\alpha)}\right)\
\left[1\ +\ {3 (1 + \lambda)\over G(\alpha)}\right]\ =\nonumber\\
\vspace{0.3cm}\nonumber\\
&&\hspace{-0.5cm}
=\ \sqrt{-2\ s_0 }\ \alpha\ G(\alpha)\
\left\{\ \pm 1\ +\ \sqrt{-s_0}\
\left[1\ +\ {3 (1 + \lambda)\over G(\alpha)}\right]\right\}
\left(1+s_0^{-1}\right)^{1/4}\ \cdot\nonumber\\
\vspace{0.3cm}\nonumber\\
&&\hspace{-0.5cm}
\ \ \ \cdot\ \left[\ \sqrt{1+s_0^{-1}}\
P_{-5/2}\left(\left(1+s_0^{-1}\right)^{-1/2}\right)\ -\
P_{-3/2}\left(\left(1+s_0^{-1}\right)^{-1/2}\right)\ \right]\ \ .
\end{eqnarray}

Again, in general solutions $s_0(\alpha)$ of this
equation can be studied numerically only (see fig.\ 3). However,
for very small $\alpha$ ($\alpha \ll 1$) where $s_0$ is very close
to -1 it can also be investigated analytically and one finds (choose the
upper sign in eq.\ (\ref{DR30}))

\begin{eqnarray}
\label{DR31}
\hspace{-0.7cm}
\sqrt{\alpha}\ \left[1\ +\ O\left(\sqrt{\alpha} \ln \alpha\right)\right]&=&
\sqrt{{2\pi\over (1-\lambda/3)}}\ {3(1+s_0)\over 32}\
\left[\ \ln {-(1+s_0)\over 64}\ +\ 3\ \right]\ ,\ \\
\vspace{0.5cm}\nonumber\\
&&\hspace{6.cm}\alpha\ \ll\ 1\ \ .\nonumber
\end{eqnarray}

It should be noted that for eq.\ (\ref{DR30})
a critical value $\alpha = \alpha_c$ exists
which separates the $\alpha$ regions in which the
upper and lower signs in eq.\ (\ref{DR30}) apply.
For $\alpha < \alpha_c$ only in case of the upper sign a
solution $s_0$ exists \footnote{It is clear that for small
$\alpha$ (i.e., $\alpha \rightarrow 0$) a smooth transition from
$\sqrt{-s_0}\ \tilde a_g(s_0) = \tilde b_g(s_0)$
to $\sqrt{-s_1}\ \tilde a(s_1) = \pm \tilde b(s_1)$ must exist ,
consequently the upper sign holds.}
while for $\alpha > \alpha_c$
only the lower sign admits to find a solution $s_0$. This
critical value $\alpha_c$ corresponds to the singularity
$s_0(\alpha\rightarrow \alpha_c) \longrightarrow -\infty$.
Consequently, we find from (\ref{DR30}) the equation for
determining $\alpha_c$ by considering
$s_0 \longrightarrow -\infty$. It reads

\begin{eqnarray}
\label{DR32a}\hspace{-0.5cm}
1\ +\ \left[\ 1\ +\ {1\over w(\alpha_c)}\ -\
{\alpha_c\over 2 \sqrt{2}}\ G(\alpha_c)\ \right]\
\left[1\ +\ {3\ (1 + \lambda)\over G(\alpha_c)}\right]&=&0\ \ \ .
\end{eqnarray}

Numerically, one finds $\alpha_c \simeq 0.70$ (see fig.\ 3).
Furthermore, there exists a maximal value
$\alpha = \alpha_{max} > \alpha_c$ beyond which
no solution $s_0$ can be found. The value of
$\alpha_{max}$ corresponds to the limit
$s_0(\alpha\rightarrow \alpha_{max}) \longrightarrow -1$.
The corresponding equation for $\alpha_{max}$ reads

\begin{eqnarray}
\label{DR32b}\hspace{-0.5cm}
&&\hspace{-1.cm}
\left[\ 1\ +\ {4\over 3 \pi}\ \alpha_{max}\ G(\alpha_{max})\ \right]\ +
\nonumber\\
\vspace{0.3cm}\nonumber\\
&&\hspace{-0.4cm}+\ \left[\ 1\ +\ {1\over w(\alpha_{max})}\ -\
{4\over 3 \pi}\ \alpha_{max}\ G(\alpha_{max})\ \right]\
\left[1\ +\ {3\ (1 + \lambda)\over G(\alpha_{max})}\right]\ =\ 0\ .
\end{eqnarray}

The numerical calculation yields $\alpha_{max} \simeq 2.64$ (see fig.\ 3).\\

\parindent1.5em
 From above considerations it is clear that to
find a consistent IR solution of the integral
equation (\ref{DK1}) requires to understand
the parameter $\beta$ of our Ansatz (\ref{DG5})
as some function of $\alpha$ and therefore it
cannot be left arbitrary up to the point where
we are going to impose the fixed point condition
for the kernel of the gauge field action. It will be
true in general that one parameter of any Ansatz
(containing, say, $n$ parameters) for the kernel
of the gauge field action needs to be reserved to allow
to find a consistent IR solution of the integral
equation (\ref{DK1}). We have only one parameter
at hand and from eq.\ (\ref{DR17}) we immediately
find its dependence on $\alpha$ (for a plot see fig.\ 4).

\begin{eqnarray}
\label{DR33}
\beta &=\ \beta(\alpha)\ =& -\ {\alpha^2\ w(\alpha)^2\over 4\ s_0(\alpha)}
\end{eqnarray}

\parindent0.em
Here, $w(\alpha)$, $s_0(\alpha)$ are solutions of
eqs.\ (\ref{DR16}), (\ref{DR30}) respectively. One easily
recognizes (cf.\ fig.\ 4) that for small $\alpha$ the parameter $\beta$
assumes unrealistic large values what underscores the point
that the present approximative calculation has to be understood
as a model calculation only.\\

\parindent1.5em
After having applied the normalization condition
(\ref{DR26}) and having fixed the parameters $G$, $s_0$, $\beta$
the functions $\tilde a = \tilde a_s$,
$\tilde b = \tilde b(\infty) + \tilde b_s$
can be written in the low $s$ region as follows ($s_0 \le -1$).

\begin{eqnarray}
\label{DR34}
&&\hspace{-1.3cm}\tilde a(s)\ =\
-\ {1\over s}\ \left(1-{s\over s_0}\right)^{1/4}\
\left(1+{1\over s_0}\right)^{-1/4}\ \cdot\nonumber\\
\vspace{0.3cm}\nonumber\\
&&\ \cdot\ \left[\ \sqrt{1-{s\over s_0}}\
P_{-5/2}\left(\left(1-{s\over s_0}\right)^{-1/2}\right)\ -\
P_{-3/2}\left(\left(1-{s\over s_0}\right)^{-1/2}\right)\
\right]\ \cdot\nonumber\\
\vspace{0.3cm}\nonumber\\
&&\ \cdot\ \left[\ \sqrt{1+{1\over s_0}}\
P_{-5/2}\left(\left(1+{1\over s_0}\right)^{-1/2}\right)\ -\
P_{-3/2}\left(\left(1+{1\over s_0}\right)^{-1/2}\right)\ \right]^{-1}\ ,\\
\vspace{1.5cm}\nonumber\\
\label{DR35}
&&\hspace{-1.cm}\tilde b(s)\ =\
( \pm 1 -\tilde b(\infty))\ \tilde a(s)\ +\
\tilde b(\infty)\ \ \ \ .
\end{eqnarray}

\parindent0.em
The parameter $\tilde b(\infty)$ in the normalization applied reads
(for a plot see fig.\ 5)

\begin{eqnarray}
\label{DR36}
\tilde b(\infty)\ =&\pm 1\ -\ \sqrt{-s_0}\ \ \displaystyle{H\over G}&=\
\pm 1\ +\ \sqrt{-s_0}\left[\ 1\ +\ {3(1+\lambda)\over G(\alpha)}\right]\
\ \ \ \ \ .
\end{eqnarray}

For small $\alpha$ we immediately find from eq.\ (\ref{DR24})

\begin{eqnarray}
\label{DR37}
&&\hspace{-1.cm}\tilde b(\infty)\ =\
1\ +\ \sqrt{-s_0}
\left[\ 1\ +\ 4\ (1+\lambda)\ {\alpha\over\pi}\ +\
O\left(\alpha^{3/2}\right)\ \right]\ \ \ ,\ \\
\vspace{0.5cm}\nonumber\\
&&\hspace{7.5cm}\alpha\ \ll\ 1\ \ ,\ \
\alpha\ <\ \alpha_c \ .\nonumber
\end{eqnarray}

Taking into account eq.\ (\ref{DR31})
($s_0 \simeq -1$, $\alpha \ll 1$) we recognize
that for small $\alpha$ ($\alpha \ll 1$, $\alpha < \alpha_c$)
it holds $\tilde b(\infty) \simeq 2$.
 From a physical point of view this might be interpreted
such a way that at low energies the fermion action
merely describes individual real fermions
($\tilde b \simeq 1$), i.e., a single particle
interpretation is possible, while at high energies
it reflects collective properties of the vacuum
which are related to fermion (electron-positron) pairs, consequently
$\tilde b \sim \tilde b(\infty) \simeq 2$. Apparently, such an
interpretation breaks down at stronger coupling.\\

\parindent1.5em
Now, the appropriately normalized $\tilde a_g(s)$
(eq.\ (\ref{DR15})) reads in the low $s$ region

\begin{eqnarray}
\label{DR38}
&&\hspace{-1.cm}\tilde a_g(s)\ =\nonumber\\
\vspace{0.3cm}\nonumber\\
&&=\ {C_g\over 2\sqrt{\pi}}\ \
\Gamma\left(-{1\over 2}\ +\ \alpha\ {(3-\lambda)\over 2\pi}\ \right)\
\left(-s_0 - 1\right)^{-1/4}\ \cdot\nonumber\\
\vspace{0.3cm}\nonumber\\
&&\ \ \ \ \ \cdot\ \left[\ \sqrt{1+{1\over s_0}}\
P_{-5/2}\left(\left(1+{1\over s_0}\right)^{-1/2}\right)\ -\
P_{-3/2}\left(\left(1+{1\over s_0}\right)^{-1/2}\right)\ \right]^{-1}\
\cdot\nonumber\\
\vspace{0.3cm}\nonumber\\
&&\ \ \ \ \ \cdot\ {1\over s}\
\left[\ (\sqrt{-s_0}\ +\ \alpha/2\sqrt{\beta})^2\ +\ s\ \right]^{1/4\
-\ \alpha (3-\lambda)/4\pi}\ \cdot\nonumber\\
\vspace{0.3cm}\nonumber\\
&&\ \ \ \ \ \cdot\ \left[\ \sqrt{\ 1\ +\
{s\over (\sqrt{-s_0}\ +\ \alpha/2\sqrt{\beta})^2}}\
\right.\cdot\nonumber\\
\vspace{0.3cm}\nonumber\\
&&\ \ \ \ \ \ \ \ \cdot\
P_{-5/2\ +\ \alpha (3-\lambda)/2\pi}\left(\left(
1\ +\ {s\over (\sqrt{-s_0}\ +\ \alpha/2\sqrt{\beta})^2}\right)^{-1/2}
\right) \ -\nonumber\\
\vspace{0.3cm}\nonumber\\
&&\ \ \ \ \ \ \ \ -\ \left. P_{-3/2\ +\ \alpha (3-\lambda)/2\pi}\left(\left(
1\ +\ {s\over (\sqrt{-s_0}\ +\ \alpha/2\sqrt{\beta})^2}\right)^{-1/2}
\right)\ \right]\ \ \ \ .\
\end{eqnarray}

\parindent0.em
And eq.\ (\ref{DR19}) can be written as

\begin{eqnarray}
\label{DR39}
\tilde b_g(s)&=&\left( \pm 1-\tilde b(\infty)\right)\
\left(\ 1\ +\ {1\over w(\alpha)}\ \right)\
\tilde a_g(s)\ +\ \tilde b(\infty)\ \ \ .\
\end{eqnarray}

Clearly, $s_0$, $\beta$, $\tilde b(\infty)$ are functions
of $\alpha$ ($\lambda = 0$ as explained in chapter 3).\\

\parindent1.5em
Finally, the correctly normalized IR tails of
$a$, $b$ characterizing the kernel of the fermion
action are

\begin{eqnarray}
\label{DR40}
&&\hspace{-3.cm}a(x_E)\stackrel{m^2 x^2_E \rightarrow \infty}{=}
\nonumber\\
\vspace{0.3cm}\nonumber\\
&&\hspace{-2.cm}=\ {m^4\over\sqrt{2}\ (2\pi)^{5/2}}\ \
\left(-s_0 - 1\right)^{-1/4}\ \cdot\nonumber\\
\vspace{0.3cm}\nonumber\\
&&\hspace{-1.cm} \cdot\ \left[\ \sqrt{1+{1\over s_0}}\
P_{-5/2}\left(\left(1+{1\over s_0}\right)^{-1/2}\right)\ -\
P_{-3/2}\left(\left(1+{1\over s_0}\right)^{-1/2}\right)\ \right]^{-1}\
\cdot\nonumber\\
\vspace{0.3cm}\nonumber\\
&&\hspace{-1.cm} \cdot \ (m \vert x_E\vert)^{-7/2}\ \
{\rm e}^{\displaystyle\ -\ \sqrt{-s_0}\ m \vert x_E\vert}\ \
\Big[\ 1\ +\ \ldots\ \Big]\hspace{1.cm},\\
\vspace{0.6cm}\nonumber\\
\label{DR41}
b(x_E)&\stackrel{m^2 x^2_E \rightarrow \infty}{=}&
\left(\pm 1-\tilde b(\infty)\right)\
a(m^2 x^2_E \rightarrow \infty)\hspace{1.cm},
\end{eqnarray}

\parindent0.em
where $s_0$ and $\tilde b(\infty)$ are to be
considered as functions of $\alpha$. Clearly, in
qualitative respect eqs.\ (\ref{DR40}), (\ref{DR41})
agree with the long distance representation of the
1-loop fermion self-energy calculated in standard
QED perturbation theory.\\

\parindent1.5em
To conclude this subsection it should be emphasized
that in analyzing the integral equation (\ref{DK1})
for the kernel of the fermion action in the asymptotic
UV and IR regions respectively based on certain
reasonable assumptions we have obtained a
qualitative and nonperturbative understanding of
the behaviour of its solution. Furthermore, the
IR analysis even yields approximative quantitative,
nonperturbative results which combined with the
information about the UV behaviour of the kernel of
the fermion action obtained admits to attempt the
approximative calculation of the QED coupling
constant $\alpha$. This we will study now.\\

\subsubsection{\label{DS33}\label{SH2}The Fixed Point Condition for the
Kernel of the Gauge Field Action and the
Approximative Calculation of the QED Coupling
Constant $\alpha$}

 From eq.\ (\ref{DI12}) we recognize that the
functional integration induces a change
$\Delta\Gamma^G_I [A]$ to be added to the
gauge field action $\Gamma^G_I [A]$ to obtain
$\Gamma^G_{II} [A]$. In accordance with
our approximation strategy we display only those
terms that match our Ansatz (\ref{DG5}).

\begin{eqnarray}
\label{DL1}
\hspace{-0.5cm}\Delta\Gamma^G_I [A]\ &=&\nonumber\\
\vspace{0.3cm}\nonumber\\
\hspace{-0.7cm}
= \ {\alpha\over 4\pi} &\displaystyle{\int} d^4x &A^\mu(x)\
\left[ g_{\mu\nu} \Box\ -\ \partial_\mu \partial_\nu\right]\
\left[\ C_{1a}\ +\ C_{2a}\ {\Box\over m^2}\ +\ \ldots\ \right]
\ A^\nu(x)
\end{eqnarray}

\parindent.0em
Because $a_g$, $b_g$ respect conditions (\ref{DF9}),
(\ref{DF10}) (cf.\ eqs.\ (\ref{DV14}), (\ref{DV15}))
no terms violating gauge invariance occur
and eq.\ (\ref{ZA10}) applies. $C_{1a}$ reads (see Appendix A;
as explained in section \ref{SH1} we confine ourselves to
1-loop contributions)

\begin{eqnarray}
\label{DL2a}
C_{1a}\ &=&\ {2\over 3}\ \ln\ \left[
{\tilde b (\infty)\over\tilde b_g(0)} \right]^2\ -
\ \int\limits_0^\infty ds\ M(s)\hspace{0.5cm},\\
\vspace{0.3cm}\nonumber\\
\label{DL2b}
&&\ M(s)\ =\ \
{1\over s \tilde a^2_g + \tilde b^2_g }\
\left[\ {s\ \tilde a^2_g \over s \tilde a^2_g + \tilde b^2_g}\
\left[\ s\ \tilde a_g \tilde a^\prime_g +
\tilde b_g \tilde b^\prime_g\ \right]
\ +\right.\nonumber\\
\vspace{0.2cm}\nonumber\\
&&\hspace{2cm}
+\ {2\over 3}\ s^3\ \tilde a_g \tilde a^{\prime\prime\prime}_g\ +\
3\ s^2\ \tilde a_g \tilde a^{\prime\prime}_g\ +\
{2\over 3}\ s^2\ \tilde b_g \tilde b^{\prime\prime\prime}_g\ +\nonumber\\
&&\hspace{2cm}+\ 2\ s\ \tilde a_g \tilde a^\prime_g\ +\
3\ s\ \tilde b_g \tilde b^{\prime\prime}_g\ -\
s\ (\tilde b^\prime_g)^2\ +\
3\ \tilde b_g \tilde b^\prime_g\ \Bigg] \ \ .
\end{eqnarray}

 From above expression one recognizes that $C_{1a}$
is a RG invariant quantity, i.e., it is invariant
against (finite) mass and (fermion) wave function renormalizations.
$C_{2a}$ has not yet been calculated in terms of
$\tilde a_g$, $\tilde b_g$ but it will have an
analogous representation. Because $\tilde a_g$,
$\tilde b_g$ exclusively depend on $\alpha$ the coefficients
$C_{1a}$, $C_{2a}$ can both be understood as
functions of this parameter. Then, the fixed
point condition $d_I = d_{II}$ according to our
approximation strategy reads (cf.\ subsection 4.2.3)

\begin{eqnarray}
\label{DL3}
C_{1a}(\alpha)\ &=&\ 0\hspace{2.cm} ,\\
\vspace{-0.1cm}\nonumber\\
\label{DL4}
C_{2a}(\alpha)\ &=&\ 0\hspace{2.cm} .
\end{eqnarray}

It is clear that within our approximative approach
we do not have enough parameters left to satisfy
both of these equations (if they are not degenerate,
perhaps by accident). We decide to choose eq.\ (\ref{DL3})
as fixed point equation because we require that at
least in the asymptotic IR (long distance, long
wavelength) region the fixed point condition for the map $f$ should
be fulfilled. Consequently, to determine the QED coupling
constant $\alpha$ we have to find the zero(s) of
$C_{1a}(\alpha)$.\\

\parindent1.5em
The explicit calculation of $C_{1a}$ has of course
to be based on information obtained in the preceding
sections. The first point to be made is that we will
take eq.\ (\ref{DL2a}) as it stands. In principle, one
could identically reformulate it by exploiting partial
integrations for functions that obey conditions
(\ref{DF9}), (\ref{DF10}) (or the even somewhat
weaker conditions
$a_g(s) = O(s^\kappa),\ \kappa < -1/2,\ b_g(s) = O(1),
\ s \rightarrow \infty$).
We choose the present representation for its 'minimal'
shape (Of course, this is merely a matter of taste.).
Let us also emphasize that it turns out advantageous because a certain piece
is already integrated out and it therefore depends
on the boundary values of $\tilde b_g$ only. This term contains
certain nonperturbative information from the solution
of the integral equation (\ref{DK1}) for the kernel
of the fermion action not easily incorporated otherwise.
Finally, one should keep in mind that although different
representations of eq.\ (\ref{DL2a}) are equivalent in
a rigorous mathematical sense, they may lead
to different answers if approximative information is
taken into account only (and this is what we will do).\\

Now, the first guess might be simply to insert
into eq.\ (\ref{DL2a}) the IR representation found
for $\tilde a_g$, $\tilde b_g$ (eqs.\ (\ref{DR38}),
(\ref{DR39})). But, as comes as no surprise the
integral in eq.\ (\ref{DL2a}) is not convergent for
$\alpha \le \pi/3$ (it is logarithmically UV
divergent then). In other words, this
approximation would be so crude as to even
not deliver finite results. So, in the
parameter region $\alpha \le \pi/3$ at least one has
to proceed differently. Without any problem we
may always insert the value of $\tilde b(\infty)$
determined by the normalization conditions
applied within the IR analysis. For $\tilde b_g(0)$
and in the low $s$ integration region of the integral
we will insert $\tilde a_g$, $\tilde b_g$ as given
by eqs.\ (\ref{DR38}), (\ref{DR39}). In the large $s$
region $\tilde a_g$, $\tilde b_g$ will be taken from
eqs.\ (\ref{DV14}), (\ref{DV15}). One immediately
recognizes that this is a better approximation
because the integral in eq.\ (\ref{DL2a}) then gives
finite results. Now, of course, the
practical question arises which intermediate value
of the integration variable $s$ in eq.\ (\ref{DL2a}) should
split the application regions of the IR and UV
representation of $\tilde a_g$, $\tilde b_g$.
Perhaps, one could choose to fit together the
IR and UV representations at some value of $s$
to be determined by a certain condition.
For the purpose of the present numerical calculation
we select another way. The UV tail of the integrand $M(s)$
in eq.\ (\ref{DL2a}) will not contribute significantly
and we therefore ignore it by simply cutting the
integration over the IR representation of the
integrand at some upper value $s = s_x$. This value
is determined as follows. Observe that the exact integrand $M(s)$
in eq.\ (\ref{DL2a}) is positive for $s\rightarrow\infty$.
To see this one may insert eqs.\ (\ref{DV14}),
(\ref{DV15}) into (\ref{DL2b}) and one finds to
leading order

\begin{eqnarray}
\label{DL6}
M(s)&=&
11\ \ {C^2_{\tilde a}\over s^4}\ \ +\ \ \ldots\ \ \ >\ 0\ \ ,
\ \ \ s\ \longrightarrow\ \infty\ \ \ .
\end{eqnarray}

\parindent.0em
On the other hand, one may easily convince oneself
that for $\alpha \le \pi/3$ the integrand $M(s)$ of
eq.\ (\ref{DL2a}) turns negative for
$s \longrightarrow \infty$ if the low $s$
representations (\ref{DR38}), (\ref{DR39}) are inserted.
One now detects that the integrand with the low $s$
representation inserted is positive for $s = 0$.
Consequently, there exists a zero of the integrand taken in the
IR representation (cf.\ fig.\ 6).
Obviously, this zero determines the point beyond
which the IR (low $s$) representation starts to strongly
misrepresent the true integrand and we therefore
choose this zero as upper cut-off $s_x$ of the numerical
integration (See fig.\ 7 for the dependence of $s_x$ on $\alpha$.)
\footnote{Another choice might be to fit the IR and UV
representations of the integrand together at some $s_y < s_x$. Here,
one way is to require continuity of the integrand at
$s = s_y$ and to determine $s_y$ by extremizing the value of the
integral. However, in doing so one detects that the contribution
of the UV tail is negligible numerically.}.
It is clear that this recipe leads to
a certain slightly lower value of the integral than
if the UV region was not neglected.\\

\parindent1.5em
Now, the result of the numerical calculation of
$C_{1a}(\alpha)$ is shown in fig.\ 8, while fig.\ 9
displays the behaviour of the two contributions
$C_{1a}(\alpha)$ derives from (cf.\ eq.\ (\ref{DL2a})).
Unfortunately, within the approximation applied we do
not find any zero of $C_{1a}(\alpha)$, but from fig.\ 9
one recognizes that both contributions to be taken
into account are indeed comparable numerically. We
believe that the contribution of the integral in
eq.\ (\ref{DL2a}) is underestimated within the
approximation applied compared with the exact
one which relates to the exact solution of the
integral equation (\ref{DK1}). The contribution
of the first term in eq.\ (\ref{DL2a}) is probably
determined to a more reliable degree because only
the boundary values of $\tilde b_g(s)$ contribute
to it. Furthermore, the smaller $\alpha$ the
more the approximation applied for the second term
in eq.\ (\ref{DL2a}) miscalculates it. This can
easily be seen from fig.\ 6 (and fig.\ 7). The true
integrand (the exact solution of eq.\ (\ref{DK1}),
which we do not know presently, inserted) would likely
contribute more because we expect the integrand
$M(s)$ to be positive for large $s$. This would
shift curve 2 in fig.\ 9 to larger values and
consequently a zero of $C_{1a}(\alpha)$ might
occur.\\

To conclude, the mechanism proposed has explicitly
been shown capable to attempt the calculation of
the QED coupling constant $\alpha$. However, the
approximation applied turns out too simple yet to obtain
any specific value of $\alpha$. In particular, for
small values of $\alpha$ where most of the
approximations applied within the calculation
given in the present chapter appear to be most
justified no zero of $C_{1a}(\alpha)$ is found.
But it is clear that more advanced approximations
may lead to a different picture. This needs to be
studied in the future. We postpone further discussion of
this issue to chapter 6.\\

\newpage

\section{\label{ES}The Vacuum Energy, and Related Problems}
\setcounter{equation}{0}

In this chapter we discuss the vacuum energy
issue and some related problems we did not mention
so far. The consideration will not be aimed at the
most general theoretical set-up eventually possible
which very likely would turn out fruitless, but we
restrict consideration to QED and in particular to
that approximative approach to it studied in chapter 4.
It might be hoped that this special case yields certain
new insight into the problem useful at least for gauge
field theories in general.\\

In standard QED in 4D Minkowski space the vacuum
energy density originating from fermion as well
as from photon fluctuations and their interactions
is a divergent quantity
but it is considered as
unimportant because it can either be removed
by applying normal ordering (in operator
quantization) or by appropriately normalizing
the functional integral defining the theory.
No physical quantity depends on it. But, it
is also known that modifications of the vacuum
energy density as occurring when external
conditions are applied (boundary conditions,
temperature, external fields) do matter and in
certain cases consequences are even observable in experiment
(so, the Casimir effect) \cite{k}-\cite{kapu}. Few changes
of the vacuum energy density turn out to be finite
immediately (e.g., the Casimir energy density, or the
free energy density for QED at finite temperature).
Others require renormalization, like the QED
effective potential for (say) a constant magnetic
field. Even more care is needed in the study of
QED in a gravitational background field we will
return to later. However, large part of the motivation
for studying the vacuum energy density derives from
this situation because it gives rise to the concept
of induced (classical) gravity \cite{l} understood as some kind
of gravitational (metric) Casimir effect (for a
review of recent work and further references see
\cite{adle},\cite{novo}, also note \cite{davi1},\cite{davi2}).\\

First, let us compare the calculation of the vacuum
energy density in standard QED and within the present approach. We
restrict ourselves to the 1-loop level which contains
all important features. We apply the simplest
regularization possible, namely cut-off regularization
(with a (radial) momentum space UV cut-off at $\Lambda$),
which is most suited for our purposes. The vacuum energy
density $\rho_{vac}$ is given by

\parindent0.em
\begin{eqnarray}
\label{E1}
\Gamma_{II} [0,0,0]\ &=&-\ V_4\ \rho_{vac}\nonumber\\
\vspace{0.3cm}\nonumber\\
&=&\ const.\ -
\ i\ln\ {\rm Det}_\Lambda\ \left( S^{-1}_I\right)\ -\nonumber\\
&&\hspace{0.8cm} -\ i\ln\ {\rm Det}_\Lambda\ \left( D^{-1}_{gh\ I}\right)\ +
{i\over 2}\ \ln\ {\rm Det}_\Lambda\ \left( D^{-1}_{I\ \mu\nu}\right)\ \ \ .
\end{eqnarray}

Here,

\begin{eqnarray}
\label{E2}
S^{-1}_I(x-x^\prime)\ &=&\
i\not\hspace{-0.07cm}\partial_x\ a_I\left(x-x^\prime\right)\
-\ m\ b_I\left(x-x^\prime\right)\\
\vspace{0.3cm}\nonumber\\
\label{E3}
D^{-1}_{gh\ I}(x-x^\prime)\ &=&{1\over\sqrt{\lambda}}\
_x\partial_\mu\ n^\mu(x-x^\prime)\ \\
\vspace{0.3cm}\nonumber\\
\label{E4}
D^{-1}_{I\ \mu\nu}(x-x^\prime)\ &=&
\left[ g_{\mu\nu}\ _x\Box - \ _x\partial_\mu \ _x\partial_\nu\right]
d_I\left(x-x^\prime\right)\ -\nonumber\\
\vspace{0.3cm}\nonumber\\
&&\ -\ {1\over\lambda}\ \int d^4 y\ \  n_\mu(y-x)\ n_\nu(y-x^\prime)
\end{eqnarray}

are the quadratic kernels of the fermion, ghost
(contributing in QED to the vacuum energy only), and gauge field
actions respectively \footnote{ $n_\mu$ can be here any
vector-valued distribution, e.g., perhaps a derivative
$\partial_\mu$ acting on some scalar function
leading to a Lorentz type gauge, or any constant
vector times a scalar function yielding an axial
type gauge.}. From eq.\ (\ref{E1}) we find accordingly

\begin{eqnarray}
\label{E5}
&&\hspace{-1.4cm}\Gamma_{II} [0,0,0]\ =\nonumber\\
\vspace{0.3cm}\nonumber\\
&&\hspace{-1.2cm}
=\ const.\ -\ 2i\ V_4 \int\limits_\Lambda {d^4p\over (2\pi)^4}\ \
\ln\left[ \ -p^2\ \tilde a_I (p)^2\ +\ m^2\ \tilde b_I (p)^2\ \right]\ -
\nonumber\\
\vspace{0.3cm}\nonumber\\
&&\hspace{-0.7cm} -\ i\ V_4 \int\limits_\Lambda {d^4p\over (2\pi)^4}\ \
\ln\left[\ i\ \lambda^{-1/2}\ p\tilde n(p)\ \right]\ +
\nonumber\\
\vspace{0.3cm}\nonumber\\
&&\hspace{-0.7cm} +\ {i\over 2}\ V_4 \int\limits_\Lambda
{d^4p\over (2\pi)^4}\ \
\ln\left[ \det\left[ (g_{\mu\nu}\ p^2\ -\ p_\mu p_\nu)\ \tilde d_I(p)\
-\ \lambda^{-1}\ \tilde n_\mu(p)\ \tilde n_\nu(p)\
 \right]\ \right]\ .
\end{eqnarray}

Taking into account the relation

\begin{eqnarray}
\label{E6}
\det\left[ (g_{\mu\nu}\ p^2\ -\ p_\mu p_\nu)\ \tilde d\
-\ \lambda^{-1}\ \tilde n_\mu\tilde n_\nu \right]\ &=&
-\ {\tilde d^{\ 3}\over\lambda}\ \
\left[p\tilde n\right]^2 \left[p^2\right]^2
\end{eqnarray}

and applying a Wick rotation one finds after some
manipulations \footnote{We have absorbed certain $\ \ln m$
terms into the first (normalization) constant on the r.h.s.\ of
eq.\ (\ref{E5}).}

\begin{eqnarray}
\label{E7}
&&\hspace{-0.8cm}\Gamma_{II} [0,0,0]\ =\ const.\ +\nonumber\\
\vspace{0.3cm}\nonumber\\
&&\hspace{-0.4cm}+\ {V_4\over 8\pi^2}\ m^4\
\int\limits_0^{\Lambda^2\over m^2} ds\ s\ \left\{\
\ln\left[ \ s\ \tilde a_I (s)^2\ +\ \tilde b_I (s)^2\ \right]\
-\ {1\over 2}\ \ln\left[\ s\ \tilde d_I(s)^{3/2}\ \right]\
\right\}\ .
\end{eqnarray}

There is no trace left of the gauge condition because
we have correctly included in the kernel of the ghost
action (\ref{E3}) the gauge parameter $\lambda$ (For
a related discussion see \cite{alle},\cite{niel}.).
One immediately recognizes the well-known fact that in standard QED
($\tilde a_I = \tilde b_I = \tilde d_I \equiv 1$)
the vacuum energy density $\rho_{vac}$ diverges
\footnote{Incidentally, one may always formally (by ignoring
finiteness/convergence requirements of properly applied
mathematics)
transform eq.\ (\ref{E5}) in the
'sum over the spectrum' formula for the vacuum
energy density by exploiting the cut in the
appropriate variable (i.e., $p_0$) connected with the
logarithms, starting at the lowest energy eigenvalue
of the spectrum, and extending to infinity.}.
Now, QED in a background field (electromagnetic
or gravitational; we restrict consideration
to these external conditions most interesting in view of standard QED
difficulties) will change the quantity $s$ (stemming from
differential operators in configuration space) appearing in
the argument of the logarithms above to some
$s + \Delta s$ where for large $s$ the change
$\Delta s$ behaves like
$\Delta s \stackrel{s \longrightarrow \infty}{\sim} const.$
\footnote{Considering a connection in the covariant derivatives
this naively yields
$\Delta s \stackrel{s \longrightarrow \infty}{\sim} \sqrt{s}$,
but symmetry reasons finally lead to the somewhat weaker behaviour
$\Delta s \stackrel{s \longrightarrow \infty}{\sim} const.$.}.
Of course, as already mentioned one can always
absorb the divergent terms characteristic for 4D
Minkowski space and displayed in eq.\ (\ref{E7}) on the r.h.s.\
into the normalization constant of the functional
integral. But, for QED in a background field
the logarithm in the integrand
of eq.\ (\ref{E7}) then reads for large $s$

\begin{eqnarray}
\label{E8}
\ln\left[\ 1\ +\ {\Delta s\over s}\ +\ \ldots\ \right]&
\stackrel{s \longrightarrow \infty}{=}&
\ln\left[\ 1\ +\ O(s^{-1})\ \right]\ =\ O(s^{-1})
\end{eqnarray}

and the vacuum energy density depending on the background
field is still divergent (This even holds up to
$\Delta s \stackrel{s \longrightarrow \infty}{\sim} 1/s$.).\\

\parindent1.5em
Now, compare this with our approximative approach to the
equation for the complete effective action of QED. From
eqs.\ (\ref{DV14}), (\ref{DV15}) we know that
it holds

\begin{eqnarray}
\label{E9}
s\ \tilde a_I(s)^2\ +\ \tilde b_I(s)^2\
&\stackrel{s \longrightarrow \infty}{=}
&\tilde b(\infty)^{2}\
\left[\ 1\ -\ {C^2_{\tilde a}\over s^3}\ +\ \ldots\ \right]\hspace{1.5cm}.
\end{eqnarray}

\parindent0.em
Absorbing a $\ \ln \tilde b(\infty)$ term into the
normalization constant of the functional integral
we see that the part of the vacuum energy density
originating from fermion fluctuations (the first
term in the integrand of eq.\ (\ref{E7})) is even finite
without any further appeal to this constant.
As we have explained in subsection
\ref{SH4} this is true irrespectively of the
particular approximation applied (i.e., whether
we first perform the gauge field integration or
the fermionic integration). Consequently, any change
of the fermionic part of the vacuum energy density
under the influence of external (electromagnetic
as well as gravitational) fields will
also be finite. But, in view of condition
(\ref{DG4}) the part of the vacuum energy density
originating from photon fluctuations (the second
term in the integrand of eq.\ (\ref{E7})) is still divergent and
equally as in standard QED we need to absorb this divergency
for 4D Minkowski space into the normalization
constant of the functional integral
in order to properly define the equation
for the complete effective action of QED. This can be done
without any problem. The only
concern remaining is the behaviour of the gauge field
determinant in the presence of a gravitational background field.
We do not have any quick answer on this, but let
us speculate for a moment. Assume we had for 4D Minkowski
space absorbed the UV divergency stemming from the gauge field
determinant into the normalization constant of the
functional integral by using a certain power of
the determinant of the d'Alembertian \footnote{One may well
imagine that $d_I$ behaves for high energies such
a way that this recipe removes all divergencies.
If not, perhaps the determinant of $d_I$ in whole has to be
included in the normalization constant.}. If one now
generalizes the 4D Minkowski space functional integral
to an arbitrary gravitational background this has to be
done for the whole functional integral measure, i.e.,
also the (normalization) determinant of the d'Alembertian
has to be generalized covariantly. Then of course, using
this recipe the vacuum energy density of QED would be
finite in electromagnetic as well as in gravitational
background fields. If one is to reject above recipe one
has to further discuss the determinant of the d'Alembertian
in the presence of a gravitational background field what
is a problem of long standing concern, in particular
the gauge field conformal anomaly and its
regularization dependence \cite{birr}. Finally, it appears not
unreasonable to expect that above discussion persists to
apply also if further contributions (higher loops)
are taken into account.\\

\parindent1.5em
Above consideration now admits to compare
standard QED in a gravitational
background field and the present approach.
In standard QED the structure
of the first few terms of the effective gravitational action
(i.e., up to a minus sign the (time integrated) vacuum energy) is
known \cite{k},\cite{birr},\cite{full}.

\parindent0.em
\begin{eqnarray}
\label{E10}
&&\hspace{-1.5cm}\Gamma_{II} [0,0,0]\ =\nonumber\\
\vspace{0.3cm}\nonumber\\
&&=\ \int d^4x\ \sqrt{-g}\
\left\{\ m^4\ c_1\ +\ m^2\ c_2\ R\ +\ c_3\ \Box\ R\ +\
c_4\ R^2\ +\right.\nonumber\\
\vspace{0.3cm}\nonumber\\
&&\hspace{2.5cm}\left.+\ c_5\ R_{\mu\nu}\ R^{\mu\nu}\ +\
c_6\ R_{\mu\nu\alpha\beta}\ R^{\mu\nu\alpha\beta}\ +\ \ldots\ \right\}
\end{eqnarray}

$c_1$ to $c_6$ are certain divergent dimensionless
constants. We have already discussed above $c_1$
(i.e., $-\rho_{vac}$ for 4D Minkowski space),
$c_2$ is a quadratically (in the cut-off $\Lambda$)
divergent quantity while $c_3$ to $c_6$ diverge
logarithmically. All further terms are finite.
Consistency requires to start in the standard
QED functional integral with a certain
bare gravitational action (included in $\Gamma_I$)
containing all terms displayed in eq.\ (\ref{E10})
in order to be able to absorb the divergencies
into the bare constants in front of them. Consequently,
induced gravity is not a consistent concept within
standard QED. In contradistinction to standard QED,
by taking into account the UV behaviour of the quadratic kernel
of the fermion action (a consequence
of the equation for the complete effective action of
QED) we have demonstrated above that whatever the
technical approach to calculate $c_2$ to $c_6$ will
be in detail \footnote{In cut-off regularization they
will have representations analogous to eqs.\
(\ref{ZA6})--(\ref{ZA8}).}
these coefficients will come out finite
(at least at the 1-loop level).
The contribution from the determinant of the gauge field
kernel will depend on the choice one is willing to make
for the normalization of the functional integral.
Therefore, within the present framework induced gravity
might under certain circumstance turn out to be
a valid concept. Of course, as has been
pointed out by {\sc Sakharov} in his pioneering paper \cite{l}
the (induced) gravitational action will very likely
not be dominated by contributions stemming from QED
but from the heaviest excitations (particles) existent
in nature. If one would like to attempt the calculation
of the induced gravitational action within the concept
proposed in the present paper one would first have
to study the equation for the complete effective
action of the standard model at least. If one is
willing to do so this will require much effort and
certainly results cannot be obtained quickly. But,
in view of the possible outcome perhaps it might
be worth to be done.\\

\newpage

\section{\label{FS}Discussion and Conclusions}
\setcounter{equation}{0}

Before turning to some matters of principle let
us further discuss the approximative approach to
the functional integral equation for the complete
effective action of QED. We have seen that the
general approximative approach chosen (cf.\ section 4.1)
admits to find certain nonperturbative information
about the quadratic kernels of the QED action.
Particular emphasis deserves the fact that the
information found indicates that there exists an
unique solution to the functional integral equation
only (at least within the approximative approach
studied). Of course, this point has to be studied
further using more advanced approximations in order
to see whether for the QED coupling constant $\alpha$
only one admissible value exists (if any at all -- but
nature appears to allow for some). Furthermore, within
the approximative approach divergencies as they are
characteristic for standard QED do not show up (at
least, as far as the present study runs). It should
perhaps also be said that the nonlocal character of
the fermion action admits to employ nonperturbative
techniques which are not quickly applicable in standard
QED. For example, as we have seen this way the well-known Bloch-Nordsieck
contribution can be obtained easily and it contains
important IR (long distance) information crucial to
the further calculation.\\

\parindent1.5em
However, so far the concept proposed in the present article
has not yet successfully passed the crucial
test attempted in subsection 4.3.3, namely the
approximative calculation of the QED coupling
constant $\alpha$. As we have seen the approach
used is indeed suited for explicit calculation but
inasmuch as within the simple approximation applied
we did not find any zero of $C_{1a}(\alpha)$ the
question remains presently open. How might a better
approximation look like? First, it should be noted
that by imposing eq.\ (\ref{DR16}) independently of
the value of $\alpha$ a strong coupling condition
has been enforced which annihilates the hope that
higher loop contributions can really be neglected in
the integral equation for the quadratic kernel of
the fermion action (\ref{DK1}). But, to take into
account higher loop contributions would add complications
to the formalism not easily to be resolved in analytical
calculations. One way out of this dilemma might be to
relax for approximative purposes the fixed point
condition for the quadratic kernel of the fermion
action to $a_{II} = C\ a_I$, $b_{II} = C\ b_I$ where
$C$ is some arbitrary real constant, instead of
immediately enforcing $C = 1$. This requirement of
structural similarity perhaps could be sufficient to
keep the conceptual content alive and at the same time
admits to count indeed (not only seemingly) in any arguments
on the eventual smallness of $\alpha$. The parameter
$\beta$ then would also be unconstrained as long as
the fixed point condition $d_{II} = d_I$ is not enforced.
To finally fix both $\alpha$ and the parameter $\beta$ the conditions
(\ref{DP2}), (\ref{DP3}) can be applied simultaneously.
Whether this recipe yields a more effective approximation
remains to be seen in future investigations. It might
perhaps also be necessary to include some higher loop
contributions to $C_{1a}$ and $C_{2a}$. Certainly,
the solution of the integral equation for the kernel of the fermion
action (\ref{DK1}) has to be studied further. May be, it will
also be advisable to improve the Ansatz (\ref{DG5}).
These are few of the changes in the approximation strategy
which can be implemented most easily along the lines of chapter 4.
Perhaps, still more severe changes are required.
Finally, it should be
said that the calculation discussed in chapter 4 should
merely be understood as a first (naive) attempt to
extract information out of the functional integral
equation for the complete effective action by means
of a simple approximation which however admits mostly
analytical investigation. It is clear, of course, that
the present understanding is poor and much remains
to be learned.\\

Throughout the paper we have preliminary applied the
standard point of view that the space-time structure
is prescribed to the functional integral equation for
the complete effective action. In a certain sense, it
is considered as 'classical' and as prior to quantum
effects (at least for flat space-time). However, the
criticism spelled out in section 2.3 with respect to
the artificial distinction between classical action
and effective action also applies to this view on the space-time
structure. Therefore, more adequate the structure of space-time
should be understood as some characteristics of the
quantum field theoretic vacuum. Basically, this is
the point of view applied within the concept of
induced gravity although this aspect
is hardly discussed in the literature. But, also in
flat space-time the idea applies. Recent investigations
of propagation of light in a Casimir vacuum indicate
that this concept is already implicitly
entailed in standard QED \cite{scha}-\cite{bart2}.
As discussed in ref.\ \cite{bart2}, although lack of
appropriate nonperturbative calculational tools
leaves the question so far unsettled in the strict sense
the only conceptually viable (as far as present
knowledge is concerned) of the alternatives allowed by
the Kramers-Kronig relation for the refractive index
$n(\omega)$ of the Casimir vacuum
($\omega$ is the frequency of the test wave) is that
$n(\infty) < 1$ holds for propagation of light
perpendicular to two parallel mirrors in the
slab between them (This entails a signal velocity of light
larger than in the free space vacuum.). While the
result is often viewed as something like a paradox in standard
QED it is easily understandable by means of the
concept put forward in the present article (where
it may count as a special application). If the map
$f$ is modified such a way that it is no longer
fully Lorentz invariant \footnote{For an appropriate functional
integral formulation of standard QED in the presence
of two parallel mirrors see \cite{bord}.} then
also the solution of the functional
integral equation for the complete effective action
is no longer fully Lorentz invariant and the
dispersion analysis in accordance with the effective Maxwell
action may well reveal a change in the signal velocity
of light. The point is that only one situation can be
considered as the one where normalization is performed
(and we typically choose free Minkowski space as
reference situation and the signal velocity of light
there as reference standard, although of course also
any less symmetrical set-up could be used). But, in
view of the discussion performed in section 2.3 it makes
no sense to consider any normalized value of a certain
quantity (mass, charge, velocity of light, e.g.)
as classical because this is a concept not accessible to
experiment. We can only denote certain values defined by
a certain measurement scenario under defined circumstances
as reference values. Any changes of these values measured
under different circumstances are certainly of quantum
nature but equally well these values could have served
as initial reference values. Consequently, it appears most
sensible to consider these quantities from the very
beginning as characteristics of
the quantum field theoretic vacuum and their changes as parameterizing
changes of it with respect to some reference situation.\\

Summarizing the concept proposed in the present article
let us point out that it proposes a view on quantum field theory
which differs from the established one, but the established
standard paradigm finds it natural explanation and place
within this new approach. In particular, it incorporates
and continues in modified shape certain ideas used in
local renormalizable quantum field theory such as the
unobservability of bare quantities and the hypothesis
that the vanishing of the beta function(s) (corresponding to a
fixed point of the renormalization group) defines the
physical coupling constant(s) of a model. The functional
integral equation for the complete effective action
proposed ensures (merely by definition) that any of
its solutions is finite (It is not a solution, otherwise.).
This removes to a certain extent the concern of
divergencies standard quantum field theory is beset
by, but the price to pay for this is the present
uncertainty whether the functional integral equation
proposed has beyond free field theories any other
nontrivial solution (i.e., any nonlinear (interacting)
field theory). The most natural place to find out
whether the proposed concept is physically correct
should be QED because unlike some other model theories
it is a theory for phenomena definitely present in
nature. QED is certainly structurally more complex
than scalar model field theories, e.g., but if for QED
something new can be learned we may feel sure that
our physical understanding has advanced. The
approximative approach to the functional integral
equation for the QED effective action presented has proved
its calculational accessibility. Although the
particular approximation studied is
still quite simple it has yielded certain
nonperturbative information what indicates that
the present approach also has certain calculational
advantages. However, only further investigation will
show whether any obviously appropriate approximation can be found
which yields with reasonable calculational effort
the correct value of the fine structure constant. In
a certain sense this should be viewed as a crucial
test because in principle the present approach if really
physically correct and adequate should be able to pass it.\\

\newpage
\subsubsection*{Acknowledgements}

\parindent0.em
The author thanks D.\ Robaschik, E.\ Wieczorek and M.\ Bordag
for helpful discussions on the subject and on an earlier
version of the paper. Large part of the present investigation
has been performed in 1992 at the INTSEM, University of Leipzig, and
the author is grateful to A.\ Uhlmann who made this research possible.
Finally, kind hospitality during a stay at the
Naturwissenschaftlich-Theoretisches Zentrum (NTZ), University of Leipzig,
where this preprint has been produced is herewith
acknowledged.\\

\newpage

\setcounter{section}{1}
\section*{Appendix A}
\addcontentsline{toc}{section}{Appendix A}
\label{GS}
\renewcommand{\theequation}{\mbox{\Alph{section}.\arabic{equation}}}
\setcounter{equation}{0}

Consider the following formula

\parindent0.em
\begin{eqnarray}
\label{ZA1}
{\rm e}^{\displaystyle\ i \Delta\Gamma^G_I [A]}\ \ &=&\ C\
\int D\psi D\bar\psi\
\ {\rm e}^{\displaystyle\ i\Gamma_I^F [A,\psi,\bar\psi]}\ \ \ ,
\end{eqnarray}

where $\Gamma_I^F [A,\psi,\bar\psi]$ is given by eq.\ (\ref{D3}).
In the present Appendix we are going to calculate the
coefficients of the first two quadratic terms of the
derivative expansion of $\Delta\Gamma^G_I [A]$, i.e.,
the coefficient of the mass term $A_\mu A^\mu$ and the
coefficients of $(\partial_\mu A^\mu)^2$ and
$\partial_\mu A_\nu \partial^\mu A^\nu$. For this
purpose we rewrite $\Gamma_I^F [A,\psi,\bar\psi]$ in
the following symmetrized form.

\begin{eqnarray}
\label{ZA2}
&&\hspace{-1.cm}\Gamma_I^F [A,\psi,\bar\psi]\ =\nonumber\\
\vspace{0.3cm}\nonumber\\
&&\hspace{-1.cm}=\ {1\over2}\ \int d^4x\ d^4x^\prime\ \ \bar\psi (x)
\ \ {\rm e}^{\displaystyle\ ie \int^{x^\prime}_x dy_\mu\ A^\mu(y)}\ \
\cdot\nonumber\\
\vspace{0.3cm}\nonumber\\
&&\hspace{-0.2cm}\cdot\ \left[ a_I\left( x-x^\prime\right)
\ \left( i\stackrel{\rightarrow}{\not\hspace{-0.07cm}\partial_{x^\prime}}
- e \not\hspace{-0.13cm}A(x^\prime) \right)\
-\ m\ b_I\left( x-x^\prime\right) \right]\psi (x^\prime)\ + \nonumber\\
\vspace{0.3cm}\nonumber\\
&&\hspace{-0.3cm} +\ {1\over2}\ \int d^4x\ d^4x^\prime\ \ \bar\psi (x)
\ \left[ -\left( i
\stackrel{\leftarrow}{\not\hspace{-0.07cm}\partial_x}
+ e \not\hspace{-0.13cm}A(x) \right)\
\ a_I\left( x-x^\prime\right)\
-\ m\ b_I\left( x-x^\prime\right) \right]\ \
\cdot\nonumber\\
\vspace{0.3cm}\nonumber\\
&&\hspace{0.8cm}\cdot
\ {\rm e}^{\displaystyle\ ie \int^{x^\prime}_x dy_\mu\ A^\mu(y)}\
\psi (x^\prime)
\end{eqnarray}

We now expand the r.h.s.\ of eq.\ (\ref{ZA2}) in
powers of $A_\mu$ up to $O(A^2)$ (i.e., $O(e^2)$) and
insert following expansions (the upper obtained by using
$y_\mu(\tau)\ =\ (x^\prime - x)_\mu\ \tau + x_\mu$,
$\tau\ \in\ [0,1]$).

\begin{eqnarray}
\label{ZA3}
&&\hspace{-1.5cm}\int^{x^\prime}_x dy_\mu\ A^\mu(y)\ =\
(x^\prime - x)^\mu\ \bigg\{\ A_\mu(y)\ +\nonumber\\
&&\ +\ \left. {1\over 24}\ (x^\prime - x)^\nu
(x^\prime - x)^\lambda
\ \partial_\nu \partial_\lambda\ A_\mu(y)
\ +\ \ldots\ \right\}_{y={(x+x^\prime)\over 2}}\ \ \ ,\\
\vspace{0.3cm}\nonumber\\
\label{ZA4}
&&\hspace{-1.5cm} A_\mu (x)\ +\ A_\mu (x^\prime)\ =\ \nonumber\\
&&=\ 2\ \left\{\ A_\mu (y)\ +\
{1\over 8}\ (x^\prime - x)^\nu(x^\prime - x)^\lambda\
\partial_\nu \partial_\lambda\ A_\mu (y)
\ +\ \ldots\
\right\}_{y={(x+x^\prime)\over 2}}\  .
\end{eqnarray}

For calculating the coefficients of $A_\mu A^\mu$,
$(\partial_\mu A^\mu)^2$, and
$\partial_\mu A_\nu \partial^\mu A^\nu$ in $\Delta \Gamma_I^G [A]$
it is sufficient to keep at most
two derivatives acting on the gauge potentials
in $\Gamma_I^F [A,\psi,\bar\psi]$.
The expression obtained this way for
$\Gamma_I^F$ (we will not give
this rather long expression) now serves as the
starting point for deriving Feynman rules and
calculating the effective action terms desired.
One should take notice that $\Gamma_I^F$ also contains
terms quadratic in $A_\mu$ what leads to the
situation that besides the standard photon polarization
diagram also a tadpole contribution to the
photon self-energy is to be taken into account.\\

\parindent1.5em

The explicit calculation of the terms we are aiming
at is quite tedious and shall not be displayed here.
We only comment few points of the calculation.
Coordinate differences as occurring in eqs.\ (\ref{ZA3}),
(\ref{ZA4}) are translated into momentum space as
derivatives with respect to a corresponding momentum variable
acting on certain functions in momentum space. This of
course involves partial integrations in momentum space
for which as usual boundary contributions are assumed not to occur.
The photon polarization function is a nonlocal distribution. Therefore,
from the formal expression derived by the Feynman rules
the local structures we are interested in have to be
extracted. In order to properly define this procedure
we apply a (radial) momentum space UV cut-off at $\Lambda$
for the loop integration. The final result will be given within
this gauge non-invariant cut-off regularization.
Furthermore, a Wick rotation for the loop integration
is performed and such equivalences like (\ref{DF5}), (\ref{DF6}) are used.
Then, the final result reads
\parindent0.em

\begin{eqnarray}
\label{ZA5}
&&\hspace{-0.5cm}\Delta\Gamma^G_I [A]\ =\ const.\ +\
{e^2\over 16\pi^2}
\ \int d^4x\ \bigg\{\ C_0\ m^2 A_\mu(x) A^\mu(x)\ + \nonumber\\
\vspace{0.3cm}\nonumber\\
&&+\ \Big[\ C_{1s}\
[g_{\mu\nu} g_{\alpha\beta}\ +\ g_{\mu\alpha} g_{\nu\beta}]\ +\
C_{1a}\ [g_{\mu\nu} g_{\alpha\beta}\ -\ g_{\mu\alpha} g_{\nu\beta}]\
\Big]\ A^\mu(x) \partial^\alpha \partial^\beta A^\nu(x)\ +
\nonumber\\
\vspace{0.3cm}\nonumber\\
&&+\ \ldots \bigg\}
\end{eqnarray}

where ($h^\prime= d/ds\ h$)

\begin{eqnarray}
\label{ZA6}
C_0\ \ &=&\ -\ s^2\ h^\prime\ \ \Bigg\vert_0^{\Lambda^2\over m^2}\\
\vspace{1.2cm}\nonumber\\
\label{ZA7}
\hspace{-1.cm}C_{1s}\ &=&\ -\ {1\over 6}\ s^3\ h^{\prime\prime\prime}\ -\
{1\over 2}\ s^2\ h^{\prime\prime}\ +\nonumber\\
\vspace{0.2cm}\nonumber\\
&&\ +\ {1\over 2}\ \left(\ {\rm e}^{\displaystyle -h}\ \left[\
s^4\ \tilde a \tilde a^{\prime\prime}\ +\
2\ s^3\ \tilde a \tilde a^\prime\ +\
s^3\ \tilde b \tilde b^{\prime\prime}\ +\
s^2\ \tilde b \tilde b^\prime\ \right]\right)^\prime\ -\nonumber\\
\vspace{0.2cm}\nonumber\\
&&\ -\ {\rm e}^{\displaystyle -h}\
\left[\ {1\over 3}\ s^4\ \tilde a \tilde a^{\prime\prime\prime}\ +\
2\ s^3\ \tilde a \tilde a^{\prime\prime}\ +\
{1\over 3}\ s^3\ \tilde b \tilde b^{\prime\prime\prime}\ +\
2\ s^2\ \tilde a \tilde a^\prime\ +\right.\nonumber\\
&&\hspace{1.8cm}
+\left. {3\over 2}\ s^2\ \tilde b \tilde b^{\prime\prime}\ +\
s\ \tilde b \tilde b^\prime\ \right]\ \ \ \Bigg\vert_0^{\Lambda^2\over m^2}\\
\vspace{1.2cm}\nonumber\\
\label{ZA8}
C_{1a}\ &=&\ {1\over 18}\ s^3\ h^{\prime\prime\prime}\ -\
{1\over 6}\ s^2\ h^{\prime\prime}\ -\
{2\over 3}\ s\ h^\prime\ +\
{2\over 3}\ h\ +\nonumber\\
\vspace{0.2cm}\nonumber\\
&&\ +\ {1\over 2}\ \left(\ {\rm e}^{\displaystyle -h}\ \left[\
-\ {1\over 3}\ s^4\ \tilde a \tilde a^{\prime\prime}\ +\
{2\over 3}\ s^3\ \tilde a \tilde a^\prime\ -\
{1\over 3}\ s^3\ \tilde b \tilde b^{\prime\prime}\ +\
s^2\ \tilde b \tilde b^\prime\ \right]\right)^\prime\ +\nonumber\\
\vspace{0.2cm}\nonumber\\
&&\ +\ {\rm e}^{\displaystyle -h}\
\left[\ {1\over 9}\ s^4\ \tilde a \tilde a^{\prime\prime\prime}\ +\
{4\over 3}\ s^3\ \tilde a \tilde a^{\prime\prime}\ +\
{1\over 9}\ s^3\ \tilde b \tilde b^{\prime\prime\prime}\ +\
2\ s^2\ \tilde a \tilde a^\prime\ +\right.\nonumber\\
&&\hspace{1.8cm}+\left.\
{7\over 6}\ s^2\ \tilde b \tilde b^{\prime\prime}\ +\
2\ s\ \tilde b \tilde b^\prime\ \right]\ \ \
\Bigg\vert_0^{\Lambda^2\over m^2}\ - \nonumber\\
&&\ -\ \int\limits_0^{\Lambda^2\over m^2} ds\
{1\over s \tilde a^2 + \tilde b^2}\
\left[\ {s\ \tilde a^2 \over s \tilde a^2 + \tilde b^2}\
\left[\ s\ \tilde a \tilde a^\prime + \tilde b \tilde b^\prime\ \right]
\ +\right.\nonumber\\
\vspace{0.2cm}\nonumber\\
&&\hspace{1.5cm}
+\ {2\over 3}\ s^3\ \tilde a \tilde a^{\prime\prime\prime}\ +\
3\ s^2\ \tilde a \tilde a^{\prime\prime}\ +\
{2\over 3}\ s^2\ \tilde b \tilde b^{\prime\prime\prime}\ +\nonumber\\
&&\hspace{1.5cm}+\ 2\ s\ \tilde a \tilde a^\prime\ +\
3\ s\ \tilde b \tilde b^{\prime\prime}\ -\
s\ (\tilde b^\prime)^2\ +\
3\ \tilde b \tilde b^\prime\ \Bigg] \ \ \ ,\\
\vspace{0.3cm}\nonumber\\
h\ &=&\ h(s)\ =\ \ln\left[ s \tilde a^2 +\tilde b^2 \right]
\hspace{0.5cm},\hspace{0.5cm}\tilde a\ = \tilde a (s)
\hspace{0.5cm},\hspace{0.5cm}\tilde b\ = \tilde b (s)\ \ \ .\nonumber
\end{eqnarray}

For convenience, in the equations we have omitted
the index $I$ for $\tilde a$ and $\tilde b$. The
result given above is exact for any value of the cut-off $\Lambda$,
so far no term vanishing at removing the cut-off has been
neglected. A comparison of the mass term with eq.\ (\ref{DF7})
shows that both results although obtained by different methods
agree as expected. Also the first line of eq.\ (\ref{ZA7}) can
be re-identified in eq.\ (\ref{DF7}). For
$\tilde a\ =\ \tilde b\ \equiv\ 1$ the standard QED result
is reproduced (cf.\ \cite{ach}; \cite{jau},
eq.\ (9-64), for $\Lambda \longrightarrow \infty$ the coefficient $C(0)$
there is related to our expressions by the equation
$C(0) = - e^2\ (5 C_{1s} + 3 C_{1a})/24\pi^2$).\\
\parindent1.5em

Now, if conditions (\ref{DF9}), (\ref{DF10}) are fulfilled
above result significantly simplifies. Then, the UV cut-off
can be lifted without any problem ($\Lambda\longrightarrow\infty$),
the coefficients $C_0$ and $C_{1s}$ connected with terms
spoiling gauge invariance are vanishing and the final
completely gauge invariant result reads
\parindent0.em

\begin{eqnarray}
\label{ZA9}
\hspace{-0.5cm}\Delta\Gamma^G_I [A]\ &=&\ const.\ +\nonumber\\
&&\ \ \ +\ \ C_{1a}\ \ {e^2\over 16\pi^2}
\ \int d^4x \  A^\mu(x)\
[g_{\mu\nu} \Box\ -\ \partial_\mu \partial_\nu]\ A^\nu(x)\ +
\ \ldots\ ,\
\end{eqnarray}

with

\begin{eqnarray}
\label{ZA10}
C_{1a}\ &=&\ {2\over 3}\ \ln\ \left[
{\tilde b (\infty)\over\tilde b(0)} \right]^2\ - \nonumber\\
\vspace{0.3cm}\nonumber\\
&&\ \ \ -\ \int\limits_0^\infty ds\
{1\over s \tilde a^2 + \tilde b^2 }\
\left[\ {s\ \tilde a^2 \over s \tilde a^2 + \tilde b^2}\
\left[\ s\ \tilde a \tilde a^\prime + \tilde b \tilde b^\prime\ \right]
\ +\right.\nonumber\\
\vspace{0.2cm}\nonumber\\
&&\hspace{1.5cm}
+\ {2\over 3}\ s^3\ \tilde a \tilde a^{\prime\prime\prime}\ +\
3\ s^2\ \tilde a \tilde a^{\prime\prime}\ +\
{2\over 3}\ s^2\ \tilde b \tilde b^{\prime\prime\prime}\ +\nonumber\\
&&\hspace{1.5cm}+\ 2\ s\ \tilde a \tilde a^\prime\ +\
3\ s\ \tilde b \tilde b^{\prime\prime}\ -\
s\ (\tilde b^\prime)^2\ +\
3\ \tilde b \tilde b^\prime\ \Bigg] \ \ \ .
\end{eqnarray}

It is worth noting that the coefficient $C_{1a}$ is finite
due to conditions (\ref{DF9}), (\ref{DF10}). Gauge invariance
and UV finiteness are closely related here \footnote{Were it
not for the first term ($\sim (s \tilde a^2 + \tilde b^2)^{-2}$)
in the integral in eqs.\ (\ref{ZA8}),
(\ref{ZA10}), also the weaker condition given in footnote \ref{foot1} on
p.\ \pageref{foot1} then replacing (\ref{DF9})
would lead to gauge invariance and
UV finiteness at the same time.}
(For a further discussion see \cite{scharn2}.).\\

\newpage

\setcounter{section}{2}
\section*{Appendix B}
\addcontentsline{toc}{section}{Appendix B}
\label{HS}
\setcounter{equation}{0}

\parindent0.em
In this Appendix we explicitly calculate the function

\begin{eqnarray}
\label{ZB1}
g(x - x^\prime)\ =\
{\rm e}^{\displaystyle\ -{i\over 2}\int d^4y\ d^4y^\prime\
\bar{J}_\mu (x,x^\prime;y)\ D_I^{\mu\nu}(y - y^\prime)\
\bar{J}_\nu (x,x^\prime;y^\prime)}
\end{eqnarray}

for the Ansatz (\ref{DG5})

\begin{eqnarray}
\label{ZB2}
d_I(x)\ &=&\
\left[\ 1 + \beta\ {\Box\over m^2}\ \right]\ \delta^{(4)}(x)
\ \ \ .
\end{eqnarray}

Eq.\ (\ref{ZB1}) can easily be rewritten as
\footnote{Of course, this transformation is not specific
to the Ansatz (\ref{ZB2}).
To obtain eq.\ (\ref{ZB3}) a gauge fixing term $\Gamma_{gf}$
with $\tilde n_\mu = i p_\mu\ \tilde d_I(p)^{1/2}$ has been
added to the gauge field action $\Gamma^G_I$.}

\begin{eqnarray}
\label{ZB3}
g(x - x^\prime)\ &=&\ \ \
{\rm e}^{\displaystyle\ -\ i e^2\ (x - x^\prime)^2\
\int_0^1 d\tau\ (1-\tau)\ D_I((x - x^\prime)\ \tau)}\ \cdot\nonumber\\
\vspace{0.3cm}\nonumber\\
&&\ \ \cdot\ {\rm e}^{\displaystyle\ \ i e^2\ (1-\lambda)\
\left[\ D^\ast_I(x - x^\prime ) - D^\ast_I(0)\ \right]}\ \ \ \ ,
\end{eqnarray}

where \footnote{The IR divergency can
be regularized and drops then out for $g(x)$. The
spurious pole generated by the model Ansatz
$\tilde d_I(p)$ is understood as also supplied
with the $i\epsilon$-prescription.}

\begin{eqnarray}
\label{ZB4}
D^\ast_I(x)\ &=&\ \int {d^4p\over(2\pi)^4}\
{{\rm e}^{\ ipx}\over (p^2\ + \ i\epsilon)^2}\
{1\over \tilde d_I(p)}\ \ \ ,\ \\
\vspace{0.3cm}\nonumber\\
&&\hspace{3.cm}\tilde d_I(p)\ =\ 1\ -\ \beta\ {p^2\over m^2}\ \ \ ,\
\nonumber
\end{eqnarray}

and

\begin{eqnarray}
\label{ZB5}
D_I(x)\ &=&\ \Box\ D^\ast_I(x)\hspace{2.cm}.
\end{eqnarray}

For simplicity, let us perform the calculation
for $g$ in Euclidean space. Results then can
be read off for Minkowski space whenever needed
by rotating back the fourth coordinate. In
Euclidean space $D^\ast_I$ and
$D_I$ read

\begin{eqnarray}
\label{ZB6}
D^\ast_I(x_E)\ &=&\
-\ {i\over 16\pi^2}\ \ln\left(\mu^2 x_E^2\right)\ -\
{\beta\over m^2}\ D_I(x_E)
\end{eqnarray}

($\mu^2$ is the temporary IR cut-off applied), and

\begin{eqnarray}
\label{ZB7}
D_I(x_E)\ &=&\ {i\over 4\pi^2\ x_E^2}\ -\
{i\ m\over 4\pi^2\ \sqrt{\beta}\ \vert x_E \vert}\
K_1\left( {m \vert x_E \vert\over\sqrt{\beta}}\right)\ \ \ .
\end{eqnarray}

For the further calculation following integral turns out
to be useful (${\bf L}_\nu$ are Struve functions) \cite{g}, vol.\ 2.

\begin{eqnarray}
\label{ZB8}
&&\hspace{-1.2cm}\int {d\tau\over \tau}\ K_1(\tau)\ =\nonumber\\
\vspace{0.3cm}\nonumber\\
&&\hspace{-0.3cm}=\
-\ K_1(\tau)\ -\ \tau\ K_0(\tau)\ -\ {\pi\over 2}\ \tau\
\left[\ K_1(\tau)\ {\bf L}_0(\tau)\ +\ K_0(\tau)\ {\bf L}_1(\tau)\ \right]
\end{eqnarray}

Consequently, we find ($\gamma$ is the Euler constant)

\begin{eqnarray}
\label{ZB9}
&&\hspace{-1.2cm}-\ x_E^2\ \int_0^1 d\tau\
(1-\tau)\ D_I(x_E\ \tau)\ =\nonumber\\
\vspace{0.3cm}\nonumber\\
&&\hspace{-0.3cm}=\ {i\ m\over 4\pi^2}\ \left\{\ 1\ +\ \gamma\ +\
{1\over 2}\ \ln\left({m^2 x_E^2\over 4\beta}\right)\ +
\right.\nonumber\\
\vspace{0.3cm}\nonumber\\
&&\hspace{-0.1cm}\ \ +\ \left( 1\ -\ {m^2 x_E^2\over \beta}\right)\
K_0\left( {m \vert x_E \vert\over\sqrt{\beta}}\right)\ -\
{m \vert x_E \vert\over\sqrt{\beta}}\
K_1\left( {m \vert x_E \vert\over\sqrt{\beta}}\right)\
-\ {\pi\over 2}\ {m^2 x_E^2\over\beta}\ \cdot\nonumber\\
\vspace{0.3cm}\nonumber\\
&&\hspace{0.4cm}\ \cdot\ \left.\left[\
K_1\left( {m \vert x_E \vert\over\sqrt{\beta}}\right)\
{\bf L}_0\left( {m \vert x_E \vert\over\sqrt{\beta}}\right)\ + \
K_0\left( {m \vert x_E \vert\over\sqrt{\beta}}\right)\
{\bf L}_1\left( {m \vert x_E \vert\over\sqrt{\beta}}\right)\
\right]\right\}\ \ .\
\end{eqnarray}

The final result for $g(x_E)$ is then
($\ t = m \vert x_E \vert /\sqrt{\beta}\ $)

\begin{eqnarray}
\label{ZB10}
&&\hspace{-1.3cm}g(x_E)\ =\ \nonumber\\
\vspace{0.3cm}\nonumber\\
&&\hspace{-0.9cm}=\ \exp\left\{\ {\alpha\over\pi}\
\left[\ 1\ +\ \gamma\ +\ {1\over 2}\ \ln\ {t^2\over 4}\ +\
(1-t^2)\ K_0(t)\ -\ t\ K_1(t)\ - \right.\right.\nonumber\\
\vspace{0.3cm}\nonumber\\
&&\hspace{2.5cm}\left. -\ {\pi\over 2}\ t^2\
\left[\ K_1(t)\ {\bf L}_0(t)\ +\ K_0(t)\ {\bf L}_1(t)\ \right]\ \right]\
\ +\nonumber\\
\vspace{0.3cm}\nonumber\\
&&\hspace{0.4cm}\left. +\ {\alpha\over\pi}\
(1-\lambda)\ \left[\ {1\over t^2}\ -\ {1\over t}\ K_1(t)\ +\
{1\over 4}\ (2\gamma\ -\ 1)\ +\ {1\over 4}\ \ln\ {t^2\over 4}\
\right]\ \right\} \ .\
\end{eqnarray}

In the long distance limit ($t \gg 1$) eq.\ (\ref{ZB10}) reads

\begin{eqnarray}
\label{ZB11}
g(x_E)\ &=&\ \exp\left\{\ {\alpha\over 2\pi}\
\left[\ -\ \pi\ t\ +\ {(3-\lambda)\over 2}\ \ln\ {t^2\over 4}\
\right.\right. +\nonumber\\
\vspace{0.3cm}\nonumber\\
&&\hspace{3.cm}\left.\left. +\ {(3+\lambda)\over 2}\ +\
(3-\lambda)\ \gamma\ +\ \ldots\ \right]\right\}\ \ \ .\
\end{eqnarray}

So, in the long distance region we are mainly
interested in the function $g(x_E)$ can be written
as follows.

\begin{eqnarray}
\label{ZB12}
g(x_E)&=&C_g\ \left(m^2 x_E^2\right)^{\displaystyle\
\alpha (3-\lambda)/4\pi}\ \
{\rm e}^{\displaystyle\ -\ {\alpha\over 2 \sqrt{\beta}}
\ m \vert x_E \vert}\ \ \Big[\ 1\ + \ \ldots\ \Big]\\
\vspace{0.3cm}\nonumber\\
C_g&=&\left( 4\beta\right)^{\displaystyle\
- \alpha (3-\lambda)/4\pi}\ \
\exp\left\{\ {\alpha\over 4\pi}\left[\ (3+\lambda)\ +\
2\ (3-\lambda)\ \gamma\ \right]\ \right\}
\end{eqnarray}

One easily recognizes in eq.\ (\ref{ZB12}) the
well-known exponent of the (power-like) Bloch-Nordsieck
contribution (cf.\ \cite{h},\cite{i} and references therein).\\

\newpage
\addcontentsline{toc}{section}{Figures}
\label{JS}
\parindent0.em

$w(\alpha)$

\vspace{16cm}
\hspace{14.cm}
$\alpha$

\vspace{1.5cm}

{\bf Fig.\ 1}: Solution $w$ of eq.\ (\ref{DR16}) as function of $\alpha$

\vfill
\newpage

$G(\alpha)$

\vspace{16cm}
\hspace{14.cm}
$\alpha$

\vspace{1.5cm}

{\bf Fig.\ 2}: Solution $G$ of eq.\ (\ref{DR24}) (with $w$ as solution
of eq.\ (\ref{DR16}) inserted) as function of $\alpha$

\vfill
\newpage

$- s_0(\alpha)$

\vspace{16cm}
\hspace{14.cm}
$\alpha$

\vspace{1.5cm}

{\bf Fig.\ 3}: Solution $s_0$ of eq.\ (\ref{DR30}) as function of $\alpha$.
The dashed line is located at $\alpha_c \simeq 0.70$ corresponding
to the singularity
$- s_0 \rightarrow \infty$. It separates the regions where the
upper ($\alpha < \alpha_c$) and the lower ($\alpha > \alpha_c$)
sign in the normalization condition (\ref{DR26}) is applied
respectively. Beyond $\alpha_{max} \simeq 2.64$ eq.\
(\ref{DR30}) does not have any solution as one recognizes from the
dotted line drawn at $- s_0 = 1$.

\vfill
\newpage

$\beta(\alpha)$

\vspace{16cm}
\hspace{14.cm}
$\alpha$

\vspace{1.5cm}

{\bf Fig.\ 4}: The parameter $\beta$ as function of $\alpha$ (cf.\ eq.\
(\ref{DR33})). For further comments see fig.\ 3.

\vfill
\newpage

$\tilde b(\infty)$

\vspace{16cm}
\hspace{14.cm}
$\alpha$

\vspace{1.5cm}

{\bf Fig.\ 5}: The parameter $\tilde b(\infty)$ as function of $\alpha$
(cf.\ eq.\ (\ref{DR36})).
Note, that $\tilde b(\infty)$ is close to $2$ for small $\alpha$.
For further comments see fig.\ 3.

\vfill
\newpage

$M(s)$

\vspace{16cm}
\hspace{14.cm}
$s$

\vspace{1.5cm}

{\bf Fig.\ 6}: Typical behaviour of the integrand $M(s)$ in eq.\
(\ref{DL2a}) for small arguments where eqs.\ (\ref{DR38}),
(\ref{DR39}) are inserted ($M(s)$ is drawn here for
$\alpha = 0.05$.). The zero (we denote it by $s_x$) of
the function $M(s)$ is understood as defining the
applicability region of the low $s$ representation
(\ref{DR38}), (\ref{DR39}). $s_x$ as function of $\alpha$
is shown in fig.\ 7.

\vfill
\newpage

$s_x(\alpha)$

\vspace{16cm}
\hspace{14.cm}
$\alpha$

\vspace{1.5cm}

{\bf Fig.\ 7}: Zero $s_x$ of $M(s)$ as function of $\alpha$ (see
fig.\ 6). The dashed line is drawn at $\alpha_c$ while the
dotted line is located at $\alpha \simeq 1.55$. In both
cases one finds numerically $s_x \rightarrow \infty$.

\vfill
\newpage

$C_{1a}(\alpha)$

\vspace{16cm}
\hspace{14.cm}
$\alpha$

\vspace{1.5cm}

{\bf Fig.\ 8}: The coefficient $C_{1a}$ (see eq.\ (\ref{DL2a}))
as function of $\alpha$. Above curve is the difference
of the contributions represented by the
curves 1 and 2 shown in fig.\ 9 (Curve 1 stands for the
first term in eq.\ (\ref{DL2a}) while curve 2 is the
contribution of the integral.). Please note that $C_{1a}$
is a completely smooth function at $\alpha = \alpha_c$
(dashed line) although certain parameters involved
(see figs.\ 3, 5) are singular there.

\vfill
\newpage

\ \\

\vspace{16cm}
\hspace{14.cm}
$\alpha$

\vspace{1.5cm}

{\bf Fig.\ 9}: Contributions to $C_{1a}$ as functions of $\alpha$.
For a further explanation see fig.\ 8.

\vfill

\end{document}